\documentclass[paperpaper,superscriptaddress,english,floatfix, showpacs, amsfonts, amssymb]{revtex4}
\usepackage{epsfig,graphicx,psfrag,amsmath,amssymb,float}
\usepackage[T1]{fontenc}
\usepackage[latin9]{inputenc}
\usepackage{amsmath}
\usepackage{amssymb}
\usepackage{bbold}
\usepackage{amscd}
\usepackage{bm}
\usepackage{psfrag}
\usepackage{epsfig}
\usepackage{graphicx}
\usepackage{dcolumn}
\usepackage{bm}
\usepackage{subfigure}

\usepackage{babel}

\begin{document}


\title{Bethe strings in the dynamical structure factor of the spin-$1/2$ Heisenberg $XXX$ chain}
\author{Jos\'e M. P. Carmelo}
\affiliation{Center of Physics of University of Minho and University of Porto, P-4169-007 Oporto, Portugal}
\affiliation{Department of Physics, University of Minho, Campus Gualtar, P-4710-057 Braga, Portugal}
\affiliation{Boston University, Department of Physics, 590 Commonwealth Ave, Boston, MA 02215, USA}
\author{Tilen \v{C}ade\v{z}}
\affiliation{Center for Theoretical Physics of Complex Systems, Institute for Basic Science (IBS), Daejeon 34126, Republic of Korea}
\author{Pedro D. Sacramento}
\affiliation{CeFEMA, Instituto Superior T\'ecnico, Universidade de Lisboa, Av. Rovisco Pais, P-1049-001 Lisboa, Portugal}

\date{11 August 2020}


\begin{abstract}
Recently there has been a renewed interest in
the spectra and role in dynamical properties of excited states of the spin-$1/2$ Heisenberg antiferromagnetic chain
in longitudinal magnetic fields associated with Bethe strings. The latter are bound states of elementary magnetic excitations
described by Bethe-ansatz complex non-real rapidities. Previous studies on this problem referred to
finite-size systems. Here we consider the thermodynamic limit and study it for the isotropic spin-$1/2$ 
Heisenberg $XXX$ chain in a longitudinal magnetic field. We confirm that also in that limit the most significant spectral 
weight contribution from Bethe strings leads to $(k,\omega)$-plane gapped continua in the spectra of the 
spin dynamical structure factors $S^{+-} (k,\omega)$ and $S^{xx} (k,\omega)=S^{yy} (k,\omega)$. 
The contribution of Bethe strings to $S^{zz} (k,\omega)$ is found to be small at low spin 
densities $m$ and to become negligible upon increasing that density above $m\approx 0.317$. For
$S^{-+} (k,\omega)$, that contribution is found to be negligible at finite magnetic field. We derive analytical expressions 
for the line shapes of $S^{+-} (k,\omega)$, $S^{xx} (k,\omega)=S^{yy} (k,\omega)$, and $S^{zz} (k,\omega)$ 
valid in the $(k,\omega)$-plane vicinity of singularities located at and just above the gapped lower thresholds 
of the Bethe-string states's spectra. As a side result and in order to provide an overall physical picture that includes the
relative $(k,\omega)$-plane location of all spectra with a significant amount
of spectral weight, we revisit the general problem of the line-shape 
of the transverse and longitudinal spin dynamical structure factors at finite magnetic field and excitation 
energies in the $(k,\omega)$-plane vicinity of other singularities. This includes those located at and just 
above the lower thresholds of the spectra that stem from excited states described by only real Bethe-ansatz rapidities.
\end{abstract}


\maketitle

\section{Introduction}
\label{SECI}

Recently, there has been a renewed interest in bound states of elementary magnetic excitations named Bethe strings
known long ago \cite{Bethe_31,Takahashi_71,Gaudin_71,Taka-AN,Gaudin_14}. In spite of Bethe strings being
a rather theoretical issue, as they were first found and identified within the Bethe ansatz solution of spin and electronic integrable models
in some classes of energy eigenstates described by complex non-real spin rapidities \cite{Bethe_31,Takahashi_71}, 
this renewed interest is actually due, in part, to their experimental identification and realization in spin-chain compounds
whose magnetic properties are described by the spin-$1/2$ Heisenberg chain in longitudinal magnetic fields \cite{Bera_20,Wang_18,Kohno_09,Kohno_10,Stone_03}.
This applies to that model isotropic point in the case of experimental studies of some classes of such compounds \cite{Kohno_09,Kohno_10,Stone_03,Heilmann_78}.

The present paper addresses only theoretical issues of that interesting physical problem.
Most previous studies on the spin dynamical properties of the spin-$1/2$ $XXX$ chain 
in a longitudinal magnetic field focused on the contribution 
from energy eigenstates described by real Bethe-ansatz rapidities, which are 
associated with most spectral weight of the spin dynamical structure factors. Several such studies considered finite-size systems 
and relied on different methods. This includes for instance numerical diagonalizations \cite{Lefmann-96} and evaluation of matrix 
elements between Bethe-ansatz states \cite{Muller,Karbach_00,Karbach_02}. Previous studies that
considered the thermodynamic limit \cite{Carmelo_15A}, were also limited to the contribution to
the spin dynamical structure factors from energy eigenstates described by real Bethe-ansatz rapidities.
Concerning the specific issue of the contribution of Bethe strings to the spin dynamical properties of 
spin-$1/2$ $XXX$ chain in a longitudinal magnetic field, the few previous studies considered finite-size 
systems \cite{Kohno_09,Kohno_10}. 

In the case of that spin-$1/2$ chain, Bethe strings \cite{Bethe_31}, which here we call $n$-strings,
have for $n>1$ and in the thermodynamic limit \cite{Takahashi_71} been shown to be bound states of $n=2,...,\infty$ singlet pairs 
of the model physical spins $1/2$ \cite{Carmelo_15,Carmelo_17,Carmelo_18}. (The physical meaning of the
form of the spin-$1/2$ $XXX$ chain's $n$-strings in that limit is an issue shortly further discussed below 
in Sec. \ref{SECIIIA}.) Energy eigenstates described by only real Bethe-ansatz rapidities lack such bound pairs and 
are populated by unbound singlet pairs of such physical spins \cite{Carmelo_15,Carmelo_17,Carmelo_18}. On the other hand, 
there are predictions according to which for the {\it large} spin-$S$ Heisenberg $XXX$ chain in longitudinal magnetic fields, 
Bethe strings could rather be bound states of spin-$1$ magnons \cite{Johnson_86,Dally_20}. 

In this paper we address the problem of the contribution of $n$-strings to the spin dynamical properties of 
spin-$1/2$ $XXX$ chain in a longitudinal magnetic field, in the thermodynamic limit. Based on a relation between the
level of negativity of the momentum dependent exponents that control the $(k,\omega)$-plane line shape of
the spin dynamical structure factors near singularities and
the amount of spectral weight existing in their vicinity, respectively, 
we confirm that in the thermodynamic limit, as in the case of finite-size systems 
\cite{Kohno_09,Kohno_10}, the only contribution from excited energy eigenstates populated by $n$-strings that
leads to a $(k,\omega)$-plane gapped continuum in the spectrum of the spin dynamical structure factors refers to
$S^{+-} (k,\omega)$ and thus also to $S^{xx} (k,\omega)=S^{yy} (k,\omega)$. 
On the other hand, the contribution from $n$-strings states to $S^{zz} (k,\omega)$ is found to 
be small at low spin densities and to become negligible upon increasing it beyond a spin density, $\tilde{m}\approx 0.317$.
For the spin dynamical structure factor $S^{-+} (k,\omega)$, that contribution is found to be negligible at 
any finite magnetic field. 

The main goal of this paper is thus the study of the line shape of the spin dynamical structure factors 
$S^{+-} (k,\omega)$, $S^{xx} (k,\omega)$, and $S^{zz} (k,\omega)$
at and just above singularities located at the $(k,\omega)$-plane gapped lower thresholds 
of the spectra associated with $n$-string states. To reach that goal, we extend the dynamical 
theory of Ref. \cite{Carmelo_15A} to a larger subspace, which allows to 
account for the contribution from the latter states to the spin dynamical structure factors.
We then derive analytical expressions valid in the thermodynamic limit for line shape of these 
factors in the $(k,\omega)$-plane vicinity of the singularities under consideration. 

Complementarily and as a side result, in order to to provide an overall physical picture that includes the
relative $(k,\omega)$-plane location of all features with a significant amount
of spectral weight, we account for the contributions from all
types of states that lead to gapped and gapless lower threshold singularities in the spin dynamical structure factors.
This includes both excited states with and without $n$-strings.
(As mentioned above, the contribution from the latter states, shortly revisited in this paper, is known to lead to the largest amount of
spin dynamical structure factors's spectral weight \cite{Lefmann-96,Muller,Karbach_00,Karbach_02,Carmelo_15A}.)

The paper is organized as follows. The model and the spin dynamical structure factors are the subjects
of Sec. \ref{SECII}. In Sec. \ref{SECIII} the spectral functionals that control the extended dynamical theory's general 
expressions of the dynamical structure factors are introduced. Such factors's spectra are studied
in Sec. \ref{SECIV}. The line shape near their singularities is the
issue addressed in Sec. \ref{SECV}. The subject of Sec. \ref{SECVI} is the limiting behaviors of the spin dynamical 
structure factors. Finally, the discussion and concluding remarks are
presented in Sec. \ref{SECVII}. Two Appendices provide useful side information needed for
the studies of this paper.

\section{The model and the spin dynamical structure factors}
\label{SECII}

The spin-$1/2$ Heisenberg $XXX$ chain with exchange integral $J$ and 
length $L\rightarrow\infty$ in a longitudinal magnetic field $h$ for spin densities 
$m\in ]0,1[$, which describes $N=\sum_{\sigma =\uparrow,\downarrow}N_{\sigma}$ physical spins $1/2$ of projection
$\sigma$, is a paradigmatic example of an integrable strongly correlated system \cite{Bethe_31,Takahashi_71}. 
Its Hamiltonian is given by,
\begin{equation}
\hat{H} = J\sum_{j=1}^{L}\sum_{a =x,y,z}{\hat{S}}_j^{a}{\hat{S}}_{j+1}^{a} 
+ 2\mu_B h\,{\hat{S}}^{z} \, .
\label{HXXX}
\end{equation}
For simplicity, we have taken here $g=2$, $\hat{\vec{S}}_{j}$ is the spin-$1/2$ operator at site $j=1,...,N$ with 
components $\hat{S}_j^{x,y,z}$, $\mu_B$ is the Bohr magneton, and ${\hat{S}}^{z}=\sum_{j=1}^{N}\hat{S}^{z}_j$ is the diagonal 
generator of the global spin $SU(2)$ symmetry algebra. We denote the energy eigenstate's spin projection by 
$S^{z}=-(N_{\uparrow} -N_{\downarrow})/2\in [-S,S]$ where $S\in [0,N/2]$ is their spin. Units of lattice 
spacing and Planck constant one are used in this paper.

Due to the rotational symmetry in spin space, off-diagonal components of the spin dynamical structure factors vanish, 
$S^{aa'} (k,\omega) = 0$ for $a\neq a'$ where $a$ and $a'$ are given by $x,y,z$. In addition, 
the two transverse components are identical, $S^{xx} (k,\omega)=S^{yy} (k,\omega)$.
In the present case of finite magnetic fields $0<h<h_c$, one has that $S^{zz} (k,\omega)\neq S^{xx} (k,\omega)$.
Here $h_c=J/\mu_B$ is the magnetic field above which there is fully polarized ferromagnetism. 
The corresponding magnetic energy scale, $2\mu_B\,h_c = 2J$, is associated with the quantum phase 
transition to fully polarized ferromagnetism. In the opposite limit of zero magnetic field, one has 
that $S^{zz} (k,\omega)=S^{xx} (k,\omega)$.

The dynamical structure factors $S^{aa} (k,\omega)$ are given by,
\begin{eqnarray}
S^{aa} (k,\omega) & = & \sum_{j=1}^N e^{-ik j}\int_{-\infty}^{\infty}dt\,e^{-i\omega t}\langle GS\vert\hat{S}^{aa}_j (t)\hat{S}^{a}_j (0)\vert GS\rangle 
\nonumber \\
& = & \sum_{\nu}\vert\langle \nu\vert\hat{S}^{a}_k\vert GS\rangle\vert^2
\delta (\omega - \omega^{\tau}_{\nu} (k)) \hspace{0.20cm}{\rm for}\hspace{0.20cm}a =x,y,z \,  .
\label{SDSF}
\end{eqnarray}
Here the spectra read $\omega^{aa}_f (k) = (E_{\nu}^{aa} - E_{GS})$, $E_{\nu}^{aa}$ refers to
the energies of the excited energy eigenstates that contribute to the $aa= xx,yy,zz$
dynamical structure factors, $\sum_{\nu}$ is the sum over such states,
$E_{GS}$ is the initial ground state energy, and $\hat{S}^{a}_k$ are for $a =x,y,z$ the 
Fourier transforms of the usual local $a =x,y,z$ 
spin operators $\hat{S}^{a}_j$, respectively. 

The spin dynamical structure factor $S^{xx} (k,\omega)$ can be expressed as,
\begin{equation}
S^{xx} (k,\omega) = {1\over 4}\left(S^{+-} (k,\omega)+S^{-+} (k,\omega)\right) \, .
\label{xxPMMP}
\end{equation}
One can then address the dynamical properties of $S^{xx} (k,\omega)$
in terms of those of $S^{+-} (k,\omega)$ and $S^{-+} (k,\omega)$.

Since $S^{aa} (k,\omega)=S^{aa} (-k,\omega)$ for $a=x,y,z$ and thus also
$S^{+-} (k,\omega)=S^{+-} (-k,\omega)$ and $S^{-+} (k,\omega)=S^{-+} (-k,\omega)$,
in this paper we consider excitation momentum values $k>0$ in the first Brillouin zone, $k\in [0,\pi]$. 
Another useful symmetry relating the spin density intervals $m\in ]-1,0]$ and $m\in ]0,1[$ is such that,
\begin{eqnarray}
S^{-+} (k,\omega)\vert_m & = & S^{+-} (k,\omega)\vert_{-m}
\hspace{0.20cm}{\rm and}
\nonumber \\
S^{+-} (k,\omega)\vert_m & = & S^{-+} (k,\omega)\vert_{-m}
\hspace{0.20cm}{\rm for}\hspace{0.20cm}m \in ]0,1[ \,  .
\label{SPMMPmm}
\end{eqnarray}
Hence, as mentioned above, we only consider explicitly the spin density interval $m = 2S^z/N \in ]0,1[$.
The subspace defined below in Sec. \ref{SECIIIA} of the quantum problem studied in this
paper is spanned by some classes of energy eigenstates with spin $S\in ]0,N/2[$ 
and magnetic fields $0<h<h_c$ for which the spin density belongs to the interval $m \in ]0,1[$.
($N$ is even and odd when the states spin $S$ is an integer and half-odd integer number, respectively.
In the latter case, the minimal spin value is $1/2$, rather than $0$.) 

Some useful selection rules tell us which classes of energy eigenstates have nonzero matrix elements with the 
ground state. Let $\vert S,\alpha\rangle$, $\vert S^z,\beta\rangle$, and $\vert S,S^z,\gamma\rangle$
denote energy eigenstates where $S \in [0,N/2]$ is their spin, $S^z$ their spin projection, and
$\alpha$, $\beta$ and $\gamma$ represent all other quantum numbers needed to uniquely specify these 
states, respectively. The selection rules given in the following are derived from the properties
of the operators $\hat{S}^{z}_k$ and $\hat{S}^{\pm}_k$ by straightforward manipulations involving
their operator algebra \cite{Muller}.

At vanishing magnetic field, $h=0$, the following selection rules hold in the thermodynamic limit,
\begin{eqnarray}
\langle S,\alpha\vert\hat{S}^a_k\vert S'\alpha'\rangle & = & 0 
\hspace{0.20cm}{\rm for}\hspace{0.20cm}S=S'=0\hspace{0.20cm}{\rm and}\hspace{0.20cm}
a = z, \pm
\nonumber \\
\langle S,\alpha\vert\hat{S}^a_k\vert S'\alpha'\rangle & = & 0 
\hspace{0.20cm}{\rm for}\hspace{0.20cm}\vert S-S'\vert \neq 0,1 \hspace{0.20cm}{\rm and}\hspace{0.20cm}
a = z, \pm
\nonumber \\
\langle S^z,\beta\vert\hat{S}^{\pm}_k\vert S^{z'},\beta'\rangle & = & 0 
\hspace{0.20cm}{\rm for}\hspace{0.20cm}S^{z'}\neq S^z \pm 1
\nonumber \\
\langle S^z,\beta\vert\hat{S}^z_k\vert S^{z'},\beta'\rangle & = & 0 
\hspace{0.20cm}{\rm for}\hspace{0.20cm}S^{z'}\neq S^z \, .
\label{SRh0}
\end{eqnarray}

On the other hand, for finite magnetic fields $0<h<h_c$ of most interest for our study, the following selection rules 
are valid in that limit,
\begin{eqnarray}
\langle S,S,\gamma\vert\hat{S}^{\pm}_k\vert S',S^{z'},\gamma'\rangle & = & 0 
\hspace{0.20cm}
{\rm for}\hspace{0.20cm}S'\neq S\pm 1 \hspace{0.20cm}{\rm and}\hspace{0.20cm}S^{z'}\neq S\pm 1
\nonumber \\
\langle S,S,\gamma\vert\hat{S}^z_k\vert S',S^{z'},\gamma'\rangle & = & 0 
\hspace{0.20cm}
{\rm for}\hspace{0.20cm}S'\neq S\hspace{0.20cm}{\rm and}\hspace{0.20cm}S^{z'}\neq S \, .
\label{SRhfinite}
\end{eqnarray}

The spin dynamical structure factors satisfy the following sum rules,
\begin{eqnarray}
{1\over 2\pi^2}\int_{-\pi}^{\pi}dk\int_{0}^{\infty}d\omega
\,S^{+-} (k,\omega) & = & (1+m)
\nonumber \\
{1\over 2\pi^2}\int_{-\pi}^{\pi}dk\int_{0}^{\infty}d\omega
\,S^{-+} (k,\omega) & = & (1-m) 
\nonumber \\
{1\over 2\pi^2}\int_{-\pi}^{\pi}dk\int_{0}^{\infty}d\omega
\,S^{zz} (k,\omega) & = & {1\over 2}(1-m^2) \,  .
\label{SRDSF}
\end{eqnarray}

The selection rules in Eq. (\ref{SRh0}) reveal that at $h=0$ and thus $m=0$ when
$S^{xx} (k,\omega) = S^{yy} (k,\omega) = S^{zz} (k,\omega)$, the longitudinal dynamical structure factor
$S^{zz} (k,\omega)$ is fully controlled by transitions from the ground state for which $S^z =S=0$ to excited states with
spin numbers $S^z=0$ and $S=1$. That according to such rules the transverse dynamical structure factors
are at $h=0$ controlled by transitions from that ground state to excited states with
spin numbers $S^z=\pm 1$ and $S=1$, does not prevent the equality $S^{zz} (k,\omega) = S^{xx} (k,\omega)$
imposed by the spin $SU(2)$ symmetry.

This is different from the case for magnetic fields $0<h<h_c$ considered in this paper.
According to the selection rules, Eq. (\ref{SRhfinite}), the factor
$S^{zz} (k,\omega) \neq S^{xx} (k,\omega)$ is then controlled by transitions from the ground state 
with spin numbers $S^z = -S$ to excited states with
the same spin numbers $S^z = -S$. According to the same selection rules, the dynamical structure factors
$S^{+-} (k,\omega)$ and $S^{-+} (k,\omega)$ are controlled by transitions from the ground state 
with spin numbers $S^z = -S$ to excited states with spin numbers $S^z \pm 1 = -S \pm 1$.

\section{The spectral functionals of the extended dynamical theory}
\label{SECIII}

As reported in Sec. \ref{SECI}, the general goal of this paper is the study of the contribution 
from $n$-string states to the spin dynamical structure
factors given in Eq. (\ref{SDSF}) within the spin-$1/2$ $XXX$ chain in a longitudinal magnetic field, Eq. (\ref{HXXX}).
The dynamical theory used in our studies refers to an extension of that introduced for the
present model in Ref. \cite{Carmelo_15A}. In that reference, only the contribution to the
spin dynamical structure factors from energy eigenstates described by real Bethe-ansatz
rapidities was considered.

The theory of that reference is directly related to that 
introduced for the one-dimensional Hubbard model in Ref. \cite{Carmelo_05}.
The related dynamical theories of Refs. \cite{Carmelo_15A,Carmelo_05,Carmelo_08} are equivalent to and account for the
same microscopic processes \cite{Carmelo_18} as the mobile quantum impurity model scheme 
of Refs. \cite{Imambekov_09,Imambekov_12} in the case of integrable models.

The main difference of such an extended theory to that considered in Ref. \cite{Carmelo_15A}, 
refers to the Hamiltonian, Eq. (\ref{HXXX}), acting onto an extended subspace, including $n$-string states.
This involves different new forms for the spectral functionals that control the momentum dependent exponents in the
spin dynamical structure factors's expressions obtained in this paper
for $(k,\omega)$-plane regions near specific types of spectral features. For simplicity, we
do not provide here the details of the extended dynamical theory that are common to those
already given in Ref. \cite{Carmelo_15A}, and rather focus on the
differences associated with its extension to the contribution from $n$-string states.

\begin{figure}
\begin{center}
\centerline{\includegraphics[width=16cm]{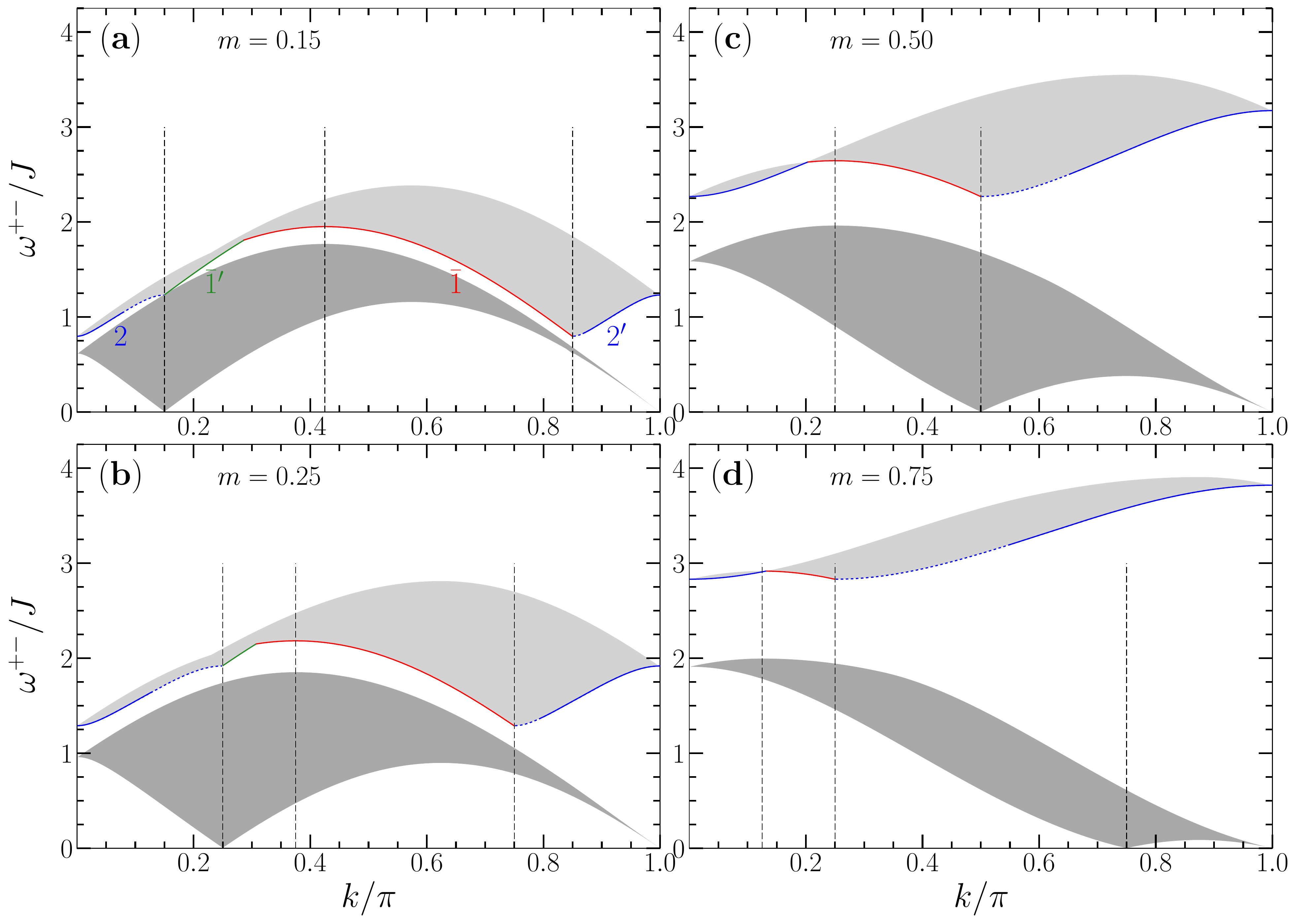}}
\caption{The two $(k,\omega)$-plane lower and upper continuum regions 
where for spin densities (a) $m=0.15$, (b) $m =0.25$, (c) $m=0.50$, and (d) $m =0.75$  
there is in the thermodynamic limit more spectral weight in $S^{+-} (k,\omega)$. 
The sketch of the $(k,\omega)$-plane distributions represented here and
in Figs. \ref{figure2} and \ref{figure3} does
not provide information on the relative amount of spectral weight contained within each spectrum's 
grey continuum. The three reference vertical lines mark the momenta (a) [$k=k_{F\uparrow} - k_{F\downarrow}=3\pi/20$,
$k=k_{F\downarrow}=17\pi/40$, $k=2k_{F\downarrow}=17\pi/20$];
(b) [$k=k_{F\uparrow} - k_{F\downarrow}=\pi/4$,
$k=k_{F\downarrow}=3\pi/8$, $k=2k_{F\downarrow}=3\pi/4$];
(c) [$k=k_{F\downarrow}=\pi/4$, $k=k_{F\uparrow} - k_{F\downarrow}=2k_{F\downarrow}=\pi/2$]; 
(d) [$k=k_{F\downarrow}=\pi/8$, 
$k=2k_{F\downarrow}=\pi/4$, $k=k_{F\uparrow} - k_{F\downarrow}=3\pi/4$].
The lower and upper continuum spectra are associated with excited energy eigenstates without 
and with $n$-strings, respectively. In the thermodynamic limit, the $(k,\omega)$-plane region between the upper threshold of
the lower continuum and the gapped lower threshold of the upper $n$-string continuum has nearly no spectral weight.
In the case of the gapped lower threshold of the $n$-string continuum,
the analytical expressions given in this paper refer to near and just above that threshold 
whose subintervals refer to branch lines parts represented in the figure
by solid and dashed lines. The latter refer to $k$ intervals where the momentum dependent
exponents plotted in Fig. \ref{figure4} are negative and positive, respectively. In the former intervals, $S^{+-} (k,\omega)$ 
displays singularity peaks.}
\label{figure1}
\end{center}
\end{figure}

\begin{figure}
\begin{center}
\centerline{\includegraphics[width=16cm]{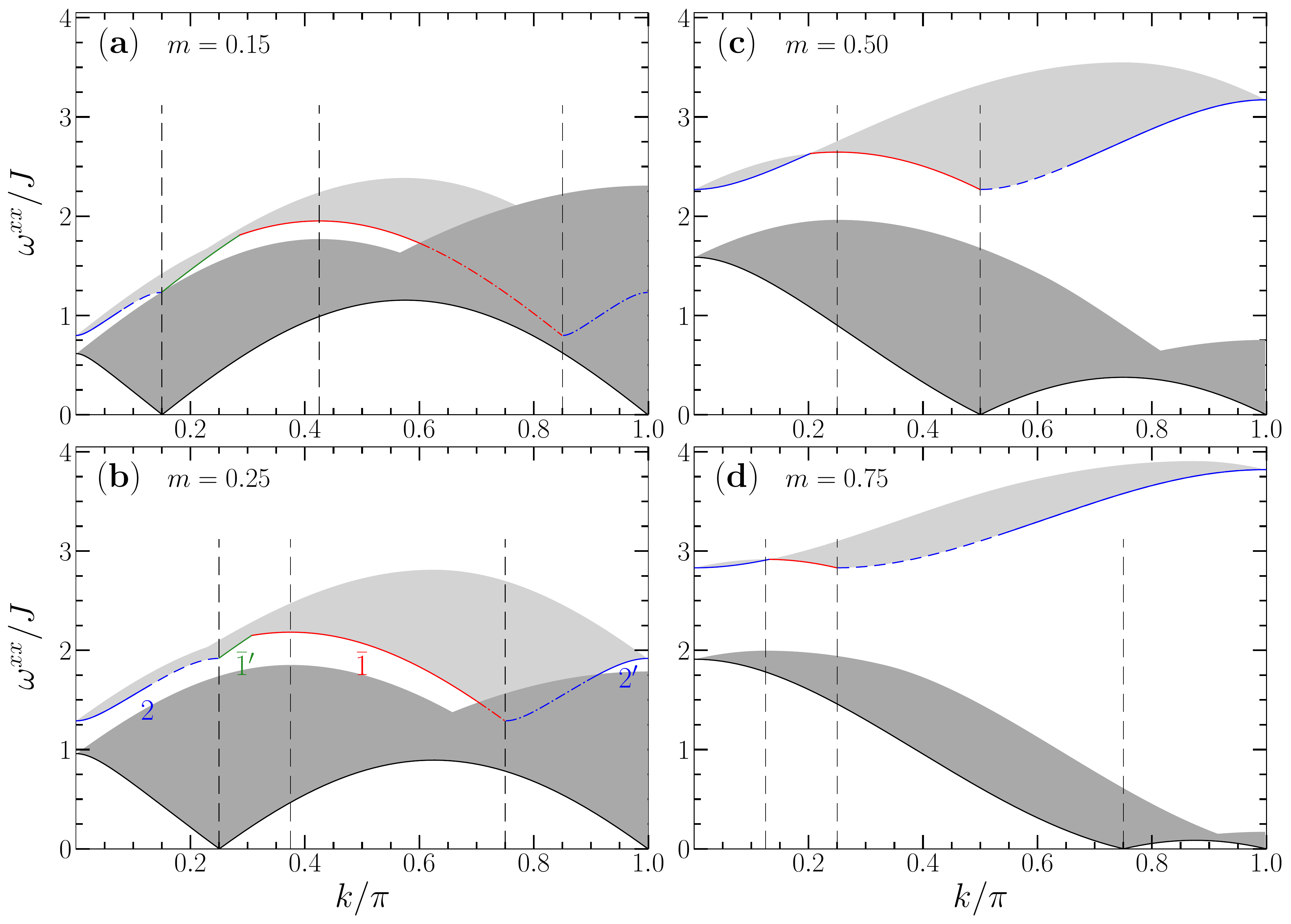}}
\caption{The two $(k,\omega)$-plane lower and upper continuum regions 
where for the same spin densities as in Fig. \ref{figure1} there is in the thermodynamic limit more 
spectral weight in $S^{xx} (k,\omega)$. The notations and the momenta associated with
the reference vertical lines are the same as in Fig. \ref{figure1}.
The additional part of the lower continuum relative to that of $S^{+-} (k,\omega)$ in Fig. \ref{figure1}
stems from the contributions from $S^{-+} (k,\omega)$. As a result, for
some $k$ intervals the upper $n$-string continuum overlaps with
the lower continuum.}
\label{figure2}
\end{center}
\end{figure}

\begin{figure}
\begin{center}
\centerline{\includegraphics[width=16cm]{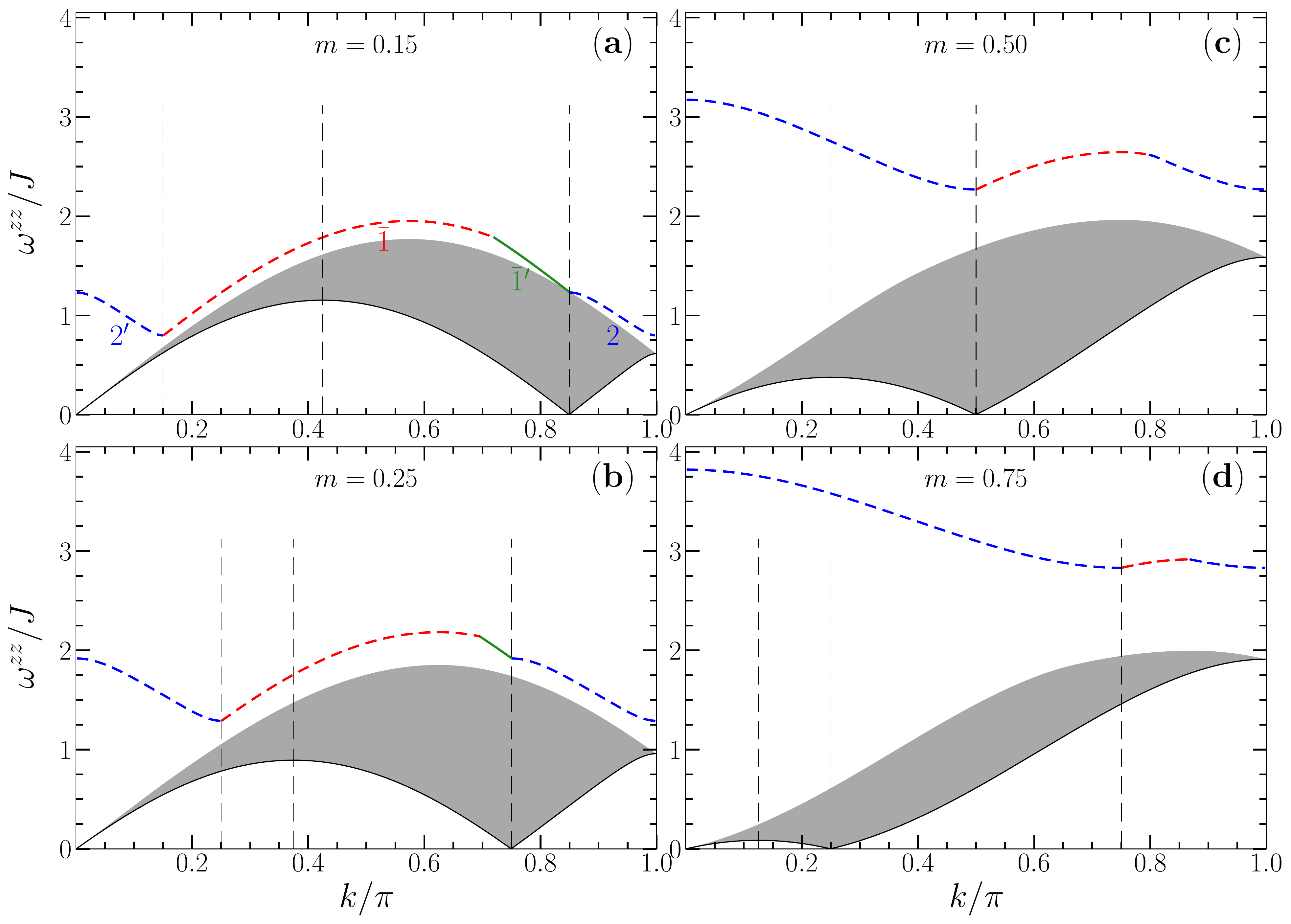}}
\caption{The $(k,\omega)$-plane continuum region
where for the same spin densities as in Figs. \ref{figure1} and \ref{figure2}
there is in the thermodynamic limit more spectral weight in $S^{zz} (k,\omega)$.
The reference vertical lines mark the same momenta as in these figures.
Contributions from excited states containing $n$-strings are much smaller
than for $S^{+-} (k,\omega)$ and $S^{xx} (k,\omega)$ and do not lead to an
upper continuum. The gapped lower threshold of such states is though
shown in the figures. Only when that threshold coincides with the $\bar{1}'$-branch
line, which only occurs for spin densities $0<m<\tilde{m}$ where $\tilde{m}\approx 0.317$, singularities
occur near and just above the $\bar{1}'$-branch line, which is represented by
a solid (green) line. In the remaining parts of the gapped lower threshold, which
for spin densities $\tilde{m}<m<1$ means all of it, the momentum dependent
exponents are positive and there are no singularities. This
is equivalent to a negligible amount of spectral weight near such lines.}
\label{figure3}
\end{center}
\end{figure}

The use of the extended dynamical theory provides
useful information on the $(k,\omega)$-plane distribution of the excited energy eigenstates's spectra that 
contain in the thermodynamic limit most spectral weight of $S^{+-} (k,\omega)$, $S^{xx} (k,\omega)$, and $S^{zz} (k,\omega)$.
Such spectra are schematically represented in Figs. \ref{figure1}, \ref{figure2}, and \ref{figure3}, respectively. 

After introducing the quantum problem's extended subspace, the general expressions of the spectral functionals
under consideration are introduced in the following. Specific expressions of the needed spectral functionals 
suitable to the line-shape near the four types of spectral features considered in our study are 
obtained. Finally, the issue concerning the $k$ intervals where the corresponding momentum dependent exponents
are valid is also addressed.

\subsection{The present quantum problem extended subspace}
\label{SECIIIA}

The quantum problem considered in this paper refers to the Hamiltonian, Eq. (\ref{HXXX}), in a subspace
spanned by two classes of energy eigenstates, populated and not populated by $n$-strings, respectively.
Our corresponding study of the spin dynamical structure factors relies on the representation of such energy eigenstates
in terms of $n$-particle occupancy configurations, which is that suitable to the dynamical theory used 
in this paper. Here $n=1,...,\infty$ is the number of singlet pairs of physical spins $1/2$ that refer to their internal degrees 
of freedom. The studies of Ref. \cite{Carmelo_15A} only involved
$n=1$ particles that in such a reference were named ``pseudoparticles''.

In the thermodynamic limit, the Bethe-ansatz rapidities have the general form given
in Eq. (\ref{Lambda-jnl-ideal}) of Appendix \ref{B} \cite{Takahashi_71}.
For $n=1$ such Bethe-ansatz rapidities are real and otherwise their imaginary part is finite. 
In that equation, the Bethe-ansatz rapidities are partitioned in a configuration of strings, where a $n$-string 
is a group of $l=1,...,n$ rapidities with the same real part $\Lambda^n (q_j)$.
The number $n$ is called in the literature the string length and the real part of the 
number $n$ of rapidities, $\Lambda^n (q_j)$, is called the string center. 

For $n>1$ the $n$-particle internal degrees of freedom refer
to a $n$-string, whereas the $n$-band momentum $q_j$ in the argument of the real part of the set 
of $l =1,...,n$ rapidities, $\Lambda^n (q_j)$ in Eq. (\ref{Lambda-jnl-ideal}) of Appendix \ref{A}, describes
its translational degrees of freedom. Each $n$-string contains a number $n=2,...,\infty$ of bound singlet pairs
of physical spins $1/2$, whose number thus equals the length of the $n$-string \cite{Carmelo_15,Carmelo_17,Carmelo_18}. 
The $l = 1,...,n$ imaginary parts, $i (n+1-2l)$, of the set of $n$ rapidities of a $n$-string describe the binding 
of the $l = 1,...,n$ pairs of physical spins $1/2$. Consistently, that imaginary part vanishes at $n=1$,
the internal degrees of freedom of the $1$-particles corresponding to a single unbound singlet pair 
of physical spins \cite{Carmelo_15,Carmelo_17,Carmelo_18}. Their translational degrees of freedom
refer again the $1$-band momentum $q_j$. Energy eigenstates that are not populated and are populated 
by $n$-particles with $n>1$ pairs, are described by only real Bethe-ansatz rapidities and both real
and complex non-real such rapidities, respectively.

The ground states with spin densities $0<m<1$ and corresponding longitudinal magnetic fields $0<h<h_c$
are not populated by $n$-strings. Concerning the amount of spectral weight of the spin dynamical structure factors 
originated from transitions from such ground states to $n$-string states, transitions to excited
energy eigenstates populated by a single $2$-particle are found to be dominant by far, as justified below.
This is consistent with results for large finite-size systems \cite{Kohno_09,Kohno_10}.

The $1$- and $2$-particles carry $1$-band and $2$-band discrete momentum values
$q_j$, respectively, Eq. (\ref{q-j}) of Appendix \ref{B}, whose spacing is $q_{j+1} - q_j=2\pi/L$.
Accounting for $1/L$ contributions, the ground state at a given spin density $0<m<1$ and 
corresponding longitudinal magnetic field $0<h<h_c$ is populated by a number $N_1 = N_{\downarrow}$ of $1$-particles that 
fill a $1$-band Fermi sea as follows,
\begin{eqnarray}
q_j & \in & [q^{-1}_{F},q^{+1}_{F}]\hspace{0.20cm}{\rm where}\hspace{0.20cm}
q_{j+1} - q_j= 2\pi/L\hspace{0.20cm}{\rm and}
\nonumber \\
q^{\iota}_F & = & \iota k_{F\downarrow} - \iota {\pi\over L} \hspace{0.20cm}{\rm for}\hspace{0.20cm}N\hspace{0.20cm}{\rm even}
\hspace{0.20cm}{\rm and}\hspace{0.20cm}\iota = \pm 1
\nonumber \\
q^{\iota}_F & = & \iota k_{F\downarrow} - (\iota \pm 1){\pi\over L} 
\hspace{0.20cm}{\rm for}\hspace{0.20cm}N\hspace{0.20cm}{\rm odd}
\hspace{0.20cm}{\rm and}\hspace{0.20cm}\iota = \pm 1 \, .
\label{qFiotaGS}
\end{eqnarray}
Here $\iota =+1$ and $\iota =-1$ refer to the $1$-band right and left Fermi points, respectively,
$k_{F\downarrow}={\pi\over 2}(1-m)$ (as given in Eq. (\ref{kkk}) of Appendix \ref{B}),
and in the case of $N$ odd, $\pm 1$ refers in $(\iota \pm 1){\pi\over L}$ to two alternative Fermi point's values.

In the present thermodynamic limit, we often use continuous momentum variables $q$ that
replace the discrete $1$- and $2$-bands momentum values $q_j$ such that $q_{j+1} - q_j=2\pi/L$.
We can then consider for the studies of some
properties that $q^{\iota}_F=\iota k_{F\downarrow}$ and thus a ground-state $1$-band occupied Fermi sea, 
$q\in[-k_{F\downarrow},k_{F\downarrow}]$. 

As reported in Sec. \ref{SECI}, there is a direct relation between the values of the momentum dependent exponents
that within the dynamical theory used here control the line shape in the $(k,\omega)$-plane vicinity
of the spin dynamical structure factors spectral features and the amount of spectral weight
located near them: Negative exponents implies the occurrence of singularities associated
with a significant amount of spectral weight in their $(k,\omega)$-plane vicinity.

The use of this criterion, reveals that in the present thermodynamic limit and for magnetic fields $0<h<h_c$,
the only significant contribution to $S^{+-} (k,\omega)$ from excited energy eigenstates populated
by $n$-particles refers to those populated by a number $N_{\downarrow}-2$ of $1$-particles and 
a single $2$-particle. There is as well a much weaker contribution at small spin densities from
states populated by a number $N_{\downarrow}-3$ of $1$-particles and a single $3$-particle.

The only significant yet weak contribution to $S^{zz} (k,\omega)$ from $n$-string states,
refers to energy eigenstates populated by a number $N_{\downarrow}-2$ of $1$-particles and 
a single $2$-particle. On the other hand, the contribution from such excited energy eigenstates
to $S^{-+} (k,\omega)$ is found to be negligible, since all relevant exponents are 
both positive and large. 

The contribution to $S^{+-} (k,\omega)$ from energy eigenstates populated
by a number $N_{\downarrow}-3$ of $1$-particles and a single $3$ particle that occurs for small values
of the spin density is very weak. It is actually inexistent in the vicinity of the $(k,\omega)$-plane 
singularities to which the analytical expressions obtained in our study refer to. Indeed, except for
very small spin densities, $m\rightarrow 0$, the latter very weak contributions occur in $(k,\omega)$-plane regions of higher 
excitation energy $\omega$, above the gapped lower threshold of the spectrum continuum associated with 
energy eigenstates populated by a number $N_{\downarrow}-2$ of $1$-particles and a single $2$-particle 
whose expression is given below in Sec. \ref{SECIIIB}. That spectrum refers to the upper continuum 
shown in Fig. \ref{figure1}.

The above spectral-weight analysis refers to the thermodynamic limit. Its results are fully consistent with 
corresponding results reached by a completely different method in the case of large finite-size systems \cite{Kohno_09}.

The subspace of the quantum problem studied in this paper is thus spanned by
an initial ground state for a given spin density $0<m<1$ and a corresponding longitudinal magnetic field
$0<h<h_c$ and its following excited energy eigenstates: States described by both real and complex non-real Bethe-ansatz rapidities 
populated by a number $N_{\downarrow}-2$ of $1$-particles and a single $2$-particle whose internal degrees
of freedom refer to a $n$-string of length $n=2$; States populated by a number $N_{\downarrow}$ of $1$-particles that are
described only by real Bethe-ansatz rapidities. 

\subsection{General expressions of the extended dynamical theory's spectral functionals}
\label{SECIIIB}

The following number and current number deviations under transitions from a ground state with $1$-band momentum distribution
given in Eq. (\ref{qFiotaGS}) to the excited energy eigenstates that span the present
subspace play an important role in the extended dynamical theory's expressions,
\begin{eqnarray}
& & \delta N_{1,\iota}^F \hspace{0.20cm}{\rm for}\hspace{0.20cm}\iota =1,-1
\hspace{0.20cm}{\rm (right,left)}\hspace{0.20cm}{\rm 1-particles}
\nonumber \\
& & \delta N_{1}^F = \sum_{\iota = \pm 1}\delta N_{1,\iota}^F
\hspace{0.20cm}{\rm and}\hspace{0.20cm}\delta J_{1}^F =  {1\over 2}\sum_{\iota = \pm 1}\iota\,\delta N_{1,\iota}^F
\nonumber \\
& & \delta J_{2} = {\iota\over 2}\delta N_{2} (q)\vert_{q=\iota\,(k_{F\uparrow}-k_{F\downarrow})}
\hspace{0.20cm}{\rm where}\hspace{0.20cm}\iota = \pm 1 \, .
\label{NcFNcFJcFJsF}
\end{eqnarray}

Under some of the transitions from the ground state to the excited energy eigenstates
of the present subspace, the number of $2$-particles and/or that of $1$-particles changes. 
This leads to number deviations $\delta N_{2}$ and/or $\delta N_{1}$, respectively. 
The specific number deviations $\delta N_{1,\iota}^F$ 
in Eq. (\ref{NcFNcFJcFJsF}) refer only to changes of the $1$-particles numbers at the left $(\iota =-1)$ or right $(\iota =1)$
$1$-band Fermi points, Eq. (\ref{qFiotaGS}). Exactly the same information is contained in the two Fermi points 
number deviations $\delta N_{1,\iota}^F$, on the one hand, and in the corresponding Fermi points 
number deviations $\delta N_{1}^F = \sum_{\iota = \pm 1}\delta N_{1,\iota}^F$ and
current number deviations $\delta J_{1}^F =  {1\over 2}\sum_{\iota = \pm 1}\iota\,\delta N_{1,\iota}^F$,
on the other hand. The overall $1$-particles number deviation $\delta N_{1}$ can be expressed as,
\begin{equation}
\delta N_{1} = \delta N_{1}^{F}+\delta N_1^{NF} \, .
 \label{dNdNFsNNF}
\end{equation}
Here $\delta N_1^{NF}$ refers to changes in the number of $1$-particles at $1$-band momenta
other than those at the Fermi points, Eq. (\ref{qFiotaGS}).

For the current subspace, the $2$-band number deviations may read $\delta N_{2}=0$ or $\delta N_{2}=1$.
The $2$-band is empty in the ground state. For that state its unoccupied momentum values $q_j$ such that
$q_{j+1} - q_j=2\pi/L$ refer to the range $q_j \in [-(k_{F\uparrow}-k_{F\downarrow}-1/L),(k_{F\uparrow}-k_{F\downarrow}-1/L)]$
where $k_{F\downarrow}={\pi\over 2}(1-m)$ and $k_{F\uparrow}={\pi\over 2}(1+m)$
(as given in Eq. (\ref{kkk}) of Appendix \ref{B}.) The $2$-band momentum range that is of
interest for our studies rather refers to excited energy eigenstates populated by a single $2$-particle. 
For these states, the available discrete momentum values belong to the interval
$q_j \in [-(k_{F\uparrow}-k_{F\downarrow}),(k_{F\uparrow}-k_{F\downarrow})]$ where $q_{j+1} - q_j = 2\pi/L$.
The $2$-particle can occupy any of such $2$-band discrete momentum values whose
number is $N_{\uparrow}-N_{\downarrow}+1$.
Only when the $2$-particle is created at one of the two $2$-band's limiting values, $q= -(k_{F\uparrow}-k_{F\downarrow})$ or
$q= (k_{F\uparrow}-k_{F\downarrow})$, that process leads to a current number deviation
$\delta J_{2} = -1/2$ and $\delta J_{2} = 1/2$, respectively, Eq. (\ref{NcFNcFJcFJsF}).

Within the extended dynamical theory, the line shape at and just above the gapped lower thresholds 
of the $n$-string states of $S^{+-} (k,\omega)$, $S^{xx} (k,\omega)$, and $S^{zz} (k,\omega)$
is for spin densities $0<m<1$ and momenta in the range $k\in ]0,\pi[$ of the following general form,
\begin{eqnarray}
S^{ab} (k,\omega) & = & C_{ab}^{\Delta}
\Bigl(\omega - \Delta_{\bar{n}}^{ab} (k))\Bigr)^{\zeta_{\bar{n}}^{ab} (k)}  
\hspace{0.20cm} {\rm for} \hspace{0.20cm}(\omega - \Delta_{\bar{n}}^{ab} (k)) \geq 0
\nonumber \\
&& {\rm where}\hspace{0.20cm}\bar{n} = 2,\bar{1},\bar{1}',2'
\hspace{0.20cm}{\rm and}\hspace{0.20cm}ab = +-,xx,zz 
 \nonumber \\
&& ({\rm valid}\hspace{0.20cm}{\rm when}\hspace{0.20cm}\Delta_{\rm gap}^{ab} > 0) \, .
\label{MPSsFMB}
\end{eqnarray}
Here $C_{ab}^{\Delta}$ is a constant that has a fixed value for the $k$ and $\omega$ ranges associated with 
small values of the energy deviation $(\omega - \Delta_{\bar{n}}^{ab} (k))\geq 0$ and
$\Delta_{\bar{n}}^{ab} (k)$ are the spectra that define the $(k,\omega)$-plane shape of the gapped lower thresholds of the $n$-string
states's in Figs. \ref{figure1}, \ref{figure2}, and \ref{figure3}. The analytical expressions of such spectra
are given below in Sec. \ref{SECIIIB} and the general expression of the exponents $\zeta_{\bar{n}}^{ab} (k)$
also appearing in Eq. (\ref{MPSsFMB}) is provided below. The indices $\bar{n}=2,\bar{1},\bar{1}',2'$ 
in such spectra and exponents label the branch lines 
or branch line sections that are part of the corresponding gapped lower thresholds in some specific $k$ intervals defined 
below in Sec. \ref{SECIVA}. Branch lines are types of spectral features that are defined as
within the dynamical theory of Ref. \cite{Carmelo_15A}.

The quantity $\Delta_{\rm gap}^{ab}$ in Eq. (\ref{MPSsFMB}) is the gap between the upper thresholds of the lower
continua associated with excited states described only by real Bethe-ansatz rapidities and the gapped lower
thresholds of the $n$-string states's spectra displayed in Figs. \ref{figure1}, \ref{figure2}, and \ref{figure3}. 
Only for $ab=xx$ there is overlap for small spin densities and some $k$ intervals between the lower 
continuum and the $n$-string states's upper continuum, as shown in
Fig. \ref{figure2} for spin densities $m=0.15$ and $m=0.25$. In the
corresponding $k$ intervals, one has that $\Delta_{\rm gap}^{xx} < 0$ and the general 
line-shape expression given in Eq. (\ref{MPSsFMB}) does not apply. 

Indeed, that expression is valid provided there is no spectral weight or nearly no spectral weight below 
the gapped lower thresholds of the $n$-string states.
In the present thermodynamic limit, the amount of spectral weight just below such thresholds either vanishes
or is extremely small. In the latter case, the very weak coupling to it leads to a higher order contribution to the line shape 
expressions given in that equation that can be neglected in such a limit. Hence,
the general expression of the spin dynamical structure factors given in 
Eq. (\ref{MPSsFMB}) is an excellent approximation for small values the energy deviation 
$(\omega - \Delta_{\bar{n}}^{ab} (k))\geq 0$.

On the other hand, the line shape of the spin dynamical structure factors $S^{ab} (k,\omega)$ 
where $ab = +-,-+,xx,zz$ at and just above their lower thresholds of the lower spectra 
that for $ab = +-,xx,zz$ are shown in Figs. \ref{figure1}, \ref{figure2}, and \ref{figure3} 
and are associated with excited energy eigenstates described only by real Bethe-ansatz 
rapidities has been derived within the dynamical theory of Ref. \cite{Carmelo_15A}.
It has the following general form, similar to that given in Eq. (\ref{MPSsFMB}),
\begin{eqnarray}
S^{ab} (k,\omega) & = & C_{ab} 
\Bigl(\omega - \omega^{ab}_{lt} (k)\Bigr)^{\zeta_{1}^{ab} (k)}  
\hspace{0.20cm}{\rm for}\hspace{0.20cm} (\omega - \omega^{ab}_{lt} (k)) \geq 0  
\nonumber \\
& & {\rm where}\hspace{0.20cm} ab = +-,-+,xx,zz \, .
\label{MPSs}
\end{eqnarray}
Again, here $C_{ab}$ are constants that have a fixed value for the $k$ and $\omega$ intervals 
for which the energy deviation $(\omega - \omega^{ab}_{lt} (k))\geq 0$ is small.
The lower thresholds under consideration refer to a single $1$-branch line that except for $S^{-+} (k,\omega)$
has two $k$ interval sections. The $ab=+-,-+,zz$ lower threshold's spectra 
$\omega^{+-} (k)$, $\omega^{-+} (k)$, and $\omega^{zz} (k)$ in that deviation
are given in Eqs. (\ref{OkPMRs}), (\ref{OkMPRs}), 
and (\ref{OkPMRsL}) of Appendix \ref{A}, respectively.

There is no spectral weight below the lower thresholds associated with
the line-shape expression, Eq. (\ref{MPSs}). The general
expression of the spin dynamical structure factors given in that
expression is thus exact for small values of the energy deviation 
$(\omega - \omega^{ab}_{lt} (k))\geq 0$.

The branch-line exponents that appear in both Eqs. (\ref{MPSsFMB}) 
and (\ref{MPSs}) have the same following general form,
\begin{equation}
\zeta^{ab}_{\bar{n}} (k) = -1 + \sum_{\iota =\pm1}(\Phi_{\iota} (q))^2
\hspace{0.20cm}{\rm for}\hspace{0.20cm} \bar{n}=2,\bar{1},\bar{1}',2',1 \, .
\label{expTS}
\end{equation}
The differences relative to the dynamical theory of Ref. \cite{Carmelo_15A},
refer to the form of the spectral functionals $\Phi_{\iota} (q)$ in this
general exponent's expression that is suitable to each type of branch line. In the following, the 
forms of such functionals specific to the {\it four} types of branch line involved in
our study are introduced. 

Consistent with the occurrence of an infinite number of conservation laws
associated with the present quantum problem integrability, there is
a representation of the $n$-particles for which they only undergo 
zero-momentum forward-scattering events. The corresponding phase shifts
fully control the spectral and dynamical properties. The extended dynamical theory
uses such a $n$-particle representation. Within it, $1$-particles carry discrete {\it canonical momentum} values
${\bar{q}}_j = {\bar{q}} (q_j)$ defined below such that ${\bar{q}}_{j+1}-{\bar{q}}_{j}= 2\pi/L + {\rm h.o.}$,
rather than $1$-band momentum values $q_j$ directly related
to Bethe-ansatz quantum numbers, Eq. (\ref{q-j}) of Appendix \ref{B}.
The higher order $(1/L)^2$ terms in the spacing ${\bar{q}}_{j+1}-{\bar{q}}_{j}$ 
have no physical meaning in the present thermodynamic limit. 
The key property of the $1$-particles canonical momentum representation is the lack
of energy interaction terms. This is what ensures the occurrence of only
zero-momentum forward-scattering events.

The initial ground state is populated by a macroscopic number
$N_1 = N_{\downarrow}$ of $1$-particles and contains no $n$-particles with $n>1$.
Therefore, $1$-particles contribute to the dynamical properties both as
scatterers and scattering centers, whereas the $2$-particle contributes to
them as a scattering center only. As justified below, in the case of scattering
centers the same results are obtained in the thermodynamic limit within the representations for
which such centers created under transitions to excited states carry momentum $q_j$ 
and canonical momentum ${\bar{q}}_j = {\bar{q}} (q_j)$,
respectively. While canonical momentum can also be introduced for the $2$-particle,
for simplicity we thus remain using $2$-band momentum $q_j$ for it.

From straightforward yet lengthly manipulations of the Bethe ansatz equations, 
Eqs. (\ref{Taps})-(\ref{Tap2}) of Appendix \ref{B},
one finds that for the excited energy eigenstates that span the present
subspace, the $1$-band rapidity functional $\Lambda_{1}(q_j)$ can be written
in terms of the corresponding ground-state rapidity function $\Lambda_{1}^0 (q_j)$ as follows,
\begin{equation}
\Lambda_{1} (q_j) = \Lambda_{1}^0({\bar{q}}_j)\hspace{0.20cm}{\rm for}\hspace{0.20cm}j = 1,...,N_{\uparrow} \, .
\label{FL}
\end{equation}
Here ${\bar{q}}_j = {\bar{q}} (q_j)$ where $j=1,...,N_{\uparrow}$ are the following
discrete canonical momentum values,
\begin{equation}
{\bar{q}}_j = {\bar{q}} (q_j) = q_j + {2\pi\over L}\Phi_{1} (q_j) = {2\pi\over L}\left(I^{1}_j + \Phi_{1} (q_j)\right) \, ,
\label{barqan}
\end{equation}
and $I_j^{1}$ are the Bethe-ansatz $1$-band quantum numbers given in Eq. (\ref{Ic-an}) of Appendix \ref{B}. The 
lack of energy interactions follows from in terms of canonical momentum values the $1$-band rapidity function 
having for the excited energy eigenstates the same form, $\Lambda_{1} (q_j) = \Lambda_{1}^0({\bar{q}}_j)$, 
as for the corresponding initial ground state. (For the ground state, $1$-band momentum
values and canonical momentum values are actually the same.)

The general expression of the $1$-band functional $\Phi_{1} (q_j)$ in Eq. (\ref{barqan}) is 
in the case of the present subspace given by,
\begin{equation}
\Phi_{1} (q_j) = \sum_{j'=1}^{N_{\uparrow}}\,\Phi_{1,1}(q_j,q_{j'})\,\delta N_{1}(q_{j'})
+ \sum_{j'=1}^{N_{\uparrow}-N_{\downarrow}+N_{2}}\,\Phi_{1,2}(q_j,q_{j'})\,\delta N_{2}(q_{j'}) \, , 
\label{qcanj}
\end{equation}
where $\Phi_{1,1}(q_j,q_{j'})$ and $\Phi_{1,2}(q_j,q_{j'})$ are as further discussed below phase shifts in units of $2\pi$.
They are defined by Eqs. (\ref{Phi-barPhi})-(\ref{Phis1s2-m}) of Appendix \ref{B}. The deviations 
$\delta N_{1}(q_{j'})$ and $\delta N_{2}(q_{j'})$ also appearing in Eq. (\ref{qcanj}) read,
\begin{equation}
\delta N_{n} (q_j)  = N_{n} (q_j) - N^0_{n} (q_j) \hspace{0.20cm}{\rm for}\hspace{0.20cm} n =1,2 \, .
\label{DNq}
\end{equation}
Here $N_{n} (q_j)$ and $N^0_{n} (q_j)$ are the $1$-band and $2$-band momentum distributions of
the excited energy eigenstate and ground state, respectively. Such momentum distributions appear
in the functional representation of the Bethe-ansatz equations, Eqs. (\ref{Taps})-(\ref{Tap2}) of Appendix \ref{B}.
The ground-state distribution $N^0_{1} (q_j)$ is associated with the $1$-band compact occupancy, Eq. (\ref{qFiotaGS}), 
whereas $N^0_{2} (q_j)=0$ for $q_j \in [-(k_{F\uparrow}-k_{F\downarrow}-1/L),(k_{F\uparrow}-k_{F\downarrow}-1/L)]$. 

In the $\Phi_{1} (q_j)$'s general expression, Eq. (\ref{qcanj}),
the momentum values $q_j$ and $q_{j'}$ are associated with scatterers
and scattering centers, respectively. As mentioned above, one could associate canonical momentum values both 
with the scatterars and with the scattering centers created under the transitions from the ground state to the
excited states. However, in the case of the scattering centers, the form of the expression on
the right-hand side of Eq. (\ref{qcanj}) reveals that this leads to contributions of order $(1/L)^2$ that have 
no physical meaning in the thermodynamic limit. (The validity of the corresponding dynamical theory 
refers to that limit.)

For $\delta N_{1}(q_{j'})=\pm 1$, the quantity 
$\pm \Phi_{1,1}(q_j,q_{j'}) = \Phi_{1,1}(q_j,q_{j'})\,\delta N_{1}(q_{j'})$ in Eq. (\ref{qcanj})
is the scattering phase shift in units of $2\pi$ acquired by a $1$-particle
of canonical momentum ${\bar{q}}_j = {\bar{q}} (q_j)$ (scatterer) upon creation of one $1$-hole 
($-\Phi_{1,1}(q_j,q_{j'})$) and one $1$-particle ($+\Phi_{1,1}(q_j,q_{j'})$) at a momentum $q_{j'}$ in the 
$1$-band unoccupied and occupied Fermi sea (scattering centers), respectively. 
(Given the one-to-one relation between the canonical momentum ${\bar{q}}_j$ and the
momentum $q_j$, Eq. (\ref{barqan}), scatterers of canonical momentum ${\bar{q}}_j$
can also be labelled by momentum $q_j$, which refers to two representations of the
same $1$-particle.)

On the other hand, the quantity $\Phi_{1,2}(q_j,q_{j'})=\Phi_{1,2}(q_j,q_{j'})\,\delta N_{2}(q_{j'})$
in Eq. (\ref{qcanj}) is for $\delta N_{2}(q_{j'})=1$ the scattering phase shift in units of $2\pi$ acquired by a $1$-particle of
canonical momentum ${\bar{q}}_j = {\bar{q}} (q_j)$ (scatterer)  
under creation of one $2$-particle at a momentum $q_{j'}$ in the $2$-band interval 
$q_{j'}\in [-(k_{F\uparrow}-k_{F\downarrow}),(k_{F\uparrow}-k_{F\downarrow})]$
(scattering center). 

Hence $\Phi_{1} (q_j)$ is in Eq. (\ref{qcanj}) the overall phase shift in units of $2\pi$
acquired by a $1$-particle of canonical momentum ${\bar{q}}_j = {\bar{q}} (q_j)$ 
under a transition from the ground state to an excited energy eigenstate belonging
to the present subspace.

Important quantities for the dynamical properties, are the following $\iota =\pm 1$ deviations $\delta q_{F}^{\iota}$ 
from the values of the $1$-band ground-state Fermi momenta, Eq. (\ref{qFiotaGS}), under transitions to
excited energy eigenstates,
\begin{equation}
\delta q_{F}^{\iota} =\iota\, {2\pi\over L}\delta N^{F}_{1,\iota} = 
{2\pi\over L}\left(\iota\,\delta N^{0,F}_{1,\iota} + \Phi_{1}^0\right) \, .
\label{qFiotaES}
\end{equation}
Here $\delta N^{0,F}_{1,\iota}$ are the deviations in the number of $1$-particles
at the Fermi points which either vanish or are positive or negative integer
numbers. On the other hand, the actual number deviations,
$\delta N^{F}_{1,\iota}=\delta N^{0,F}_{1,\iota}+ \iota\,\Phi_{1}^0$,
can as well be half-odd integers. Their extra term, $\iota\,\Phi_{1}^0$ where $\iota =\pm 1$, stems from contributions from 
a non-scattering phase shift $\Phi_{1}^0$ regulated by the boundary conditions in 
Eq. (\ref{Ic-an}) of Appendix \ref{B}. It shifts all $1$-band's discrete momentum 
values as, $q_j\rightarrow q_j + (2\pi/L)\,\Phi_{1}^0$, and is given by,
\begin{eqnarray}
\Phi_{1}^0 & = & 0 \hspace{0.20cm}{\rm for}\hspace{0.20cm}{\rm for}\hspace{0.20cm}\delta N_{1}\hspace{0.20cm}{\rm even} 
\nonumber \\
& = & \pm {1\over 2} \hspace{0.20cm}{\rm for}\hspace{0.20cm}\delta N_{1} \hspace{0.20cm} {\rm odd} \, .
\label{pican}
\end{eqnarray}

It follows from the form of Eqs. (\ref{barqan}) and (\ref{qcanj}) that for the initial ground
state the equality ${\bar{q}}_j = q_j$ holds. Canonical momentum values ${\bar{q}}_j$
are different from the corresponding $1$-band momentum values $q_j$ only for
excited energy eigenstates. Hence the initial $\iota =\pm 1$ ground-state $1$-band Fermi momenta
$q_{F}^{\iota}$ have exactly the same values, Eq. (\ref{qFiotaGS}), for the 
$1$-particle momentum $q_j$ and canonical momentum ${\bar{q}}_j$ representations. 

The corresponding $\iota =\pm 1$ Fermi canonical momentum values deviations $\delta {\bar{q}}_{F}^{\iota}$
fully control the momentum and spin density dependence of the exponents $\zeta^{ab}_{\bar{n}} (k)$, Eq. (\ref{expTS}), 
in the spin dynamical structure factors's general expressions, Eqs. (\ref{MPSsFMB}) and (\ref{MPSs}).
The deviations $\delta {\bar{q}}_{F}^{\iota}$ are obtained from the excited state's $1$-band Fermi momentum 
values $q_{F}^{\iota} + \delta q_{F}^{\iota}$ under the momentum - canonical-momentum transformation, Eq. (\ref{barqan}), 
as follows,
\begin{equation}
q_{F}^{\iota} + \delta q_{F}^{\iota} \rightarrow \bar{q} (q_{F}^{\iota} + \delta q_{F}^{\iota}) =
q_{F}^{\iota} + \delta q_{F}^{\iota} +  {2\pi\over L}\Phi_1 (q_{F}^{\iota}+\delta q_{F}^{\iota}) =
q_{F}^{\iota} + \delta q_{F}^{\iota} +  {2\pi\over L}\Phi_1 (\iota k_{F\downarrow}) + {\rm h.o.} \, .
\label{transf}
\end{equation}
Neglecting contributions of order $(1/L)^2$ and accounting for the values of $1$-band Fermi momenta
being in the case of the initial ground state the same for the momentum and canonical-momentum 
representations, this gives,
\begin{equation}
\delta {\bar{q}}_{F}^{\iota} = \delta q_{F}^{\iota} + {2\pi\over L}\Phi_1 (\iota k_{F\downarrow})
= {2\pi\over L}\left(\iota\,\delta N^{F}_{1,\iota} +
\Phi_1 (\iota k_{F\downarrow})\right) = {2\pi\over L}\left(\iota\,\delta N^{0,F}_{1,\iota}+
\Phi_{1}^0 + \Phi_1 (\iota k_{F\downarrow})\right) \, .
\label{dbarqF}
\end{equation}
Indeed, expanding ${2\pi\over L}\Phi_1 (q_{F}^{\iota}+\delta q_{F}^{\iota})$ in both
the ${\cal{O}}(1/L)$ corrections in the ground-state expression, 
$q_{F}^{\iota}=\iota k_{F\downarrow} + {\cal{O}}(1/L)$, Eq. (\ref{qFiotaGS}), and in
the deviation $\delta q_{F}^{\iota}$, leads to ${2\pi\over L}\Phi_1 (\iota k_{F\downarrow})$
plus contributions of order $(1/L)^2$. Those have no physical meaning
in the thermodynamic limit to which the validity of the dynamical theory refers. 
Here $k_{F\downarrow}={\pi\over 2}(1-m)$, Eq. (\ref{kkk}) of Appendix \ref{B}.

That there are no $n$-particles with $n>1$ in the ground state dictates why
only the $\iota =\pm 1$ Fermi canonical momentum values fluctuations play an active
role in the dynamical properties. Within the present extended dynamical theory, such fluctuations
though account for the creation of the $2$-particle through the
phase-shift $\Phi_{1,2}(q_j,q_{j'})$ in units of $2\pi$ appearing on the right-hand side of Eq. (\ref{qcanj}).

That the $\iota =\pm 1$ Fermi canonical momentum values fluctuations associated with
the deviations $\delta {\bar{q}}_{F}^{\iota}$ fully control the excitation momentum $k$ and spin density
dependence of the exponents, Eq. (\ref{expTS}), follows from the $\iota =\pm 1$ spectral functionals $\Phi_{\iota}$ 
in that equation being such deviations $\delta {\bar{q}}_{F}^{\iota}$,
Eq. (\ref{dbarqF}), in units of the quantum momentum spacing $2\pi/L$,
\begin{eqnarray}
\Phi_{\iota} = {\delta {\bar{q}}_{F}^{\iota}\over (2\pi/L)} & = & \iota\,\delta N^{F}_{1,\iota}+
\Phi (\iota k_{F\downarrow}) = \iota\,\delta N^{0,F}_{1,\iota} + \Phi_{1}^0 + \Phi (\iota k_{F\downarrow}) 
\nonumber \\
& = & \iota\,\delta N^{F}_{1,\iota} +
\sum_{j'=1}^{N_{\uparrow}}\,\Phi_{1,1}(\iota k_{F\downarrow},q_{j'})\,\delta N_{1}(q_{j'})
+ \sum_{j'=1}^{N_{\uparrow}-N_{\downarrow}+N_{2}}\,\Phi_{1,2}(\iota k_{F\downarrow},q_{j'})\,\delta N_{2}(q_{j'}) \, .
\label{functional}
\end{eqnarray}
The specific phase shifts in units of $2\pi$, $\Phi_{1,1}\left(\iota k_{F\downarrow},q\right)$ and 
$\Phi_{1,2}\left(\iota k_{F\downarrow},q\right)$ where $\iota = \pm 1$, of the $1$-particles with
momentum and canonical momentum values at the $1$-band Fermi points that appear in the last expression 
of this equation, are defined by Eq. (\ref{Phis-all-qq}) of Appendix \ref{B}. Limiting behaviors of such phase shifts are provided in Eqs.
(\ref{Phis1sn-m0})-(\ref{PhiUinfm1qF}) of that Appendix. 

In the case of the four types of branch lines considered in our study, some of the deviations 
$\delta N_{1}(q_{j'})$ and $\delta N_{2}(q_{j'})$ in Eq. (\ref{functional})
refer to $1$-band momenta $q_{j'} = \iota k_{F\downarrow}$ and $2$-band
momenta $q_{j'} = \iota (k_{F\uparrow}-k_{F\downarrow})$ and $q_{j'} = 0$ where $\iota =\pm 1$.
As a result, the expressions of the functionals $\Phi_{\iota}$, Eq. (\ref{functional}), specific to the corresponding 
branch-line exponents, Eq. (\ref{expTS}), involve the phase-shifts related parameters $\xi_{1\,1}$ and
$\xi_{1\,2}^{0}$. Those are defined by Eqs. (\ref{x-aa})-(\ref{Limxiss}) and (\ref{xis20})-(\ref{Limxis20}), respectively, 
of Appendix \ref{B}. Corresponding number and current number deviations $\delta N_{1}^F$, $\delta J_{1}^F$, $\delta N_{2}$, 
and $\delta J_{2}$, Eq. (\ref{NcFNcFJcFJsF}), then emerge in such expressions of the 
spectral functionals $\Phi_{\iota} (q)$ that control the momentum and spin density dependences
of the branch-line exponents, Eq. (\ref{expTS}).

We start by providing the three specific forms of the $\iota = \pm 1$ general spectral functionals $\Phi_{\iota}$,
Eq. (\ref{functional}), suitable to the $2$- and $2'$-branch lines, $\bar{1}$-branch lines, and $\bar{1}'$-branch lines,
respectively, that are part of the gapped lower thresholds of the upper $n$-string spectra in Figs. \ref{figure1}-\ref{figure3}.
(The spectra of such branch lines are given below in Sec. \ref{SECVA}.)

In the case of the $2$- and $2'$-branch lines, the form of the $\iota = \pm 1$ spectral functionals $\Phi_{\iota} (q)$ is,
\begin{equation}
\Phi_{\iota} (q) = {\iota\,\delta N^F_{1}\over 2\xi_{1\,1}^{1}} + \xi_{1\,1}^{1}\,\delta J^F_{1}+ \Phi_{1,2}(\iota k_{F\downarrow},q) 
\hspace{0.20cm}{\rm for}\hspace{0.20cm}s2-\hspace{0.20cm}{\rm and}\hspace{0.20cm}s2'
{\rm -branch}\hspace{0.20cm}{\rm lines} \, ,
\label{Fs2}
\end{equation}
where $\xi_{1\,1} = 1 + \sum_{\iota=\pm 1} (\iota)\,\Phi_{1,1}\left(k_{F\downarrow},\iota k_{F\downarrow}\right)$
(see Eq. (\ref{x-aa}) of Appendix \ref{B}.) For the excited energy eigenstates that contribute to the singularities
at and above the $2$- and $2'$-branch lines, 
the maximum interval of the $2$-band momentum $q$ in Eq. (\ref{Fs2})
is $q\in [0,(k_{F\uparrow}-k_{F\downarrow})[$ or
$q\in ]-(k_{F\uparrow}-k_{F\downarrow}),0]$.

For the $\bar{1}'$-branch lines, the form of the spectral functionals is,
\begin{equation}
\Phi_{\iota} (q) = {\iota\,\delta N^F_{1}\over 2\xi_{1\,1}^{1}} + \xi_{1\,1}^{1}\,(\delta J^F_{1} - 2\delta J_{2}) - 
\Phi_{1,1}(\iota k_{F\downarrow},q) 
\hspace{0.20cm}{\rm for}\hspace{0.20cm}\bar{1}'{\rm -branch}\hspace{0.20cm}{\rm lines} \, ,
\label{Fbarsl}
\end{equation}
where it was accounted for that the phase shift
$\Phi_{1,2}(\iota k_{F\downarrow},\pm (k_{F\uparrow}-k_{F\downarrow}))$ can be written as
$\mp\xi_{1\,1}^{1}$ (see Eq. (\ref{xi1Phiss2}) of Appendix \ref{B}.)

In the case of the $\bar{1}$-branch lines, the spectral functionals $\Phi_{\iota}$, Eq. (\ref{functional}),
have the following form,
\begin{equation}
\Phi_{\iota} (q) = {\iota\,\delta N^F_{1}\over 2\xi_{1\,1}^{1}} + {\iota\,\xi_{1\,2}^{0}\over 2} + \xi_{1\,1}^{1}\,\delta J^F_{1} 
- \Phi_{1,1}(\iota k_{F\downarrow},q)
\hspace{0.20cm}{\rm for}\hspace{0.20cm}\bar{1}{\rm -branch}\hspace{0.20cm}{\rm lines} \, ,
\label{Fbars}
\end{equation}
where $\xi_{1\,2}^{0} = \Phi_{1,2}(k_{F\downarrow},0)$ and it was accounted for that 
$\Phi_{1,2}(\iota k_{F\downarrow},0)= \iota\,\Phi_{1,2}(k_{F\downarrow},0)$
(see Eq. (\ref{xis20}) of Appendix \ref{B} and text below it.)
The maximum interval of the $1$-band momentum
is $q\in ]-k_{F\downarrow},k_{F\downarrow}[$ in both Eqs. (\ref{Fbarsl}) and (\ref{Fbars}). 

Finally, the expressions of the spectral functionals already considered in Ref. \cite{Carmelo_15A}
that control the momentum and spin density dependence of the exponents 
associated with the $1$-branch lines are provided. Those refer to parts of the 
lower thresholds of the lower continua in Figs. \ref{figure1}-\ref{figure3}.
The spectra of such thresholds are given in Eqs. (\ref{OkPMRs})-(\ref{OkxxRs}) of Appendix \ref{A}. 
The corresponding spectral functionals are of general form,
\begin{eqnarray}
\Phi_{\iota} (q) & = & 
{\iota\,\delta N^F_{1}\over 2\xi_{1\,1}^{1}} + \xi_{1\,1}^{1}\,\delta J^F_{1} \mp\Phi_{1,1}(\iota k_{F\downarrow},q)
\hspace{0.20cm}{\rm where}\hspace{0.20cm}{\rm for}\hspace{0.20cm}1{\rm -branch}\hspace{0.20cm}{\rm lines}
\nonumber \\
- & \rightarrow & \hspace{0.20cm}{\rm maximum}\hspace{0.20cm}{\rm interval}\hspace{0.20cm}
q\in ]-k_{F\downarrow},k_{F\downarrow}[
\nonumber \\
+ & \rightarrow & \hspace{0.20cm}{\rm maximum}\hspace{0.20cm}{\rm intervals}\hspace{0.20cm}
q \in [-k_{F\uparrow},-k_{F\downarrow}[ \hspace{0.20cm}{\rm and}\hspace{0.20cm}
q \in ]k_{F\downarrow},k_{F\uparrow}] \, .
\label{Fs}
\end{eqnarray}
Here $-$ and $+$ is the phase-shift sign in $\mp\Phi_{1,1}(\iota k_{F\downarrow},q)$ suitable
to $1$-branch lines involving $1$-hole and $1$-particle creation, respectively, at a 
$1$-band momentum $q$ belonging to the maximum intervals given in the equation.

The values of the $1$- and $2$-bands number and current number 
deviations that in the case of the transverse and longitudinal excited states
are for excitation momentum $k>0$ used in Eqs. (\ref{Fs2})-(\ref{Fbars}) 
to reach the specific expressions of the branch-line exponents given in Sec. \ref{SECVA}
are provided below in Tables \ref{table1} and \ref{table2},
respectively. Those of the $1$-band number and current number deviations that are used 
in Eq. (\ref{Fs}) to derive the expressions of the branch-line exponents provided
in Sec. \ref{SECVB} are given below in Table \ref{table3}.

\subsection{Constrains to the momentum dependent exponents's $k$ intervals}
\label{SECIIIC}

The exponents of general form, Eq. (\ref{expTS}), that
control the spin dynamical structure factors's expressions, Eqs. (\ref{MPSsFMB}) and (\ref{MPSs}),
at and above the $2$-, $\bar{1}$-, $\bar{1}'$-, $2'$-branch lines and the $1$-branch lines,
respectively, depend on the excitation momentum $k\in ]0,\pi[$ through the $1$- or $2$-band
momentum values in the arguments of the spectral functionals, Eqs. (\ref{Fs2})-(\ref{Fs}).

The $\bar{1}$- and $\bar{1}'$-branch lines's exponents and those of
the $1$-branch lines involving $1$-hole creation are valid
in $k$ ranges corresponding to a maximum $1$-band momentum
interval $q\in[-(k_{F\downarrow}-\delta q_{1}),(k_{F\downarrow}-\delta q_{1})]$.
On the other hand, the exponents of the $1$-branch lines involving $1$-particle creation are valid
in $k$ ranges corresponding to maximum $1$-band momentum intervals
such that $\vert q\vert\in[(k_{F\downarrow}+\delta q_1),k_{F\uparrow}]$.
In both cases, $\delta q_{1}$, such that $\delta q_{1}/k_{F\uparrow} \ll 1$ for $0<m<1$,
is for the different branch lines very small or vanishes in the thermodynamic limit. 

In the very small $k$ intervals corresponding to the $1$-band intervals
$q\in[-(k_{F\downarrow}+\delta q_{1}),-(k_{F\downarrow}-\delta q_{1})]$
and $q\in [(k_{F\downarrow}-\delta q_{1}),(k_{F\downarrow}+\delta q_{1})]$, the exponents that control the line shape of
the spin dynamical structure factors have a different form, as given in Ref. \cite{Carmelo_15A}.
(See Eqs. (77)-(82) of that reference.)

Similarly, the $2$- and $2'$-branch lines's exponents that control the line shape
at and above some parts of the gapped lower thresholds refer to
$k$ ranges corresponding to $2$-band momentum maximum intervals 
$q \in [-(k_{F\uparrow}-k_{F\downarrow}-\delta q_{2}),0]$ or $q \in [0,(k_{F\uparrow}-k_{F\downarrow}-\delta q_{2})]$.
Here $\delta q_{2}$, such that $\delta q_{2}/(k_{F\uparrow}-k_{F\downarrow}) \ll 1$ for $0<m<1$, is
for the different branch lines again very small or vanishes in the thermodynamic limit. (And again, the spin dynamical structure factors
expressions are different and known for 
$q \in [-(k_{F\uparrow}-k_{F\downarrow}),-(k_{F\uparrow}-k_{F\downarrow}-\delta q_{2})]$ 
and $q \in [(k_{F\uparrow}-k_{F\downarrow}-\delta q_{2}),(k_{F\uparrow}-k_{F\downarrow})]$
yet are not of interest for our study.)

In the present thermodynamic limit, the maximum $1$-band $q$ intervals corresponding
to $k$ intervals for which the exponents, Eq. (\ref{expTS}), are valid, are represented in the following as 
$q\in ]-k_{F\downarrow},k_{F\downarrow}[$ and $\vert q\vert\in ]k_{F\downarrow},k_{F\uparrow}]$.
Similarly, the $2$-band $q$ intervals corresponding to such $k$ intervals are represented 
as $q \in ]-(k_{F\uparrow}-k_{F\downarrow}),0]$ and $q \in [0,(k_{F\uparrow}-k_{F\downarrow})[$. 

Around the specific excitation momentum $k$ values where along a gapped lower threshold
or a lower threshold two neighboring branch lines or branch line sections
cross, there are small momentum widths in which the corresponding lower threshold refers to 
a boundary line that connects the two branch lines or branch line sections under consideration.

In the thermodynamic limit, such momentum intervals are in general
negligible and the corresponding small spectra's deviations
are not visible in the spectra plotted in Figs. \ref{figure1}-\ref{figure3}.
In the cases they are small yet more extended, the two branch lines or branch line sections
run very near and just above the (gapped or gapless) lower threshold and there is very little spectral weight
between it and such lines. In this case, the singularities on the two branch lines or branch line sections
remain the dominant spectral feature. 

We again account for such negligible effects by replacing parenthesis $[$ or $]$ by $]$ or $[$, respectively, 
at the $k$'s limiting values that separate (gapped or gapless) lower thresholds's $k$ intervals associated with two neighboring 
branch lines or branch line sections.

\section{Dynamical structure factors's spectra}
\label{SECIV}

Here the spectra associated with the $(k,\omega)$-plane regions
that contain most spectral weight of the spin dynamical structure factors are introduced, with emphasis 
in those associated with $n$-string states. The $(k,\omega)$-plane distribution 
of such spectra is represented for $S^{+-} (k,\omega)$, $S^{xx} (k,\omega)$,
and $S^{zz} (k,\omega)$ in Figs. \ref{figure1}, \ref{figure2}, and \ref{figure3}, respectively, 
for spin densities (a) $m=0.15$, (b) $m =0.25$, (c) $m=0.50$, and (d) $m =0.75$.
Note that the sketch of the $(k,\omega)$-plane distributions represented in these figures does
not provide information on the relative amount of spectral weight contained within each spectrum's 
grey continuum.

In the case of the spin dynamical structure factors $S^{+-} (k,\omega)$ and $S^{xx} (k,\omega)$, the figures show both a
lower continuum $(k,\omega)$-plane region, whose spectral weight is associated with
excited states without $n$-strings, and an upper continuum whose spectral
weight stems from excited states populated by such $n$-strings. 
In the case of $S^{zz} (k,\omega)$, the contribution to spectral weight from excited states 
containing $n$-strings is much weaker than for $S^{+-} (k,\omega)$ and $S^{xx} (k,\omega)$ and does not lead to an
upper continuum. The gapped lower threshold of such states's spectrum is 
represented in Fig. \ref{figure3} by a well defined $(k,\omega)$-plane line.

At finite magnetic fields $0<h<h_c$ the contribution to the spectral weight from excited states 
containing $n$-strings is negligible in the case of $S^{-+} (k,\omega)$. Therefore, its lower continuum
spectrum is not plotted here. (It was previously studied in Ref. \cite{Carmelo_15A}.)
The additional part of the lower continuum in Fig. \ref{figure2} for 
$S^{xx} (k,\omega) = {1\over 4}\left(S^{+-} (k,\omega)+S^{-+} (k,\omega)\right)$
relative to that of $S^{+-} (k,\omega)$ represented in Fig. \ref{figure1}
stems actually from contributions from $S^{-+} (k,\omega)$. As a result of such contributions, 
for small spin densities and some $k$ intervals the upper $n$-string continuum of $S^{xx} (k,\omega)$ overlaps with
its lower continuum. 

In the case of both $S^{+-} (k,\omega)$ and $S^{zz} (k,\omega)$, 
there is in the present thermodynamic limit for spin densities $0<m<1$ and thus finite
magnetic fields $0<h<h_c$ very little spectral weight between the upper threshold 
of the lower continuum associated with excited states described only by real Berthe-ansatz
rapidities and the gapped lower threshold of the $n$-string states's spectra in Figs. \ref{figure1} and \ref{figure3},
respectively. The same applies to $S^{xx} (k,\omega)$ in the $k$ intervals of Fig. \ref{figure2}
for which there is a gap between the upper continuum associated with $n$-string states 
and the lower continuum. 

Indeed, in the thermodynamic limit nearly all the small amount of spectral weight associated with 
excited energy eigenstates described by only real Bethe-ansatz rapidities that involve the emergence 
of four $1$-band holes under transitions from the ground state (named in the literature four-spinon states), 
is contained inside the lower continuum in such figures. This also applies to large finite systems at zero magnetic field,
as shown in Fig. 4 of Ref. \cite{Caux_06}. In the present case of a finite magnetic field, it also
applies to the spin-$1/2$ Heisenberg antiferromagnetic with anisotropy $\Delta < 1$,
see Fig. 1 of Ref. \cite{Caux_05}. 

In the case of $S^{+-} (k,\omega)$ at finite magnetic fields $0<h<h_c$, due to the behavior of spin 
operators matrix elements between energy eigenstates in the selection rules, Eq. (\ref{SRhfinite}), 
the spectral weight stemming from $n$-string-less states
existing in finite systems for $k\in [2k_{F\downarrow},\pi]$ and $\omega$ values above the 
upper threshold of the lower spectrum in Fig. \ref{figure1}, is negligible for a macroscopic system. 
Such a spectral weight decreases upon increasing 
the system size, as confirmed by analysis of the spectra in the $S^{+-} (k,\omega)$'s first row frames of Figs. 3 (a) and (b) of Ref. \cite{Kohno_09} for two finite-size systems with $N=320$ and $N=2240$ spins, respectively. 

The further discussion of this important issue is presented below in the final section of this paper, Sec. \ref{SECVII}.

\subsection{The spin densities $\tilde{m}$, $\bar{m}$, and $\bar{m}_0$ and related momentum values}
\label{SECIVA}

The spectra of the gapped lower thresholds of $S^{+-} (k,\omega)$, $S^{xx} (k,\omega)$, and $S^{zz} (k,\omega)$ 
studied below have a different form for two spin density intervals, $m\in ]0,\tilde{m}]$ and $m\in [\tilde{m},1[$, 
respectively. Here $\tilde{m}\approx 0.317$ is the spin density at which the following equality holds,
\begin{equation}
W_{2}\vert_{m=\tilde{m}\approx 0.317} = -  \varepsilon_{1} (2k_{F\downarrow}-k_{F\uparrow})\vert_{m=\tilde{m}\approx 0.317} \, .
\label{mtilde}
\end{equation}
From the use of the value of $2$-band energy dispersion $\varepsilon_{2} (q)$ at $q=0$ given in 
Eq. (\ref{vares2limits}) of Appendix \ref{B}, the $2$-band energy bandwidth $W_{2}$ appearing here can be expressed as 
$W_{2} = 4\mu_B h - \varepsilon_{2} (0)$.

Some specific excitation momentum $k$'s values separate momentum intervals of the gapped lower threshold
spectra of $S^{+-} (k,\omega)$, $S^{xx} (k,\omega)$, and $S^{zz} (k,\omega)$ 
that refer to different types of momentum dependences. Such specific $k$ values
either equal a momentum denoted here by $\tilde{k}$ or their expression involves $\tilde{k}$.
The latter momentum is defined by the following relations, 
\begin{eqnarray}
W_{2} & = & \varepsilon_{1} (k_{F\uparrow}-\tilde{k}) -  \varepsilon_{1} (k_{F\downarrow}-\tilde{k})
\hspace{0.20cm}{\rm for} \hspace{0.20cm}\tilde{k}\geq (k_{F\uparrow} - k_{F\downarrow})\hspace{0.20cm}
{\rm and}\hspace{0.20cm}m\in ]0,\tilde{m}]
\nonumber \\
W_{2} & = & 4\mu_B\,h -\varepsilon_{2} (\tilde{k}) -  \varepsilon_{1} (k_{F\downarrow}-\tilde{k})
\hspace{0.20cm}{\rm for} \hspace{0.20cm}\tilde{k}\leq (k_{F\uparrow} - k_{F\downarrow})\hspace{0.20cm}
{\rm and}\hspace{0.20cm}m\in [\tilde{m},1[ \, .
\label{ktilde}
\end{eqnarray}
The momentum $\tilde{k}$ is given by $\tilde{k}=(k_{F\uparrow} - k_{F\downarrow})$ at $m = \tilde{m}$.

In the relations provided in Eqs. (\ref{mtilde}) and (\ref{ktilde}), as well as in the expressions of the spin dynamical structure factors's
spectra given below and in Appendix \ref{A}, the quantities $\varepsilon_{1} (q)$ and $\varepsilon_{2} (q)$ 
are the $1$- and $2$-band energy dispersions, respectively, defined by 
Eqs. (\ref{equA4}), (\ref{vare2}), and (\ref{equA6})-(\ref{equA10B}) of Appendix \ref{B}.
Limiting behaviors of such dispersions and corresponding $1$- and $2$-band group velocities that provide useful information 
on the  momentum and spin density dependences of the corresponding spin dynamical structure factors's spectra
studied below are given in Eqs. (\ref{vares2limits})-(\ref{vvm1}) of that Appendix.

The $(k,\omega)$-plane spectrum that contains most of $S^{xx} (k,\omega)$'s spectral weight has features 
whose form is different for specific spin densities and momentum intervals. As shown in 
Fig. \ref{figure2} (a) for $m=0.15$ and (b) $m=0.25$, for small spin densities there are $k$ intervals for which
the $S^{xx} (k,\omega)$'s gapped lower threshold of the $2$-string continuum overlaps with
the lower continuum associated with energy eigenstates described by real Bethe-ansatz rapidities.
Such $k$ intervals are given by,
\begin{eqnarray}
k & \in & [\bar{k}_0,\pi]\hspace{0.20cm}{\rm for}\hspace{0.20cm}m\in ]0,\bar{m}_0] 
\nonumber \\
k & \in & [\bar{k}_0,\bar{k}_1]\hspace{0.20cm}{\rm for}\hspace{0.20cm}m\in ]\bar{m}_0,\bar{m}] \, ,
\label{gapineq}
\end{eqnarray}
where the spin densities $\bar{m}_0$ and $\bar{m}$ are given by
$\bar{m}_0 \approx 0.239$ and $\bar{m} \approx 0.276$, respectively.

The momenta $\bar{k}_0$ and $\bar{k}_1$ appearing in Eq. (\ref{gapineq}) are such  
$k^{xx}_{ut}\leq \bar{k}_0\leq\bar{k}_1$ and $\bar{k}_0\leq \bar{k}_1\leq\pi$. 
The equality, $\bar{k}_0=\bar{k}_1$, holds at $m=\bar{m}$. At that spin density,
$\bar{k}_0=\bar{k}_1$ has a value very near and just above $2k_{F\downarrow}$.
The spectra plotted in Fig. \ref{figure2} (a) for $m=0.15$ and (b) for $m=0.25$ 
refer to the two types of $k$ intervals reported in Eq. (\ref{gapineq}), respectively.
For (a), spin density $m=0.15$, one has that $\bar{k}_0\approx 0.60\,\pi$,
whereas for (b), spin density $m=0.25$, the two limiting momenta
read $\bar{k}_0\approx 0.71\,\pi$ and $\bar{k}_1\approx 0.92\,\pi$.

\subsection{Dynamical structure factors's spectra}
\label{SECIVB}

The information on the constrains to the $k$ intervals provided in Sec. \ref{SECIIIC}
applies to the branch-line spectra studied below in Sec. \ref{SECV}. On the other hand, 
in the particular case of the small $k$ intervals corresponding to the small $1$-band intervals
$q\in[-(k_{F\downarrow}+\delta q_{1}),-(k_{F\downarrow}-\delta q_{1})]$
and $q\in [(k_{F\downarrow}-\delta q_{1}),(k_{F\downarrow}+\delta q_{1})]$ and
$2$-band intervals $q \in [-(k_{F\uparrow}-k_{F\downarrow}),-(k_{F\uparrow}-k_{F\downarrow}-\delta q_{2})]$ 
and $q \in [(k_{F\uparrow}-k_{F\downarrow}-\delta q_{2}),(k_{F\uparrow}-k_{F\downarrow})]$,
such constrains do not apply to the two-parametric spectra of the $(k,\omega)$-plane regions
where there is more spectral weight. They also do not apply to the corresponding one-parametric spectra
of the gapped lower thresholds and lower thresholds themselves.
Only the general line-shape expressions, Eqs. (\ref{MPSsFMB}) and (\ref{MPSs}), are not
valid in such small $k$ intervals.

The remaining constrains mentioned in Sec. \ref{SECIIIC} associated with 
small $k$ intervals of the (gapped and gapless) lower thresholds 
in the vicinity of crossing points of two branch lines or branch line sections 
though apply to the expressions of the one-parametric spectra of the gapped lower thresholds and lower 
thresholds given in the following and in Appendix \ref{A}.
This applies, for instance, to the small excitation momentum intervals
in the vicinity of $k=\tilde{k}$ (in the transverse case)
and other $k$ values whose expression involves 
the momentum $\tilde{k}$ (in the logitudinal case) defined by the relations given in Eq. (\ref{ktilde})

The $(k,\omega)$-plane upper continuum shown in Fig. \ref{figure1} is
associated with the gapped upper spectrum of $S^{+-} (k,\omega)$ that stems
from transitions from the ground state to excited energy eigenstates populated by a number
$N_s = N_{\downarrow} - 2$ of $1$-particles and a single $2$-particle. It is given by,
\begin{eqnarray}
\omega^{+-}_{\Delta} (k) & = & - \varepsilon_{1} (q) + \varepsilon_{2} (q') 
\hspace{0.20cm}{\rm and}\hspace{0.20cm} k = \iota k_{F\downarrow} - q + q' 
\hspace{0.20cm}{\rm for}\hspace{0.20cm}\iota = \pm 1\hspace{0.20cm}{\rm where}
\nonumber \\
& & q \in [-k_{F\downarrow},k_{F\downarrow}] \, , \hspace{0.20cm}
\hspace{0.20cm}
q' \in [0,(k_{F\uparrow}-k_{F\downarrow})]\hspace{0.20cm}{\rm for}\hspace{0.20cm}\iota = 1 
\hspace{0.20cm}{\rm and}
\nonumber \\
& & q \in [-k_{F\downarrow},k_{F\downarrow}] \, , \hspace{0.20cm}
\hspace{0.20cm}
q' \in [-(k_{F\uparrow}-k_{F\downarrow}),0]\hspace{0.20cm}{\rm for}\hspace{0.20cm}\iota = - 1 \, .
\label{dkEdPPM}
\end{eqnarray}
This spectrum has two branches corresponding to $\iota =\pm 1$, such that,
\begin{equation}
k = k_{F\downarrow} - q + q' \in [0,\pi]
\hspace{0.50cm}{\rm and}\hspace{0.50cm}
k = - k_{F\downarrow} - q + q' \in [-\pi,0] \, .
\label{dkEdPPM2}
\end{equation}
As mentioned above, $\varepsilon_{1} (q)$ and $\varepsilon_{2} (q)$ 
are here the $1$- and $2$-band energy dispersions, respectively, defined by 
Eqs. (\ref{equA4}), (\ref{vare2}), and (\ref{equA6})-(\ref{equA10B}) of Appendix \ref{B}.
 
We denote by $\Delta^{ab} (k)$ where $ab=+-,xx,zz$ the spectra of the $n$-string excited states's gapped lower 
thresholds of the spin dynamical structure factors $S^{ab} (k,\omega)$. 
These gapped lower thresholds play an important role in our study. For some of their $k$ intervals there are 
singularities of power-law form, Eq. (\ref{MPSsFMB}), in the spin dynamical structure factors at and just above them. 
(In that equation, the extra lower index $\bar{n}$ in $\Delta^{ab} (k)$ refers to the different branch lines
or branch line sections that coincide with the gapped lower threshold for well-defined $k$ intervals
given below in Sec. \ref{SECVA}.)

The spectra of the transverse gapped lower thresholds are such that,
\begin{equation}
\Delta^{xx} (k) = \Delta^{+-} (k)\hspace{0.20cm} {\rm for} \hspace{0.20cm}k\in [0,\pi] \, .
\label{GappLT}
\end{equation}
(The equality $\Delta^{-+} (k) = \Delta^{+-} (k)$ also holds, yet, as reported previously, the amount of $S^{-+} (k,\omega)$'s 
spectral weight produced by $n$-string states is negligible in the thermodynamic
limit at finite magnetic fields.) The spectrum of the longitudinal gapped lower threshold is also related to $\Delta^{+-} (k)$ as follows,
\begin{equation}
\Delta^{zz} (k) = \Delta^{+-} (\pi - k)\hspace{0.20cm} {\rm for} \hspace{0.20cm}k\in [0,\pi] \, .
\label{GappLTlong}
\end{equation}

For smaller spin densities in the range $m\in ]0,\tilde{m}]$, the spectrum of the lower threshold is of the form,
\begin{eqnarray}
\Delta^{+-} (k) & = & \varepsilon_{2} (k) \hspace{0.20cm}{\rm for}\hspace{0.20cm}k\in [0,(k_{F\uparrow} - k_{F\downarrow})]
\nonumber \\
& = & 4\mu_B\,h - \varepsilon_{1} (k_{F\uparrow}-k) \hspace{0.20cm}{\rm for}\hspace{0.20cm}
k\in [(k_{F\uparrow} - k_{F\downarrow}),{\tilde{k}}[
\nonumber \\
& = & 4\mu_B\,h - W_{2} - \varepsilon_{1} (k_{F\downarrow}-k) 
\hspace{0.20cm}{\rm for}\hspace{0.20cm}k \in ]{\tilde{k}},2k_{F\downarrow}]
\nonumber \\
& = & \varepsilon_{2} (k - 2k_{F\downarrow}) \hspace{0.20cm}{\rm for}\hspace{0.20cm}k\in [2k_{F\downarrow},\pi]  \, .
\label{Dxx03}
\end{eqnarray}

For larger spin densities $m\in [\tilde{m},1[$, the spectrum of the lower threshold is slightly different
and reads,
\begin{eqnarray}
\Delta^{+-} (k) & = & \varepsilon_{2} (k) \hspace{0.20cm}{\rm for}\hspace{0.20cm}k\in [0,{\tilde{k}}[
\nonumber \\
& = & 4\mu_B\,h - W_{2} - \varepsilon_{1} (k_{F\downarrow}-k) 
\hspace{0.20cm}{\rm for}\hspace{0.20cm}k \in ]{\tilde{k}},2k_{F\downarrow}]
\nonumber \\
& = & \varepsilon_{2} (k - 2k_{F\downarrow}) \hspace{0.20cm}{\rm for}\hspace{0.20cm}k\in [2k_{F\downarrow},\pi]  \, .
\label{Dxx31}
\end{eqnarray}

The expressions of two-parametric transverse gapless spectra $\omega^{-+} (k)$ and $\omega^{+-} (k)$ 
previously studied in Ref. \cite{Carmelo_15A}, whose superposition 
gives $\omega^{xx} (k)$, and that of the longitudinal gapless spectrum $\omega^{zz} (k)$ 
that (except for $\omega^{-+} (k)$) refer to the lower continua in Figs. \ref{figure1}-\ref{figure3},
are given in Eqs. (\ref{dkEdPxxMP})-(\ref{dkEdPl}) of Appendix \ref{A}. 
The corresponding excited energy eigenstates are described by only real Bethe-ansatz rapidities. 
The related expressions of the one-parametric spectra of their upper thresholds 
$\omega^{-+}_{ut} (k)$, $\omega^{+-}_{ut} (k)$, $\omega^{xx}_{ut} (k)$, and $\omega^{zz}_{ut} (k)$
and lower thresholds $\omega^{-+}_{lt} (k)$, $\omega^{+-}_{lt} (k)$, $\omega^{xx}_{lt} (k)$,
and $\omega^{zz}_{lt} (k)$ are also provided in that Appendix.

We consider the following energy gaps that refer to $(k,\omega)$-plane regions with a negligible amount
of spectral weight in the thermodynamic limit,
\begin{eqnarray}
\Delta_{\rm gap}^{+-} (k) & = & \Delta^{+-} (k) - \omega^{+-}_{ut} (k) \geq 0
\nonumber \\
\Delta_{\rm gap}^{xx} (k) & = & \Delta^{xx} (k) - \omega^{xx}_{ut} (k) 
\nonumber \\
\Delta_{\rm gap}^{zz} (k) & = & \Delta^{zz} (k) - \omega^{zz}_{ut} (k) \geq 0 \, ,
\label{gapPMMP}
\end{eqnarray}
where,
\begin{eqnarray}
\Delta_{\rm gap}^{xx} (k) & = & \Delta^{+-} (k) - \omega^{+-}_{ut} (k)
\hspace{0.20cm}{\rm for}\hspace{0.20cm}k\in [0,k^{xx}_{ut}]
\nonumber \\
\Delta_{\rm gap}^{xx} (k) & = & \Delta^{+-} (k) - \omega^{-+}_{ut} (k)
\hspace{0.20cm}{\rm for}\hspace{0.20cm}k\in [k^{xx}_{ut},\pi] \, ,
\label{gap}
\end{eqnarray}
and
\begin{equation}
\Delta_{\rm gap}^{zz} (k) = \Delta_{\rm gap}^{+-} (\pi - k)  
\hspace{0.20cm}{\rm for}\hspace{0.20cm}k \in [0,\pi] \, .
\label{gapL}
\end{equation}
The momentum $k^{xx}_{ut}>k_{F\uparrow}-k_{F\downarrow}$ in Eq. (\ref{gap}) is that at which the equality, 
$\omega^{-+}_{ut} (k^{xx}_{ut}) = \omega^{+-}_{ut} (k^{xx}_{ut})$, holds.
As confirmed from analysis of Figs. \ref{figure1}-\ref{figure3},
one has that $\Delta_{\rm gap}^{+-} (k)\geq 0$ and $\Delta_{\rm gap}^{zz} (k)\geq 0$
for $k\in ]0,\pi[$, whereas $\Delta_{\rm gap}^{xx} (k)$ is negative for small spin densities $m\in ]0,\bar{m}]$, where
$\bar{m} \approx 0.276$, in the $k$ intervals given in Eq. (\ref{gapineq}). 

It follows from the $k$ intervals given in that equation that at $k=\pi$ the energy gap $\Delta_{\rm gap}^{xx} (k)$ 
has negative and positive values for spin densities $m<\bar{m}_0$ and $m>\bar{m}_0$, respectively,
where $\bar{m}_0 \approx 0.239$. Its corresponding negative, vanishing, and positive limiting values are provided
in Eq. (\ref{gaMPpuL}) of Appendix \ref{A}.

The upper threshold spectra $\omega^{-+}_{ut} (k)$, $\omega^{+-}_{ut} (k)$,
$\omega^{xx}_{ut} (k)$, $\omega^{zz}_{ut} (k)$ in Eqs. (\ref{gapPMMP})-(\ref{gapL})
are given in Eqs. (\ref{Omxxut1})-(\ref{Omlut})  of Appendix \ref{A}. The upper thresholds of
the lower continua in Figs. \ref{figure1}, \ref{figure2}, and \ref{figure3}
refer to the spectra $\omega^{+-}_{ut} (k)$, $\omega^{xx}_{ut} (k)$, and $\omega^{zz}_{ut} (k)$,
respectively.

In Appendix \ref{A}, the $k$ dependent expressions and the
limiting values of the energy gaps considered here are provided.

\section{The line shape near the singularities of the dynamical structure factors}
\label{SECV}

The power-law singularities, Eq. (\ref{MPSsFMB}), in the spin dynamical structure factors
$S^{+-} (k,\omega)$, $S^{xx} (k,\omega)$, and $S^{zz} (k,\omega)$'s expressions occur in the $k$ intervals for which
the corresponding momentum dependent exponents are negative. This applies to $(k,\omega)$-plane regions at and just above 
the $\bar{n}$-branch lines that are part of the corresponding gapped lower thresholds 
of the $n$-string states's spectra. As discussed in Sec. \ref{SECIIIA},
the main contributions to the line shape in these $(k,\omega)$-plane regions
stems from transitions to excited states populated by a number $N_{\downarrow}-2$ of $1$-particles 
and a single $2$-particle. The momentum interval, $k\in ]0,\pi[$, 
of the gapped lower thresholds under consideration is divided into several subintervals, which correspond to a set of 
$2$-, $\bar{1}'$-, $\bar{1}$-, and $2'$-branch lines, respectively. (In the case of $S^{zz} (k,\omega)$,
the order of the $\bar{1}'$- and $\bar{1}$-branch lines's $k$ subintervals is the reverse, $\bar{1}$- and $\bar{1}'$-.)
\begin{table}
\begin{center}
\begin{tabular}{|c|c|c|c|c|c|c|} 
\hline
branch line& $k$ in terms of $q$ & $\delta N_1^F$ & $\delta J_1^F$ & $\delta N_1^{NF}$ & $\delta J_{2}$ & $\delta N_{2}$ \\
\hline
$2$ & $k=q$ (for $2$-band $q$) & $-1$ & $0$ & $0$ & $0$ & $1$ \\
\hline
$\bar{1}'$ & $k=k_{F\uparrow} - q$ (for $1$-band $q$) & $0$ & $1/2$ & $-1$ & $1/2$ & $1$ \\
\hline
$\bar{1}$ & $k=k_{F\downarrow} - q$ (for $1$-band $q$) & $0$ & $1/2$ & $-1$ & $0$ & $1$ \\
\hline
$2'$ & $k=2k_{F\downarrow} + q$ (for $2$-band $q$) & $-1$ & $1$ & $0$ & $0$ & $1$ \\
\hline
\end{tabular}
\caption{The excitation momenta $k>0$ of $S^{+-} (k,\omega)$ and $S^{xx} (k,\omega)$
expressed in terms of $1$- and $2$-bands momenta $q$ and corresponding number and current number 
deviations for excited states populated by a number $N_{\downarrow}-2$ 
of $1$-particles and a single $2$-particle and thus described by both real and complex non-real rapidities 
in the case of the $2$-branch line, $\bar{1}'$-branch line, $\bar{1}$-branch line, and $2'$-branch line. For the 
momentum intervals given in Eqs. (\ref{Ds2})-(\ref{Ds2p}), such branch lines are part of the corresponding gapped lower thresholds.
(In the case of $S^{xx} (k,\omega)$, the data given here do not apply to branch-line intervals overlapping with
those given in Eq. (\ref{gapineq}.)}
\label{table1}
\end{center}
\end{table}

On the other hand, the lower thresholds of the gapless spectra associated with the lower continua
in Figs. \ref{figure1}-\ref{figure3} that were already studied in Ref. \cite{Carmelo_15A}, which stem from transitions to excited states that
are populated by a number $N_{\downarrow}$ of $1$-particles, either correspond to a single
$1$-branch line or to two sections of such a line.

Following the information provided in Sec. \ref{SECIIIC}, the constrains to the $k$ intervals where the 
expressions of the exponents in the general line-shape expressions, Eqs. (\ref{MPSsFMB}) and (\ref{MPSs}),
are valid imply replacements of parenthesis $[$ and/or $]$ by $]$ and/or $[$, respectively, 
in the limits of such intervals. 

\subsection{The line shape at and just above the gapped lower threshold of $S^{+-} (k,\omega)$, $S^{xx} (k,\omega)$, 
and $S^{zz} (k,\omega)$}
\label{SECVA}

In the case of $S^{xx} (k,\omega)$, the present study of the line shape at and just above the gapped lower 
thresholds of the spectra plotted in Fig. \ref{figure2} refers to $k$ intervals for which the corresponding
energy gap $\Delta_{\rm gap}^{xx} (k)$ obeys the inequality
$\Delta_{\rm gap}^{xx} (k)>0$. At small spin densities, this excludes the $k$ intervals given in Eq. (\ref{gapineq}).

Otherwise, the line shape near the gapped lower thresholds of $S^{+-} (k,\omega)$, $S^{xx} (k,\omega)$, 
and $S^{zz} (k,\omega)$ is given by Eq. (\ref{MPSsFMB}). The exponents in the
expression given in that equation have a general form provided in Eq. (\ref{expTS}). The $\iota = \pm 1$ spectral functionals
$\Phi_{\iota} (q)$ in that exponent's expression are in the case of the (i) $2$- and $2'$-branch lines, (ii) 
$\bar{1}'$-branch lines, and (iii) $\bar{1}$-branch lines given in Eqs. (\ref{Fs2}), (\ref{Fbarsl}), and (\ref{Fbars}), respectively.

The relation of the excitation momentum $k$ to the $1$-band momentum $q$ or
$2$-band momentum $q$ in the $\Phi_{\iota} (q)$'s argument is branch-line dependent. 
Hence it is useful to revisit the expressions of the spectra of the gapped lower thresholds,
Eqs. (\ref{GappLT})-(\ref{Dxx31}) and (\ref{GappLTlong}), for each 
of their branch lines or branch line sections, including information on the
relation between $k$ and the $1$- or $2$-band momenta $q$. 
Additional $1$-band and $2$-band constrains reported in Sec. \ref{SECIIIC}
associated with the validity of the corresponding expressions of the
momentum dependent exponents are also accounted for. This ensures that the 
branch-line spectra's expressions are given in the following for the $k$ intervals for 
which the dynamical structure factor's expression is of the form, Eq. (\ref{MPSsFMB}). 
\begin{figure}
\begin{center}
\centerline{\includegraphics[width=16cm]{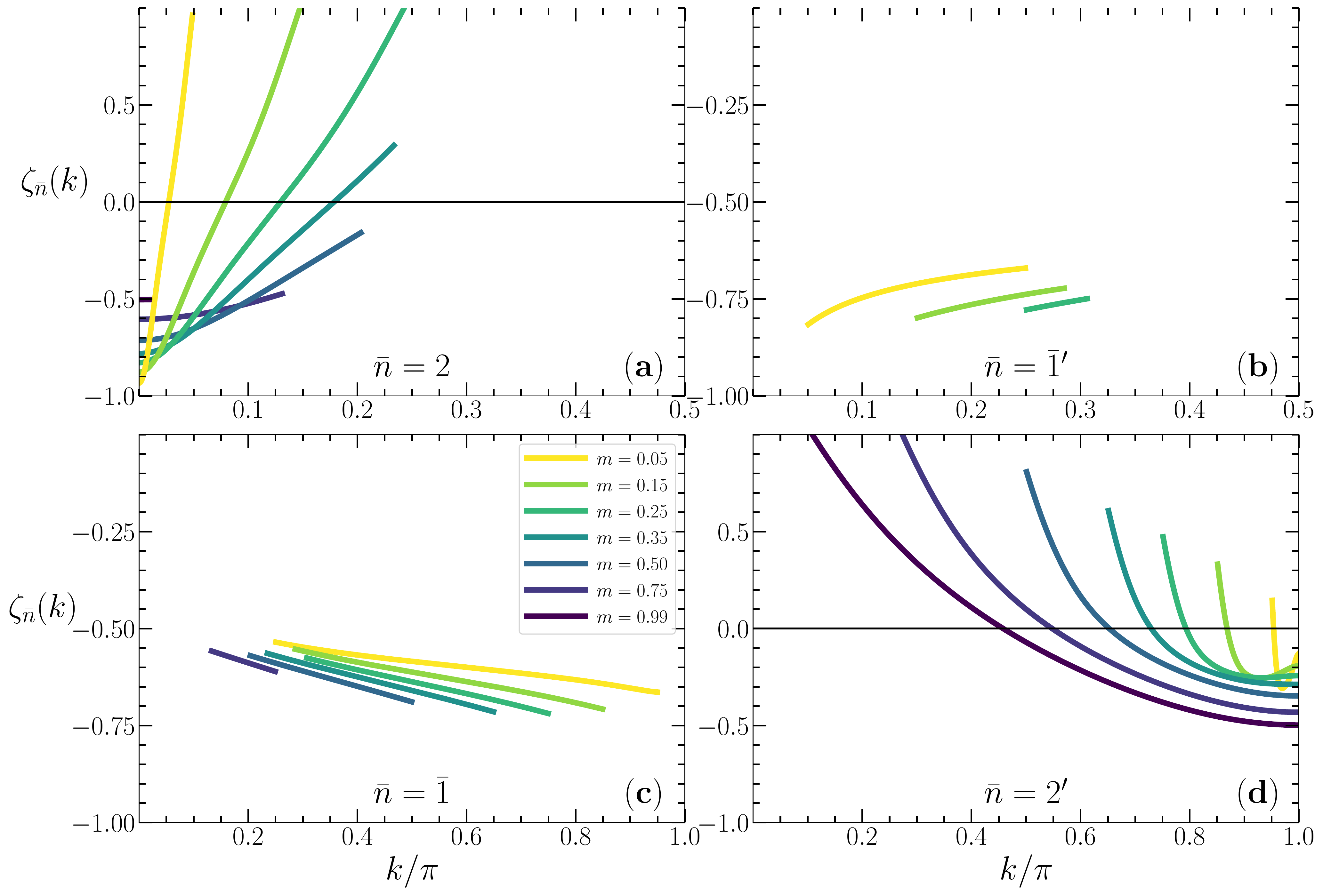}}
\caption{The momentum dependence of the exponents that 
in the $k$ intervals for which they are negative control the
$S^{+-} (k,\omega)$ line shape near and just above the 
(a) $\bar{n}=2$, (b) $\bar{n}=\bar{1}'$, (c) $\bar{n}=\bar{1}$,
and (d) $\bar{n}=2'$ branch lines for several spin densities. The present branch lines are part of the
gapped lower threshold of the $n$-strings continuum displayed in
Fig. \ref{figure1}. The same exponents, in the $k$ intervals
for which they are negative, also control the
$S^{xx} (k,\omega)$ line shape near and just above the 
corresponding branch lines 
in the $n$-strings continuum displayed in Fig. \ref{figure2}.
In the case of the $\bar{n}=\bar{1}$ 
and $\bar{n}=2'$ branch lines, this only applies to the $k$ intervals 
for which there is a gap between them and the upper threshold of the lower
continuum (See Fig. \ref{figure2}).}
\label{figure4}
\end{center}
\end{figure}

In the case of $S^{+-} (k,\omega)$, the gapped lower threshold spectrum $\Delta^{+-} (k)$ is divided
into the following branch-line intervals,
\begin{eqnarray}
\Delta_{2}^{+-} (k) & = & \varepsilon_{2} (k) \hspace{0.20cm}{\rm and}\hspace{0.20cm}k = q
\hspace{0.20cm}{\rm where}
\nonumber \\
k & \in & [0,(k_{F\uparrow}-k_{F\downarrow})[\hspace{0.20cm}{\rm and}
\hspace{0.20cm}q\in [0,(k_{F\uparrow}-k_{F\downarrow})[
\hspace{0.20cm}{\rm for}\hspace{0.20cm}m\in ]0,\tilde{m}]
\nonumber \\
k & \in & [0,{\tilde{k}}[\hspace{0.20cm}{\rm and}\hspace{0.20cm} q\in [0,{\tilde{k}}[\hspace{0.20cm}
\hspace{0.20cm}{\rm for}\hspace{0.20cm}m\in[\tilde{m},1[ \, ,
\label{Ds2}
\end{eqnarray}
\begin{eqnarray}
\Delta_{\bar{1}'}^{+-} (k) & = & 4\mu_B\,h - \varepsilon_{1} (k_{F\uparrow}-k)
\hspace{0.20cm}{\rm and}\hspace{0.20cm}k = k_{F\uparrow} - q
\hspace{0.20cm}{\rm where} 
\nonumber \\
k & \in & ](k_{F\uparrow}-k_{F\downarrow}),\tilde{k}[\hspace{0.20cm}{\rm and}
\hspace{0.20cm}q\in ](k_{F\uparrow}-\tilde{k}),k_{F\downarrow}[
\hspace{0.20cm}{\rm for}\hspace{0.20cm}m\in ]0,\tilde{m}] \, ,
\label{Dsppp1}
\end{eqnarray}
\begin{eqnarray}
\Delta_{\bar{1}}^{+-} (k) & = & 4\mu_B\,h - W_{2} - \varepsilon_{1} (k_{F\downarrow}-k) 
\hspace{0.20cm}{\rm and}\hspace{0.20cm}k = k_{F\downarrow}- q
\hspace{0.20cm}{\rm where} 
\nonumber \\
k & \in & \in ]{\tilde{k}},2k_{F\downarrow}[\hspace{0.20cm}{\rm and}
\hspace{0.20cm}q\in ]-k_{F\downarrow},(k_{F\downarrow} - {\tilde{k}})[
\hspace{0.20cm}{\rm for}\hspace{0.20cm}m\in ]0,\tilde{m}]
\nonumber \\
k & \in & ]{\tilde{k}},2k_{F\downarrow}[\hspace{0.20cm}{\rm and}\hspace{0.20cm}
q\in ]-k_{F\downarrow},(k_{F\downarrow} - {\tilde{k}})[
\hspace{0.20cm}{\rm for}\hspace{0.20cm}m\in[\tilde{m},1[ \, ,
\label{Dsppp2}
\end{eqnarray}
and
\begin{eqnarray}
\Delta_{2'}^{+-} (k) & = & \varepsilon_{2} (k-2k_{F\downarrow}) \hspace{0.20cm}{\rm and}\hspace{0.20cm}
k = 2k_{F\downarrow} + q\hspace{0.20cm}{\rm where}
\nonumber \\
k & \in & ]2k_{F\downarrow},\pi[\hspace{0.20cm}{\rm and}
\hspace{0.20cm}q\in ]0,(k_{F\uparrow}-k_{F\downarrow})[
\hspace{0.20cm}{\rm for}\hspace{0.20cm}m\in ]0,1[ \, .
\label{Ds2p}
\end{eqnarray}
\begin{table}
\begin{center}
\begin{tabular}{|c|c|c|c|c|c|c|} 
\hline
branch line & $k$ in terms of $q$ & $\delta N_1^F$ & $\delta J_1^F$ & $\delta N_1^{NF}$ & $\delta J_{2}$ & $\delta N_{2}$ \\
\hline
$2$ & $k=(k_{F\uparrow} - k_{F\downarrow}) + q$ (for $2$-band $q$) & $-2$ & $-1$ & $0$ & $0$ & $1$ \\
\hline
$\bar{1}$ & $k=k_{F\uparrow} - q$ (for $1$-band $q$) & $-2$ & $-1/2$ & $-1$ & $0$ & $1$ \\
\hline
$\bar{1}'$ & $k=k_{F\uparrow} - q$ (for $1$-band $q$) & $1$ & $-1/2$ & $-1$ & $-1/2$ & $1$ \\
\hline
$2'$ & $k=\pi + q$ (for $2$-band $q$) & $-2$ & $0$ & $0$ & $0$ & $1$ \\
\hline
\end{tabular}
\caption{The same information as in Table \ref{table1} for the $2$-branch line, $\bar{1}$-branch line, 
$\bar{1}'$-branch line, and $2'$-branch line that for the momentum
intervals given in Eqs. (\ref{Ds2pL})-(\ref{Ds2L}) are part of the gapped lower threshold of $S^{zz} (k,\omega)$.}
\label{table2}
\end{center}
\end{table} 

The corresponding $k$ dependent exponents of general form,
Eq. (\ref{expTS}), that appear in the expression, $S^{+-} (k,\omega) = C_{+-}^{\Delta}
(\omega - \Delta_{\bar{n}}^{+-} (k))^{\zeta_{\bar{n}}^{+-} (k)}$,
Eq. (\ref{MPSsFMB}) for $ab=+-$ and $\beta=2,\bar{1}',\bar{1},2'$, are given by,
\begin{eqnarray}
\zeta_{2}^{+-} (k) & = & -1 + \sum_{\iota=\pm 1}\left(- {\iota\over 2\xi_{1\,1}} 
+ \Phi_{1,2}(\iota k_{F\downarrow},q)\right)^2  
\hspace{0.20cm}{\rm for}\hspace{0.20cm}q=k
\hspace{0.20cm}{\rm where}
\nonumber \\
k & \in & ]0,(k_{F\uparrow}-k_{F\downarrow})[
\hspace{0.20cm}{\rm for}\hspace{0.20cm}m\in ]0,\tilde{m}]
\hspace{0.20cm}{\rm and}\hspace{0.20cm}
k \in ]0,{\tilde{k}}[
\hspace{0.20cm}{\rm for}\hspace{0.20cm}m\in[\tilde{m},1[ 
\nonumber \\
\zeta_{\bar{1}'}^{+-} (k) & = & -1 +
\sum_{\iota=\pm 1}\left(- {\xi_{1\,1}\over 2} 
- \Phi_{1,1}(\iota k_{F\downarrow},q)\right)^2
\hspace{0.20cm}{\rm for}\hspace{0.20cm}q=k_{F\uparrow}-k
\hspace{0.20cm}{\rm where}
\nonumber \\
k & \in & ](k_{F\uparrow}-k_{F\downarrow}),\tilde{k}[
\hspace{0.20cm}{\rm for}\hspace{0.20cm}m\in ]0,\tilde{m}] 
\nonumber \\
\zeta_{\bar{1}}^{+-} (k) & = & -1 + \sum_{\iota=\pm 1}
\left(\iota {\xi_{1\,2}^0\over 2} + {\xi_{1\,1}\over 2} 
- \Phi_{1,1}(\iota k_{F\downarrow},q)\right)^2
\hspace{0.20cm}{\rm for}\hspace{0.20cm}q=k_{F\downarrow}-k
\hspace{0.20cm}{\rm where}
\nonumber \\
k & \in & ]{\tilde{k}},2k_{F\downarrow}[
\hspace{0.20cm}{\rm for}\hspace{0.20cm}m\in ]0,\tilde{m}]
\hspace{0.20cm}{\rm and}\hspace{0.20cm}
k \in ]{\tilde{k}},2k_{F\downarrow}[
\hspace{0.20cm}{\rm for}\hspace{0.20cm}m\in ]0,\tilde{m}] 
\nonumber \\
\zeta_{2'}^{+-} (k) & = & -1 + \sum_{\iota=\pm 1}
\left(- {\iota\over 2\xi_{1\,1}} + \xi_{1\,1}
+ \Phi_{1,2}(\iota k_{F\downarrow},q)\right)^2 
\hspace{0.20cm}{\rm for}\hspace{0.20cm}q=k-2k_{F\downarrow}
\hspace{0.20cm}{\rm where}\hspace{0.20cm}k\in ]2k_{F\downarrow},\pi[ \, .
\label{expG+-}
\end{eqnarray}

The three $\iota = \pm 1$ spectral functionals $\Phi_{\iota} (q)$ in the general expression, Eq. (\ref{expTS}),
specific to the exponents given in Eq. (\ref{expG+-}) for the $S^{+-} (k,\omega)$'s 
$2$- and $2'$-branch lines, $\bar{1}'$-branch line,
and $\bar{1}$-branch line are provided in Eqs. (\ref{Fs2}), (\ref{Fbarsl}), and
(\ref{Fbars}), respectively. The corresponding suitable specific values
of the number and current number deviations, Eq. (\ref{NcFNcFJcFJsF}), used in such functionals are
for the present branch lines given in Table \ref{table1}.

The $S^{+-} (k,\omega)$'s $2$-, $\bar{1}'$-, $\bar{1}$-, and $2'$-branch line exponents whose expressions
are given in Eq. (\ref{expG+-}) are plotted as a function of $k$ in Fig. \ref{figure4}.
In the $k$ intervals of the gapped lower threshold of the $n$-string
continuum in Fig. \ref{figure1} for which they are negative, 
which are represented by solid lines, there are singularities
at and just above the corresponding $\bar{n}=2,\bar{1}',\bar{1},2'$ branch lines
in the expression $S^{+-} (k,\omega) = C_{+-}^{\Delta}
(\omega - \Delta_{\bar{n}}^{+-} (k))^{\zeta_{\bar{n}}^{+-} (k)}$,
Eq. (\ref{MPSsFMB}) for $ab=+-$.

The related $S^{xx} (k,\omega)$'s expression, Eq. (\ref{MPSsFMB}) for $ab=xx$, 
also valid at and just above the gapped lower threshold of the $n$-string
continuum in Fig. \ref{figure2}, is similar to that of $S^{+-} (k,\omega)$ and involves exactly the same exponents.
This though applies provided that in the corresponding $k$ intervals there is no overlap between that
continuum and the lower continuum associated with excited states described by only real Bethe-ansatz rapidities. 
For small spin densities, this thus excludes the $k$ intervals given in Eq. (\ref{gapineq}).

In the case of $S^{zz} (k,\omega)$, the expressions
of the gapped lower threshold spectrum $\Delta^{zz} (k)$, Eqs. (\ref{gapPMMP}) and (\ref{gapL}), 
are for the $k$ intervals of each corresponding branch line given by,
\begin{eqnarray}
\Delta_{2}^{zz} (k) & = & \varepsilon_{2} (k - (k_{F\uparrow}-k_{F\downarrow})) 
\hspace{0.20cm}{\rm and}\hspace{0.20cm}k =  (k_{F\uparrow}-k_{F\downarrow}) + q
\hspace{0.20cm}{\rm where} 
\nonumber \\
k & \in & ]0,(k_{F\uparrow}-k_{F\downarrow})[
\hspace{0.20cm}{\rm and}\hspace{0.20cm}q\in ]-(k_{F\uparrow}-k_{F\downarrow}),0[
\hspace{0.20cm}{\rm for}\hspace{0.20cm}m\in]0,1[ \, ,
\label{Ds2pL}
\end{eqnarray}
\begin{eqnarray}
\Delta_{\bar{1}}^{zz} (k) & = & 4\mu_B\,h - W_{2} - \varepsilon_{1} \left(k_{F\uparrow} - k\right)
\hspace{0.20cm}{\rm and}\hspace{0.20cm}k = k_{F\uparrow}- q
\hspace{0.20cm}{\rm where} 
\nonumber \\
k & \in & ](k_{F\uparrow}-k_{F\downarrow}),(\pi - {\tilde{k}})[
\hspace{0.20cm}{\rm and}\hspace{0.20cm}q\in ]-(k_{F\downarrow} - {\tilde{k}}),k_{F\downarrow}[
\hspace{0.20cm}{\rm for}\hspace{0.20cm}m\in ]0,\tilde{m}]
\nonumber \\
k & \in & ](k_{F\uparrow}-k_{F\downarrow}),(\pi - {\tilde{k}})[
\hspace{0.20cm}{\rm and}\hspace{0.20cm}q\in ]-(k_{F\downarrow} - {\tilde{k}}),k_{F\downarrow}[
\hspace{0.20cm}{\rm and}\hspace{0.20cm}{\rm for}\hspace{0.20cm}m\in[\tilde{m},1[ \, ,
\label{Dsppp2L}
\end{eqnarray}
\begin{eqnarray}
\Delta_{\bar{1}'}^{zz} (k) & = & 4\mu_B\,h - \varepsilon_{1} (k_{F\downarrow}-k)
\hspace{0.20cm}{\rm and}\hspace{0.20cm}k = k_{F\downarrow} - q
\hspace{0.20cm}{\rm where} 
\nonumber \\
k & \in & ](\pi - {\tilde{k}}),2k_{F\downarrow}[ 
\hspace{0.20cm}{\rm and}\hspace{0.20cm}q\in ]-k_{F\downarrow},- (k_{F\uparrow}-\tilde{k})[
\hspace{0.20cm}{\rm for}\hspace{0.20cm}m\in ]0,\tilde{m}] \, ,
\label{Dsppp1L}
\end{eqnarray}
and
\begin{eqnarray}
\Delta_{2'}^{zz} (k) & = & \varepsilon_{2} (k-\pi) \hspace{0.20cm}{\rm and}\hspace{0.20cm}k = \pi + q
\hspace{0.20cm}{\rm where} 
\nonumber \\
k & \in & ]2k_{F\downarrow},\pi[
\hspace{0.20cm}{\rm and}\hspace{0.20cm}q\in ]-(k_{F\uparrow}-k_{F\downarrow}),0[
\hspace{0.20cm}{\rm for}\hspace{0.20cm}m\in ]0,\tilde{m}]
\nonumber \\
k & \in & ](\pi - {\tilde{k}}),\pi[ 
\hspace{0.20cm}{\rm and}\hspace{0.20cm}q\in ]-{\tilde{k}},0[\hspace{0.20cm}{\rm for}\hspace{0.20cm}m\in[\tilde{m},1[ \, .
\label{Ds2L}
\end{eqnarray}
\begin{figure}
\begin{center}
\centerline{\includegraphics[width=10cm]{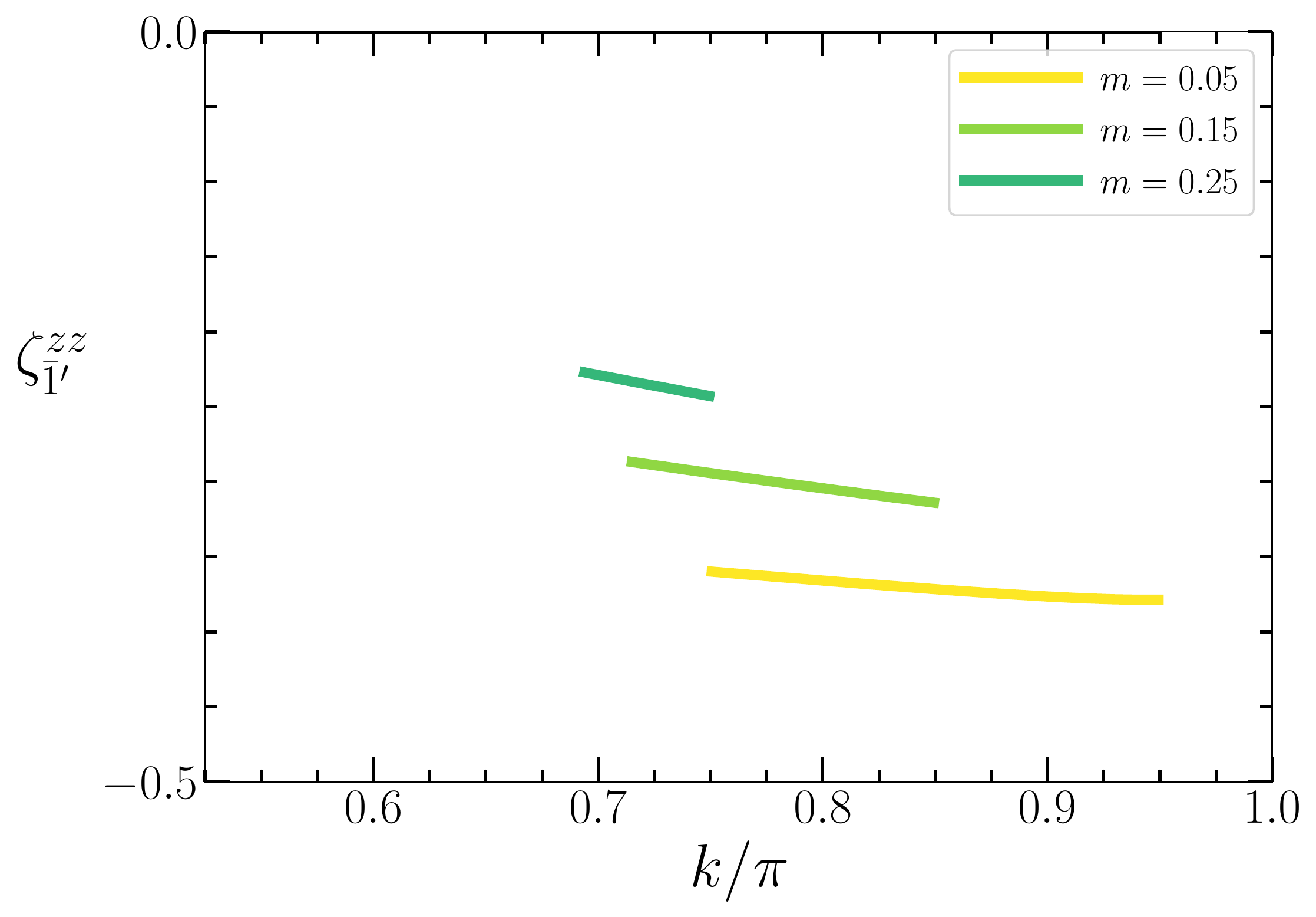}}
\caption{The same as in Fig. \ref{figure4} for the $\bar{1}'$ branch
line of $S^{zz} (k,\omega)$. For that dynamical structure factor, this
exponent is the only one that is negative and refers to singularities near
the corresponding small momentum intervals of the
gapped lower threshold of the $n$-string continuum in
Fig. \ref{figure3}. Such singularities only emerge in $S^{zz} (k,\omega)$
for spin densities $0<m<\tilde{m}$ where $\tilde{m}\approx 0.317$.}
\label{figure5}
\end{center}
\end{figure}

The corresponding $k$ dependent exponents of general form,
Eq. (\ref{expTS}), that appear in the expression, $S^{zz} (k,\omega) = C_{zz}^{\Delta}
(\omega - \Delta_{\bar{n}}^{zz} (k))^{\zeta_{\bar{n}}^{zz} (k)}$,
Eq. (\ref{MPSsFMB}) for $ab=zz$ and $\bar{n}=2,\bar{1},\bar{1}',2'$, read,
\begin{eqnarray}
\zeta_{2}^{zz} (k) & = & -1 + \sum_{\iota=\pm 1}
\left(- {\iota\over \xi_{1\,1}} - \xi_{1\,1}
+ \Phi_{1,2}(\iota k_{F\downarrow},q)\right)^2 
\hspace{0.20cm}{\rm for}\hspace{0.20cm}q=k-k_{F\uparrow}+k_{F\downarrow}
\hspace{0.20cm}{\rm where}
\nonumber \\
k & \in & ]0,(k_{F\uparrow}-k_{F\downarrow})[
\nonumber \\
\zeta_{\bar{1}}^{zz} (k) & = & -1 + \sum_{\iota=\pm 1}
\left(- {\iota\over\xi_{1\,1}} 
+ \iota {\xi_{1\,2}^0\over 2} - {\xi_{1\,1}\over 2} 
- \Phi_{1,1}(\iota k_{F\downarrow},q)\right)^2
\hspace{0.20cm}{\rm for}\hspace{0.20cm}q=k_{F\uparrow}-k
\hspace{0.20cm}{\rm where}
\nonumber \\
k & \in &  ](k_{F\uparrow}-k_{F\downarrow}),(\pi - {\tilde{k}})[
\hspace{0.20cm}{\rm for}\hspace{0.20cm}m\in ]0,\tilde{m}]
\hspace{0.20cm}{\rm and}\hspace{0.20cm}
k \in ](k_{F\uparrow}-k_{F\downarrow}),(\pi - {\tilde{k}})[
\hspace{0.20cm}{\rm for}\hspace{0.20cm}m\in ]0,\tilde{m}] 
\nonumber \\
\zeta_{\bar{1}'}^{zz} (k) & = & -1 + \sum_{\iota=\pm 1}
\left({\iota\over 2\xi_{1\,1}} + {\xi_{1\,1}\over 2} 
- \Phi_{1,1}(\iota k_{F\downarrow},q)\right)^2 
\hspace{0.20cm}{\rm for}\hspace{0.20cm}q=k_{F\downarrow}-k
\hspace{0.20cm}{\rm where}
\nonumber \\
k & \in & ](\pi - {\tilde{k}}),2k_{F\downarrow}[
\hspace{0.20cm}{\rm for}\hspace{0.20cm}m\in ]0,\tilde{m}] 
\nonumber \\
\zeta_{2'}^{zz} (k) & = & -1 + \sum_{\iota=\pm 1}\left(- {\iota\over\xi_{1\,1}} 
+ \Phi_{1,2}(\iota k_{F\downarrow},q)\right)^2 
\hspace{0.20cm}{\rm for}\hspace{0.20cm}q=k-\pi
\hspace{0.20cm}{\rm where}
\nonumber \\
k & \in & ]2k_{F\downarrow},\pi[
\hspace{0.20cm}{\rm for}\hspace{0.20cm}m\in ]0,\tilde{m}]
\hspace{0.20cm}{\rm and}\hspace{0.20cm}
k \in ](\pi - {\tilde{k}}),\pi[
\hspace{0.20cm}{\rm for}\hspace{0.20cm}m\in[\tilde{m},1[ \, .
\label{exps2pL}
\end{eqnarray}
Also in the present case of $S^{zz} (k,\omega)$, the three $\iota = \pm 1$ spectral functionals $\Phi_{\iota} (q)$ in the general
expression, Eq. (\ref{expTS}), specific to the $2$- and $2'$-branch lines, $\bar{1}$-branch line,
and $\bar{1}'$-branch line are provided in Eqs. (\ref{Fs2})-(\ref{Fbars}). The corresponding suitable 
values of the number and current number deviations, Eq. (\ref{NcFNcFJcFJsF}), 
given in Table \ref{table2} and used in such functionals to reach the expressions in Eq. (\ref{exps2pL})
are though different relative to those used for $S^{+-} (k,\omega)$.

The corresponding behaviors of the spin dynamical structure factor $S^{zz} (k,\omega)$ are also
qualitatively different from those of $S^{+-} (k,\omega)$. Except for $\zeta_{\bar{1}'}^{zz} (k)$, the exponents in Eq. (\ref{exps2pL}) 
are positive for all their $k$ intervals. The $\bar{1}'$-branch line's exponent
is plotted as a function of $k$ in Fig. \ref{figure5}. It is negative for its whole $k$ subinterval, which
is part of the $k$ interval of the gapped lower threshold in Figs. \ref{figure3} (a) and (b).
The $\bar{1}'$-branch line's $m$-dependent subinterval is though either small or 
that line is not part of the $S^{zz} (k,\omega)$'s gapped lower threshold at all. Its momentum width decreases upon increasing 
$m$ up to a spin density $\tilde{m}\approx 0.317$. For $\tilde{m}<m<1$, the $\bar{1}'$-branch line
is not part of the $S^{zz} (k,\omega)$'s gapped lower threshold spectrum.
This is why for $m=0.50>\tilde{m}$ and $m=0.75>\tilde{m}$ that line does not appear in the
gapped lower threshold in Figs. \ref{figure3} (c) and (d), respectively.

One then concludes that gapped lower threshold's singularities only emerge in $S^{zz} (k,\omega)$
for spin densities $0<m<\tilde{m}$ at and just above the $\bar{1}'$-branch line,
the corresponding line shape reading, 
$S^{zz} (k,\omega) = C_{zz}^{\Delta} (\omega - \Delta_{\bar{1}'}^{zz} (k))^{\zeta_{\bar{1}'}^{+-} (k)}$.
That branch line $k$ subinterval width though strongly decreases upon increasing $m$ up to $\tilde{m}$. 
 \begin{table}
\begin{center}
\begin{tabular}{|c|c|c|c|c|c|c|} 
\hline
$1$-branch line & $k$ in terms of $q$ & $k$ interval & $\delta N_1^F$ & $\delta J_1^F$ & $\delta N_1^{NF}$ \\
\hline
$-+$ & $k=k_{F\uparrow} - q$ & $k\in ](k_{F\uparrow}-k_{F\downarrow}),\pi[$ & $0$ & $-1/2$ & $-1$ \\
\hline
$+-$ & $k=k_{F\uparrow} + q$ & $k\in [0,(k_{F\uparrow}-k_{F\downarrow})[$ & $0$ & $-1/2$ & $1$ \\
\hline
$+-$ & $k=k_{F\uparrow} - q$ & $k\in ](k_{F\uparrow}-k_{F\downarrow}),\pi[$ & $2$ & $-1/2$ & $-1$ \\
\hline
$zz$ & $k=k_{F\downarrow} - q$ & $k\in ]0,2k_{F\downarrow}[$ & $1$ & $1/2$ & $-1$ \\
\hline
$zz$ & $k=k_{F\downarrow} + q$ & $k\in ]2k_{F\downarrow},\pi] $ & $-1$  & $1/2$ & $1$ \\
\hline
\end{tabular}
\caption{The excitation momenta $k>0$ of (i) $S^{-+} (k,\omega)$ and (ii) $S^{+-} (k,\omega)$ and $S^{zz} (k,\omega)$
expressed in terms of $1$-band momenta $q$ and corresponding number and current number 
deviations for the (i) $1$-branch line and (ii) two $1$-branch line sections that coincide with the corresponding lower thresholds
of the lower continua shown in Figs. \ref{figure1}-\ref{figure3}.}
\label{table3}
\end{center}
\end{table}

These behaviors are consistent with the $S^{zz} (k,\omega)$'s spectral 
weight stemming from $n$-string states being suppressed upon increasing the spin
density $m$ within the interval $m\in ]0,\tilde{m}]$. Within it, that weight decreases
upon increasing the spin density, becoming negligible for $\tilde{m}<m<1$.

\subsection{The line shape near the lower thresholds of the spin dynamical structure factors}
\label{SECVB}

In order to provide an overall physical picture that accounts for all gapped lower threshold's singularities and
lower threshold's singularities in the spin dynamical structure factors, here we shortly revisit their line shape
behavior at and just above the lower thresholds of the lower continua in Figs. \ref{figure1}-\ref{figure3}. 
The corresponding dominant contributions are from excited states described by only real Bethe-ansatz 
rapidities \cite{Carmelo_15A}. Such lower continua contain most spectral weight of the corresponding spin 
dynamical structure factors.

In the case of the transverse dynamical structure factor,
$S^{xx} (k,\omega) = {1\over 4}\left(S^{+-} (k,\omega)+S^{-+} (k,\omega)\right)$,
we consider the transitions to excited states that determine the line shape in the vicinity
of the lower thresholds of both $S^{+-} (k,\omega)$ and $S^{-+} (k,\omega)$, respectively. 
The lower threshold's spectrum $\omega_{lt}^{xx} (k)$ of $S^{xx} (k,\omega)$ is
given in Eq. (\ref{OkxxRs}) of Appendix \ref{A}. The corresponding two-parametric
spectrum $\omega^{xx} (k)$ results from the superposition of the corresponding two-parametric spectra,
$\omega^{+-} (k)$ and $\omega^{-+} (k)$, Eqs. (\ref{dkEdPxxMP}) and (\ref{dkEdPxxPM}) of that Appendix, respectively.
It refers to the $(k,\omega)$-plane lower continuum in Fig. \ref{figure2}.

For spin densities $0<m<1$ and momenta $k\in ]0,\pi[$, the 
line shape of the spin dynamical structure factors $S^{ab} (k,\omega)$ where $ab = +-,-+,xx,zz$ 
at and just above their lower thresholds has the general form given in Eq. (\ref{MPSs}).
In the case of $S^{xx} (k,\omega)$, this expression can be expressed as,
\begin{eqnarray}
&& S^{xx} (k,\omega) = S^{+-} (k,\omega)\hspace{0.20cm} {\rm for}\hspace{0.20cm} 
k\in [0,(k_{F\uparrow}-k_{F\downarrow})[
\nonumber \\
&& \hspace{1.25cm} = S^{-+} (k,\omega)\hspace{0.20cm} {\rm for}
\hspace{0.20cm}k \in ](k_{F\uparrow}-k_{F\downarrow}),\pi[ \, .
\label{MPSsGen}
\end{eqnarray}
\begin{figure}
\begin{center}
\centerline{\includegraphics[width=10cm]{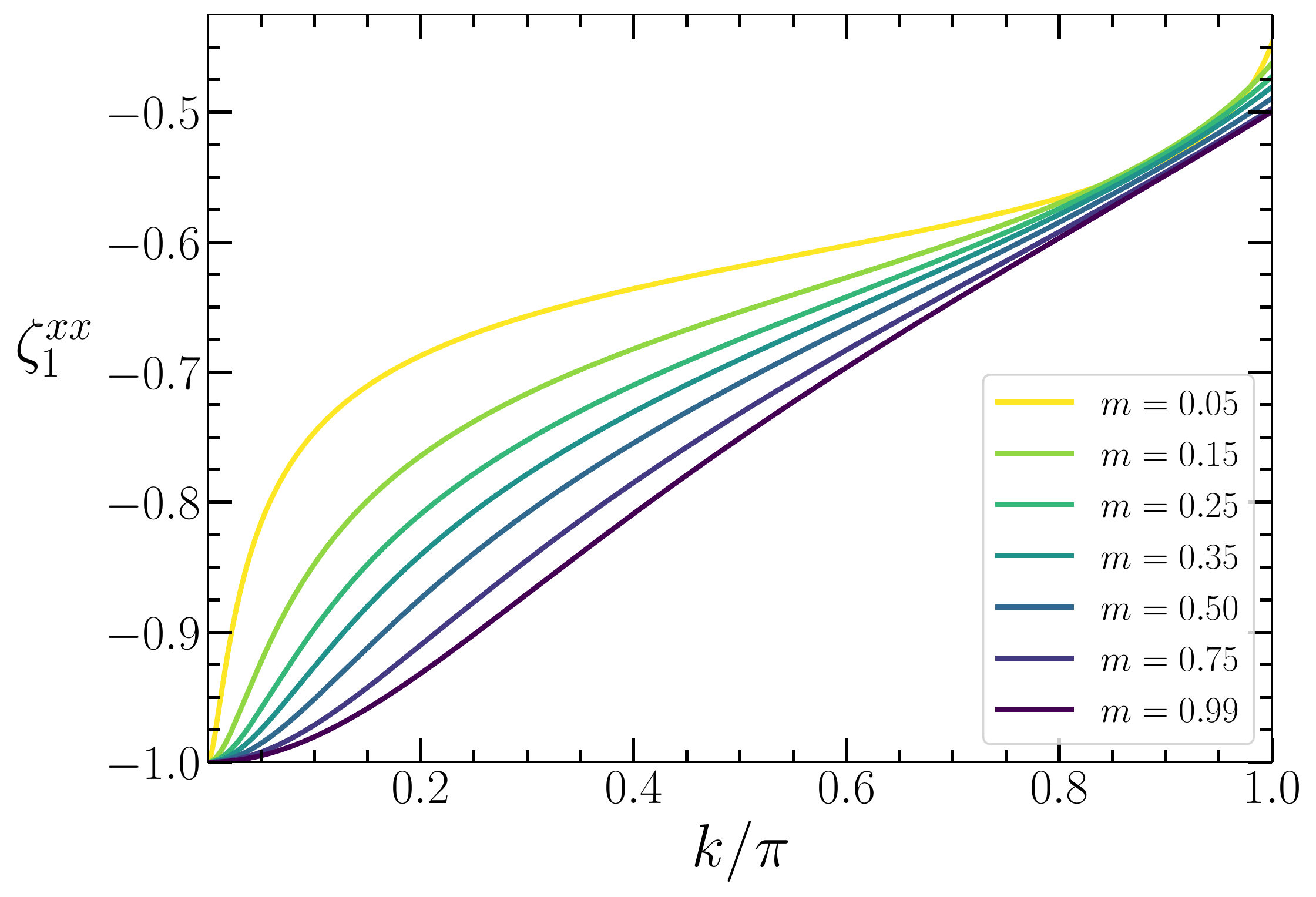}}
\caption{The momentum dependence of the exponent that controls the
$S^{xx} (k,\omega)$'s line shape near and just above the lower
threshold of the lower continuum in Fig. \ref{figure2}
for several spin densities. For $k\in ]0,(k_{F\uparrow}-k_{F\downarrow})[$ and 
$k\in ](k_{F\uparrow}-k_{F\downarrow}),\pi[$ that exponent corresponds
to that of $S^{+-} (k,\omega)$ and $S^{-+} (k,\omega)$, respectively.}
\label{figure6}
\end{center}
\end{figure}

The $k$ dependent exponents appearing in the spin dynamical factors's
expression, Eq. (\ref{MPSs}), are of general form,
Eq. (\ref{expTS}). Their specific expressions for the different $1$-branch lines
and $1$-branch line sections under consideration read,
\begin{eqnarray}
\zeta_{1}^{-+} (k) & = & -1 + \sum_{\iota =\pm 1}\left(- {\xi_{1\,1}\over 2} 
- \Phi_{1,1}(\iota k_{F\downarrow},q)\right)^2 
\nonumber \\
&& {\rm for}\hspace{0.20cm}q=k_{F\uparrow}-k
\hspace{0.20cm}{\rm and}\hspace{0.20cm}
k\in ](k_{F\uparrow}-k_{F\downarrow}),\pi[
\nonumber \\
\zeta_{1}^{+-} (k) & = & -1 + \sum_{\iota =\pm 1}\left(- {\xi_{1\,1}\over 2} 
+ \Phi_{1,1}(\iota k_{F\downarrow},q)\right)^2 
\nonumber \\
&& {\rm for}\hspace{0.20cm}q=k-k_{F\uparrow}
\hspace{0.20cm}{\rm and}\hspace{0.20cm}k\in ]0,(k_{F\uparrow}-k_{F\downarrow})[
\nonumber \\
 \zeta_{1}^{+-} (k) & = & -1 + \sum_{\iota =\pm 1}\left({\iota\over\xi_{1\,1}} - {\xi_{1\,1}\over 2} 
- \Phi_{1,1}(\iota k_{F\downarrow},q)\right)^2 
\nonumber \\
&& {\rm for}\hspace{0.20cm}q=k_{F\uparrow}-k
\hspace{0.20cm}{\rm and}\hspace{0.20cm}k\in ](k_{F\uparrow}-k_{F\downarrow}),\pi[
\nonumber \\
\zeta_{1}^{zz} (k) & = & -1 +
\sum_{\iota =\pm 1}\left({\iota\over 2\xi_{1\,1}}  + {\xi_{1\,1}\over 2} 
- \Phi_{1,1}(\iota k_{F\downarrow},q)\right)^2
\nonumber \\
&& {\rm for}\hspace{0.20cm}q=k_{F\downarrow}-k
\hspace{0.20cm}{\rm and}\hspace{0.20cm}k\in ]0,2k_{F\downarrow}[
\nonumber \\
\zeta_{1}^{zz} (k) & = & -1 + \sum_{\iota =\pm 1}\left(- {\iota\over 2\xi_{1\,1}} + {\xi_{1\,1}\over 2} 
+ \Phi_{1,1}(\iota k_{F\downarrow},q)\right)^2 
\nonumber \\
&& {\rm for}\hspace{0.20cm}q=k-k_{F\downarrow}
\hspace{0.20cm}{\rm and}\hspace{0.20cm}k\in ]2k_{F\downarrow},\pi[ \, .
\label{expsPM}
\end{eqnarray}
The corresponding $\iota = \pm 1$ spectral functionals $\Phi_{\iota} (q)$ in the exponent expression, Eq. (\ref{expTS}), 
have for the present $1$-branch lines and $1$-branch line sections the general form given in Eq. (\ref{Fs}). 
The values of the $1$-band number and current number deviations, Eq. (\ref{NcFNcFJcFJsF}), that are used 
in Eq. (\ref{Fs}) to reach the above specific exponents's expressions are provided in Table \ref{table3}.
\begin{figure}
\begin{center}
\centerline{\includegraphics[width=10cm]{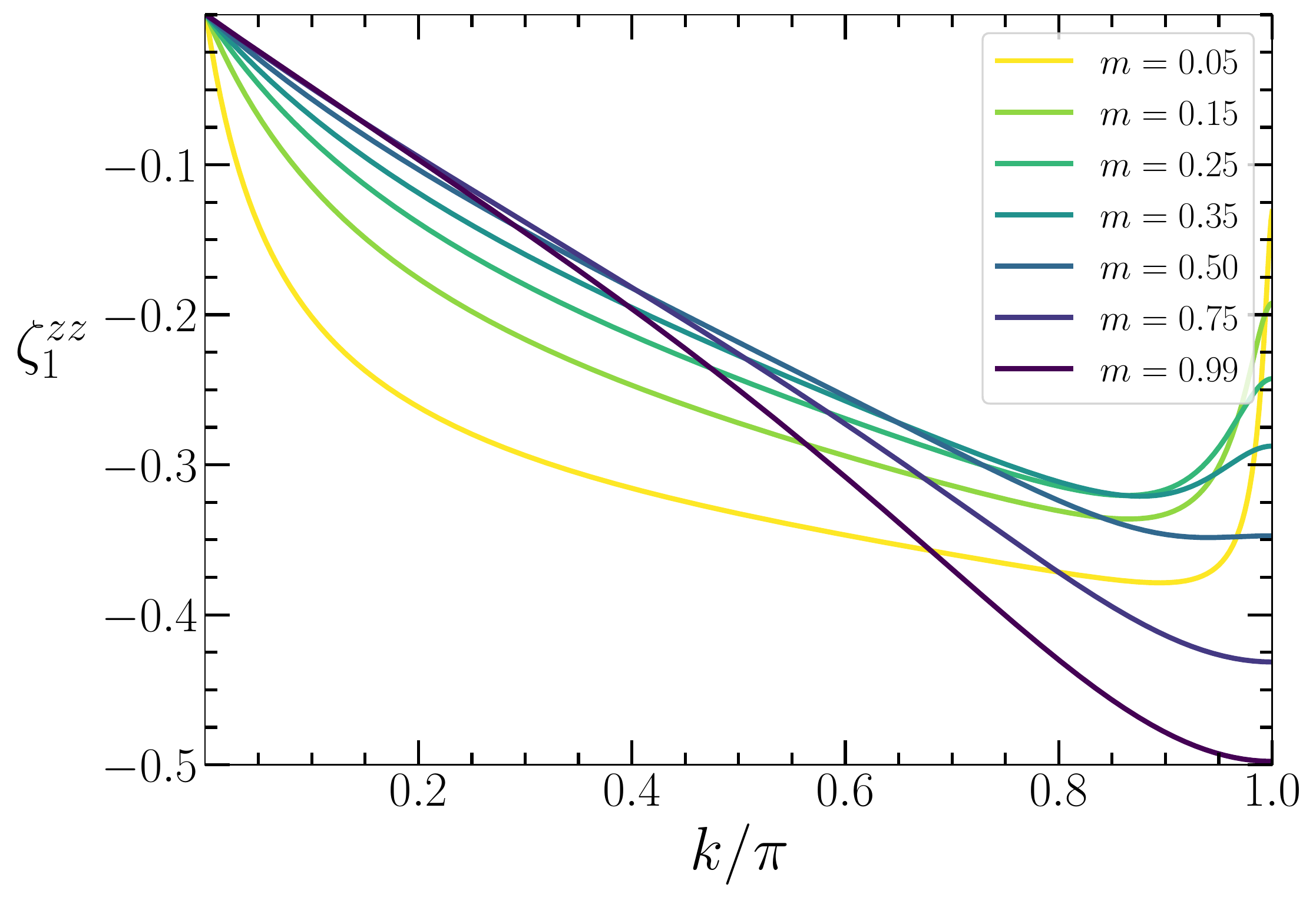}}
\caption{The momentum dependence of the exponent that controls the
$S^{zz} (k,\omega)$ line shape near and just above the lower
threshold of the lower continuum in Fig. \ref{figure3}
for several spin densities.}
\label{figure7}
\end{center}
\end{figure}  

As confirmed by the form of the expressions given in Eqs. (\ref{OkPMRs}) and (\ref{OkMPRs}) of Appendix \ref{A},
the lower threshold' spectra equality $\omega^{+-}_{lt} (k) = \omega^{-+}_{lt} (k)$ holds for $k\in [(k_{F\uparrow}-k_{F\downarrow}),\pi]$.
On the other hand, $\omega^{+-}_{lt} (k) = \omega^{xx}_{lt} (k)$ for $k\in [0,\pi]$, as confirmed by inspection of
Eqs. (\ref{OkMPRs}) and (\ref{OkxxRs}) of that Appendix. In the $k\in ](k_{F\uparrow}-k_{F\downarrow}),\pi[$ interval, the line shape of 
$S^{xx} (k,\omega) = {1\over 4}\left(S^{+-} (k,\omega)+S^{-+} (k,\omega)\right)$ is
controlled by the smallest of the exponents $\zeta_{1}^{-+} (k)$ and $\zeta_{1}^{+-} (k)$
in Eq. (\ref{expsPM}), which turns out to be $\zeta_{1}^{-+} (k)$.  In the $k\in ]0,(k_{F\uparrow}-k_{F\downarrow})[$ interval
it is rather controlled by the exponent $\zeta_{1}^{+-} (k)$. The resulting exponent $\zeta_{1}^{xx} (k)$ is thus given by,
\begin{eqnarray}
\zeta_{1}^{xx} (k) & = & -1 + \sum_{\iota =\pm 1}\left(-{\xi_{1\,1}\over 2} 
+ \Phi_{1,1}(\iota k_{F\downarrow},q)\right)^2
\hspace{0.20cm}{\rm for}
\nonumber \\
&& q=k-k_{F\uparrow}
\hspace{0.20cm}{\rm and}\hspace{0.20cm}k\in ]0,(k_{F\uparrow}-k_{F\downarrow})[
\nonumber \\
& = & -1 + \sum_{\iota =\pm 1}\left(-{\xi_{1\,1}\over 2} 
- \Phi_{1,1}(\iota k_{F\downarrow},q)\right)^2 
\hspace{0.20cm}{\rm for}
\nonumber \\
&& q=k_{F\uparrow}-k
\hspace{0.20cm}{\rm and}\hspace{0.20cm}k\in ](k_{F\uparrow}-k_{F\downarrow}),\pi[ \, .
\label{expszz}
\end{eqnarray}
This exponent is plotted as a function of $k$ in Fig. \ref{figure6}. On the other hand,
the $1$-branch line exponent $\zeta_{1}^{zz} (k)$ whose expression is given in Eq. (\ref{expsPM}) 
is plotted as a function of that excitation momentum in Fig. \ref{figure7}.

Both such exponents are for spin densities $0<m<1$ negative in the whole momentum 
interval $k\in ]0,\pi[$. It follows that there are singularities at and just above the corresponding 
lower thresholds. (Due to a sign error, the minus sign in the quantity $-\xi_{1\,1}/2$ appearing
in Eq. (\ref{expszz}) was missed in Ref. \cite{Carmelo_15A} where
the exponent $\zeta_{1}^{xx} (k)$ was named $\xi^t$. Its momentum dependence
plotted in Fig. \ref{figure6} corrects that plotted in Figs. 2-4 of that reference.)

\section{Limiting behaviors of the spin dynamical structure factors}
\label{SECVI}

At zero magnetic field, $h=0$, and thus spin density $m=0$, the spin dynamical structure factor $S^{-+} (k,\omega)$ 
equals that obtained in the $m\rightarrow 0$ limit from $m>0$ values. On the other hand,
at $h=0$ and $m=0$ the spin dynamical structure factor $S^{+-} (k,\omega)$
equals that obtained in the $m\rightarrow 0$ limit from $m<0$ values. This is
consistent with the relation, Eq. (\ref{SPMMPmm}). As required by the $SU(2)$ symmetry, this confirms as well that 
$S^{-+} (k,\omega)=S^{+-} (k,\omega)$ at $h=0$ and $m=0$. That symmetry also requires that at $h=0$ and $m=0$ 
the overall equality $S^{xx} (k,\omega) = S^{yy} (k,\omega) = S^{zz} (k,\omega)$ holds. (For the ranges
$0<h<h_c$ and $0<m<1$ considered in this paper, the equality $S^{xx} (k,\omega) = S^{yy} (k,\omega)$
remains valid, whereas one has that $S^{zz} (k,\omega) \neq S^{xx} (k,\omega)$.)

The corresponding two parametric spectrum at $h=0$ and $m=0$ of the spin dynamical
structure factors equals that obtained in the $m\rightarrow 0$ limit from the spectrum 
$\omega^{-+} (k)$ whose expression is given in Eq. (\ref{dkEdPxxMP}) of Appendix \ref{A} 
for $0<h<h_c$ and $0<m<1$. On the other hand, in that limit both the spectra $\omega^{+-} (k)$ and $\omega^{zz} (k)$ in
Eqs. (\ref{dkEdPxxPM}) and (\ref{dkEdPl}) of that Appendix become one-parametric and coincide 
with the lower threshold of the two-parametric spectrum of $S^{xx} (k,\omega) = S^{yy} (k,\omega) = S^{zz} (k,\omega)$
and thus also of $S^{-+} (k,\omega)=S^{+-} (k,\omega)$ at $h=0$ and $m=0$.

What is then the origin of the two-parametric spectrum $S^{zz} (k,\omega)$ at $h=0$ and $m=0$?
At vanishing magnetic field, $h=0$, the selection rules, Eq. (\ref{SRh0}), reveal that 
$S^{zz} (k,\omega)$ is fully controlled by transitions from the $S =S^z=0$ ground state 
to triplet excited states with spin numbers $S=1$ and $S^z=0$. 
This is different from the case considered in this paper, for which the initial ground state refers to 
$0<h<h_c$ and $0<m<1$. According to the corresponding selection rules, Eq. (\ref{SRhfinite}), 
$S^{zz} (k,\omega) \neq S^{xx} (k,\omega)$ is then controlled by transitions from the ground state 
with spin numbers $S^z = S$ or $S^z = -S$ to excited energy eigenstates with
the same spin numbers, $S^z = S$ or $S^z = -S$, respectively.

For the initial ground state at $h=0$ and $m=0$, one has more generally that (i) $S^{zz} (k,\omega)$ and
(ii) $S^{+-} (k,\omega)$ and $S^{-+} (k,\omega)$ are fully controlled by transitions to spin triplet 
$S=1$ excited states with (i) $S^z=0$ and (ii) $S^z=\pm 1$, respectively. Their $1$-band two-hole spectrum is obtained 
in the $m\rightarrow 0$ limit from that of $S^{+-} (k,\omega)$
for $m<0$ and from that of $S^{-+} (k,\omega)$ for $m>0$. One then finds that,
\begin{eqnarray}
\omega^{xx} (k) & = & \omega^{zz} (k) = - \varepsilon_{1} (q) - \varepsilon_{1} (q')
= J {\pi\over 2} (\cos (q) + \cos (q')) \hspace{0.20cm}{\rm where}\hspace{0.20cm} k = \iota\pi - q - q' \hspace{0.20cm}
{\rm and}\hspace{0.20cm}\iota = \pm 1
\nonumber \\
& & {\rm for}\hspace{0.20cm}q \in [-\pi/2,\pi/2]
\hspace{0.20cm}{\rm and}\hspace{0.20cm}q' \in [-\pi/2,\pi/2] \, .
\label{spectram0}
\end{eqnarray}

These results are consistent with spin $SU (2)$ symmetry implying that the triplet $S=1$ and $S^z=0$ excited 
energy eigenstates that control $S^{zz} (k,\omega)$ have exactly the same 
spectrum, Eq. (\ref{spectram0}), as the triplet $S=1$ and $S^z=\pm 1$ excited states
that control $S^{+-} (k,\omega)$ and $S^{-+} (k,\omega)$.

On the other hand, the two parametric spectrum given in Eq. (\ref{dkEdPPM}) refers to the $S^{+-} (k,\omega)$'s gapped continuum
of the $n$-string states shown in Fig. \ref{figure1}. It was obtained in this paper for for $0<h<h_c$ and $0<m<1$.
In the $m\rightarrow 0$ limit from such $m>0$ values, it becomes again a one-parametric line
that coincides both with its $\bar{1}$- and $\bar{1}'$-branch lines, which extend to $k\in ]0,\pi[$,
and with the lower threshold of $S^{xx} (k,\omega) = S^{yy} (k,\omega) = S^{zz} (k,\omega)$
and $S^{-+} (k,\omega)=S^{+-} (k,\omega)$ at $h=0$ and $m=0$.

It is confirmed in the following that, in spite of the singular behavior concerning the class 
of excited states that control the spin dynamical structure factors for magnetic
fields $h=0$ and $h\neq 0$, respectively, the same line shape at and above
the lower thresholds of such factors is obtained at $m=0$ and in the $m\rightarrow 0$ limit,
respectively.

\subsection{Behaviors of the spin dynamical structure factors in the $m\rightarrow 0$ limit}
\label{SECVIA}

Taking the $m\rightarrow 0$ limit from $m>0$ values, one confirms
that the lower threshold's spectrum, Eq. (\ref{OkPMRs}) of Appendix \ref{A},
of $S^{-+} (k,\omega)$ expands to $k\in [0,\pi]$. One then finds that within the present
dynamical theory the corresponding line shape at and just above the $1$-branch line becomes valid for $k\in]0,\pi[$. 
An exactly equal spectrum is obtained for the lower threshold of $S^{+-} (k,\omega)$ 
in the $m\rightarrow 0$ limit from $m<0$ values. In the $m\rightarrow 0$ limit, the 
lower threshold's spectrum of $S^{xx} (k,\omega) = {1\over 4}\left(S^{+-} (k,\omega)+S^{-+} (k,\omega)\right)$ 
is then found to read,
\begin{equation}
\omega^{xx}_{lt} (k) = - \varepsilon_1 (\pi/2-k) = J {\pi\over 2} \sin k \hspace{0.2cm}{\rm for}
\hspace{0.20cm}k \in]0,\pi[ \, .
\label{OkPMRs0}
\end{equation}

On the other hand and as reported above, the gapped continuum, Eq. (\ref{dkEdPPM}), of the $n$-string states
becomes in the $m\rightarrow 0$ limit a $(k,\omega)$-plane line whose spectrum is as well given by
$\Delta_{\bar{1}'}^{+-} (k) = \Delta_{\bar{1}}^{+-} (k) = J {\pi\over 2} \sin k$  for $k \in]0,\pi[$.
By use of the limiting behaviors $\lim_{m\rightarrow 0}\Phi_{1,1}\left(\pm k_{F\downarrow},q\right) = \pm 1/(2\sqrt{2})$ for
$q\neq \pm k_{F\downarrow}$, $\lim_{m\rightarrow 0}\Phi_{1,2}\left(\pm k_{F},0\right) = \pm 1/\sqrt{2}$,
and $\lim_{m\rightarrow 0}\xi_{1\,1} = 1/\sqrt{2}$ provided in Eqs. (\ref{Phis-all-qq-0}), (\ref{Phis2-all-qq-0}), 
and (\ref{Limxiss}) of Appendix \ref{B}, one finds that the corresponding exponents $\zeta_{\bar{1}'}^{+-} (k)$ and 
$\zeta_{\bar{1}}^{+-} (k)$, Eq. (\ref{expG+-}), as well as the exponent $\zeta_{1}^{xx} (k)$, Eq. (\ref{expszz}), become 
in the $m\rightarrow 0$ limit equal and read,
\begin{eqnarray}
\zeta_{1}^{xx} (k) & = & \zeta_{\bar{1}'}^{+-} (k) = -1 + \sum_{\iota=\pm 1}\left(- {\xi_{1\,1}\over 2} 
- \Phi_{1,1}(\iota \pi/2,q)\right)^2  = - {1\over 2} 
\nonumber \\
\zeta_{\bar{1}}^{+-} (k) & = & -1 + \sum_{\iota=\pm 1}\left(\iota {\xi_{1\,2}^0\over 2}
+ {\xi_{1\,1}\over 2} 
- \Phi_{1,1}(\iota k_{F\downarrow},q)\right)^2 = - {1\over 2} \, .
\label{expsPM0}
\end{eqnarray}
The exponent $\zeta_{1}^{xx} (k)$, Eq. (\ref{expszz}), has for $0<m<1$ 
two expressions associated with two $k$ intervals, respectively. The momentum
width of one of such intervals vanishes in the $m\rightarrow 0$ limit. This is why
only one of its two expressions contributes to the equality in this equation, which only 
occurs in that limit.

Consistent with spin $SU(2)$ symmetry requirements, we confirm that in the $m\rightarrow 0$ limit
in which the two parametric spectrum $\omega^{zz} (k)$ in Eq. (\ref{dkEdPl}) of Appendix \ref{A} 
becomes a one-parametric line that coincides with the lower threshold's spectrum of $S^{zz} (k,\omega)$, 
Eq. (\ref{OkPMRsL}) of that Appendix, and in addition becomes equal to the lower threshold's spectrum $\omega^{xx}_{lt} (k)$, 
Eq. (\ref{OkPMRs0}), and thus reads $\omega^{zz}_{lt} (k) = \omega_{1}^{zz} (k) = J {\pi\over 2} \sin k$ for $k \in]0,\pi[$.

Similarly, the $S^{zz} (k,\omega)$'s gapped $\bar{1}$- and $\bar{1}'$-branch line spectra in Eqs. (\ref{Dsppp2L}) and (\ref{Dsppp1L}), 
respectively, are in the $m\rightarrow 0$ limit found to become equal and gapless and to expand to the whole $k\in ]0,\pi[$ 
interval. In that limit, they coincide with the $S^{zz} (k,\omega)$'s lower threshold, their
spectrum thus again reading, $\Delta_{\bar{1}'}^{zz} (k) = J {\pi\over 2} \sin k$ for $k \in]0,\pi[$.

One then finds that in the $m\rightarrow 0$ limit the corresponding exponents
$\zeta_{\bar{1}}^{zz} (k)$ and  $\zeta_{\bar{1}'}^{zz} (k)$, Eq. (\ref{exps2pL}),
obey the inequality, $\zeta_{\bar{1}'}^{zz} (k)< \zeta_{\bar{1}}^{zz} (k)$. 
The smaller exponent $\zeta_{\bar{1}'}^{zz} (k)$ is associated with the spectrum in Eq. (\ref{Dsppp1L}).
That inequality has a physical meaning, at it reveals that the line shape is controlled by the 
exponents $\zeta_{\bar{1}'}^{zz} (k)$ and $\zeta_{1}^{zz} (k)$ 
such that $\zeta_{\bar{1}'}^{zz} (k)=\zeta_{1}^{zz} (k)$ in that limit.
By the use of the behaviors provided in Eqs. (\ref{Phis-all-qq-0}) and (\ref{Limxiss}) of Appendix \ref{B},
one then finds that in the same limit the exponents $\zeta_{1}^{zz} (k)$, Eq. (\ref{expsPM}),
and $\zeta_{\bar{1}'}^{zz} (k)$, Eq. (\ref{exps2pL}), equal as well those given in Eq. (\ref{expsPM0}) and read,
\begin{equation}
\zeta_{1}^{zz} (k) = \zeta_{\bar{1}'}^{zz} (k) = -1 + \sum_{\iota=\pm 1}\left({\iota\over 2\xi_{1\,1}} + {\xi_{1\,1}\over 2} 
- \Phi_{1,1}(\iota \pi/2,q)\right)^2 = - {1\over 2} \, .
\label{expsppp1L0}
\end{equation}
Also in this case, the exponent $\zeta_{1}^{zz} (k)$ given in Eq. (\ref{expsPM})
has for $0<m<1$ two expressions associated with two $k$ intervals, respectively. Only one of such expressions
contributes to the equality in this equation, which only holds in the $m\rightarrow 0$ limit.
This follows from the momentum width of one of the two $0<m<1$ excitation momentum $k$'s intervals 
considered in Eq. (\ref{expsPM}) vanishing in that limit.

The above results then confirm that in the $m\rightarrow 0$ limit the line shape
at and just above the lower threshold of the spin dynamical structure factors is 
controlled by the exponent $-1/2$, which is a result known to hold
at zero magnetic field \cite{Imambekov_09,Bougourzi_97,Caux_11}.
The corresponding line-shape expression reads,
\begin{equation}
S^{aa} (k,\omega) = C\,(\omega - \omega (k))^{-1/2}\hspace{0.50cm}{\rm where}
\hspace{0.50cm}\omega (k) = J {\pi\over 2} \sin k \hspace{0.20cm}{\rm for}\hspace{0.20cm}]0,\pi[ \, ,
\label{DSF-BL-m0}
\end{equation}
where $aa = xx,yy,zz$ and $C$ is a constant that has a fixed value for the $k$ and $\omega$ ranges corresponding 
to small values of the energy deviation $(\omega - \omega (k))$.

\subsection{Behaviors of the spin dynamical structure factors in the $m\rightarrow 1$ limit}
\label{SECVIB}

The sum rules, Eq. (\ref{SRDSF}), reveal that both the spin dynamical structure factrs $S^{-+} (k,\omega)$ and
$S^{zz} (k,\omega)$ vanish in the $m\rightarrow 1$ limit. This implies that in that limit and thus in the
$h\rightarrow h_c$ limit, only $S^{xx} (k,\omega)$ dominates the spin dynamical structure factor. 
Here $h_c = J/\mu_B$ is the critical field at which fully polarized ferromagnetism is achieved. 

At $h=h_c$ the line-shape expressions of general form given in Eqs. (\ref{MPSsFMB}) and (\ref{MPSs})
of the extended dynamical theory are not valid. Indeed, at that magnetic field the problem
refers to a different quantum phase associated with fully polarized ferromagnetism and the
line shape rather becomes of $\delta$-function like type, given by,
\begin{equation}
S^{xx} (k,\omega) = {\pi\over 2} \delta \left(\omega - \omega^{xx}_{lt} (k)\right) 
\hspace{0.50cm}{\rm where}\hspace{0.50cm}\omega^{xx}_{lt} (k) = J\,(1 + \cos k) 
\hspace{0.20cm}{\rm for}\hspace{0.20cm}[0,\pi] \, .
\label{Sxxm1}
\end{equation}

\section{Discussion and concluding remarks}
\label{SECVII}

In this paper, the contribution to the spin dynamical structure
factors $S^{+-} (k,\omega)$, $S^{xx} (k,\omega)$, and  $S^{zz} (k,\omega)$
from excited energy eigenstates populated by $n$-strings has been studied for magnetic fields $0<h<h_c$,
in the thermodynamic limit. (The contribution to $S^{-+} (k,\omega)$ from such states was found to be negligible.)
In that limit, there is nearly no spectral weight in the $(k,\omega)$-plane gap region between the upper threshold of the lower continuum
shown in Figs. \ref{figure1} and \ref{figure3} for $S^{+-} (k,\omega)$ and  $S^{zz} (k,\omega)$,
respectively, and the gapped lower threshold of the $n$-string states's spectrum. The same applies
to $S^{xx} (k,\omega)$ in Fig. \ref{figure2} for the spin densities and $k$ intervals for which
there is no overlap between the $n$-string states's upper continuum and the lower continuum.

Concerning the negligible amount of spectral weight in the $(k,\omega)$-plane gap regions, let us consider 
for instance the more involved case of $S^{+-} (k,\omega)$. Similar conclusions apply to the simpler
problems of the other spin dynamical structure factors. The behavior of spin operators matrix elements between energy eigenstates 
in the selection rules valid for magnetic fields $0<h<h_c$, Eq. (\ref{SRhfinite}), has important
physical consequences. It implies that the spectral weight stemming from 
excited energy eigenstates described by only real Bethe-ansatz rapidities
existing in finite systems in a $(k,\omega)$-plane region corresponding to the momentum interval
$k\in [2k_{F\downarrow},\pi]$ and excitation energy values $\omega$ above the 
upper threshold of the lower continuum in Fig. \ref{figure1}, whose spectrum's expression
is given in Eq. (\ref{dkEdPxxPM}) of Appendix \ref{A}, becomes negligible in the present thermodynamic limit
for a macroscopic system. 

Our thermodynamic limit's study is complementary to results obtained by completely different methods 
for finite-size systems \cite{Kohno_09,Kohno_10,Muller}.  The spectral weight located in that $(k,\omega)$-plane region
is found to decrease upon increasing the system size \cite{Kohno_09}. This is confirmed by comparing the spectra 
represented in the first row frames of Figs. 3 (a) and (b) of Ref. \cite{Kohno_09} for two finite-size systems 
with $N=320$ and $N=2240$ spins, respectively, in the case under consideration of the spin dynamical structure 
factor $S^{+-} (k,\omega)$.

More generally, the selection rules in Eqs. (\ref{SRh0}) and (\ref{SRhfinite}) are
behind in the thermodynamic limit nearly all spectral weight generated by transitions
to excited energy eigenstates described only by real Bethe-ansatz rapidities being
contained in the $(k,\omega)$-plane lower continuum shown in Fig. \ref{figure1}, whose 
spectrum is given in Eq. (\ref{dkEdPxxPM}) of Appendix \ref{A}.
Let us consider for instance the $(k,\omega)$-plane spectral weight distributions shown in Fig. 18 of Ref. \cite{Muller}
for $S^{+-} (k,\omega)$. As reported in that reference, due to the interplay of the selections rules given 
in Eqs. (\ref{SRh0}) and (\ref{SRhfinite}) for $h=0$ and $0<h<h_c$, respectively,
the spectral weight existing between the continuous lower boundary $\epsilon_{4L}$ and the upper boundary 
$\epsilon_{4U}$ at $h=0$ becomes negligible for finite magnetic fields $0<h<h_c$. 
In addition, the spectral weight existing between the continuous lower boundary $\epsilon_{5L}$ and the upper boundary 
$\epsilon_{5U}$ for small finite-size systems, becomes negligible in
the thermodynamic limit for a macroscopic system. This is indeed due to the selection rules, Eq. (\ref{SRhfinite}), as discussed
in that reference. As also reported in it, only the spectral weight below the continuous lower boundary $\epsilon_{5L} (q)$, 
located in the $(k,\omega)$-plane between the lower boundary $\epsilon_{6L}$ and the upper boundary 
$\epsilon_{6U}$ has a significant amount of spectral weight. 

This refers to the $(k,\omega)$-plane region where, according to the analysis of Ref. \cite{Muller}, for 
magnetic fields $0<h<h_c$ a macroscopic system has nearly the whole spectral weight stemming
from transitions to excited energy eigenstates described by only Bethe-ansatz rapidities. Consistent
with the spectral weight in the present gap region being negligible, the $(k,\omega)$-plane
between the continuous lower boundary $\epsilon_{6L}$ and the upper boundary 
$\epsilon_{6U}$ in Fig. 18 of that reference corresponds precisely to the
lower continuum shown in Fig. \ref{figure1}, whose spectrum is provided in Eq. (\ref{dkEdPxxPM}) of Appendix \ref{A}.

Our results have focused on the contribution from $n$-string states. This refers 
to the line shape at and just above the $(k,\omega)$-plane
gapped lower threshold's spectra $\Delta_{\bar{n}}^{ab} (k)$ where $ab = +-,xx,zz$
and ${\bar{n}}$ refers to different branch lines. In well-defined $m$-dependent $k$ subintervals, Eqs. (\ref{Ds2})-(\ref{Ds2p}) and (\ref{Ds2pL})-(\ref{Ds2L}), such branch lines coincide with the gapped lower thresholds under consideration.
In these physically important $(k,\omega)$-plane regions, the spin dynamical 
structure factors $S^{ab} (k,\omega)$ have the general analytical expression provided in 
Eq. (\ref{MPSsFMB}). In the case of $S^{+-} (k,\omega)$ and $S^{xx} (k,\omega)$, such gapped 
lower thresholds refer to the $n$-string states's upper continua shown in the $(k,\omega)$-plane 
in Figs. \ref{figure1} and \ref{figure2}, respectively.

The above results concerning the spectral weight in the gap regions being negligible
in the present thermodynamic limit, are consistent with the amount of that weight existing 
just below the $(k,\omega)$-plane gapped lower thresholds of the $n$-string states's spectra shown in 
Figs. \ref{figure1}-\ref{figure3} being vanishingly small or negligible.
This is actually behind the validity at finite longitudinal magnetic fields $0<h<h_c$ and
in the thermodynamic limit of the analytical expressions of the spin dynamical structure factors 
of general form, Eq. (\ref{MPSsFMB}), obtained in this paper. 
	
The momentum dependent exponents that control the spin dynamical structure factors's
line-shape in such expressions are given in Eq. (\ref{expG+-}) for $S^{+-} (k,\omega)$
and $S^{xx} (k,\omega)$ and in Eq. (\ref{exps2pL}) for $S^{zz} (k,\omega)$. In the former case,
the exponents associated with the $(k,\omega)$-plane vicinity of
the $2-$, $\bar{1}'-$, $\bar{1}-$, and $2'$-branch lines are plotted in Fig. \ref{figure4}. Such lines refer to different
$k$ intervals of the gapped lower threshold of the $n$-string states's spectra of 
$S^{+-} (k,\omega)$ and $S^{xx} (k,\omega)$. The solid lines in Figs. \ref{figure1} and \ref{figure2}
that belong to that gapped lower threshold correspond to $k$ intervals for which the exponents
are negative. In them, singularities occur in the spin dynamical structure factors's expression,
Eq. (\ref{MPSsFMB}), at and above the gapped lower thresholds.

In the case of $S^{xx} (k,\omega)$, the expression given in that equation does not apply for
small spin densities in the range $m \in [0,\bar{m}]$ where $\bar{m} \approx 0.276$
to the $k$ intervals given in Eq. (\ref{gapineq}). For these momentum ranges, there is
overlap between the lower continuum and upper $n$-string states's continuum, as shown in
Figs. \ref{figure2} (a) and (b). The two $k$ intervals provided in Eq. (\ref{gapineq}), $k \in [\bar{k}_0,\pi]$ 
for $m\in ]0,\bar{m}_0]$ and $k \in [\bar{k}_0,\bar{k}_1]$ for $]\bar{m}_0,\bar{m}]$ 
where $\bar{m}_0 \approx 0.239$, apply to the spectra plotted in these two figures, respectively. 
For spin density (a) $m=0.15$, the momentum $\bar{k}_0$ is given by $\bar{k}_0\approx 0.60\,\pi$ 
whereas for spin density (b)  $m=0.25$ the two limiting momenta
read $\bar{k}_0\approx 0.71\,\pi$ and $\bar{k}_1\approx 0.92\,\pi$.

On the other hand, the contribution to $S^{zz} (k,\omega)$ from excited states populated by $n$-strings
is much weaker than for $S^{+-} (k,\omega)$ and $S^{xx} (k,\omega)$. It does not lead to a
$(k,\omega)$-plane continuum. The gapped lower threshold of such states
is shown in Fig. \ref{figure3}. There the $k$ subinterval associated with the $\bar{n}=\bar{1}'$ branch line 
is the only one at and above which there are singularities.
Out of the four branch-line's exponents whose expressions are provided in Eq. (\ref{exps2pL}),
only that of the $\bar{n}=\bar{1}'$ branch line is indeed negative. That line is represented in the gapped lower threshold 
of $S^{zz} (k,\omega)$ shown in Fig. \ref{figure3} by a solid (green) line. The 
corresponding exponent is plotted in Fig. \ref{figure5}. That line's $k$ subinterval is though small. 
Its momentum width decreases upon increasing 
the spin density within the range $0<m\leq\tilde{m}$ where $\tilde{m}\approx 0.317$. 
For spin densities $\tilde{m}\leq m<1$, that line is not part of the gapped lower threshold, so that
the contribution to $S^{zz} (k,\omega)$ from excited states populated by $n$-strings 
becomes negligible. Consistent, in Fig. \ref{figure3} (c) for $m=0.50$ and (d) for $m=0.75$ that
line is lacking.

In order to to provide an overall physical picture that includes the
relative $(k,\omega)$-plane location of all spectra with a significant amount
of spectral weight, we also accounted for the contributions from all
types of excited energy eigenstates that lead to gapped and gapless lower threshold singularities in the spin dynamical structure factors.
This includes excited energy eigenstates described only by real Bethe-ansatz rapidities
and thus without $n$-strings, which are known to lead to most spectral weight
of the sum rules, Eq. (\ref{SRDSF}) \cite{Lefmann-96,Muller,Karbach_00,Karbach_02,Carmelo_15A}. Their contribution to 
$S^{+-} (k,\omega)$, $S^{xx} (k,\omega)$, and  $S^{zz} (k,\omega)$ leads to
the $(k,\omega)$-plane lower continua shown in Figs. \ref{figure1}, \ref{figure2}, and \ref{figure3}, respectively.

While the present work is purely theoretical, the singularities to which $(k,\omega)$-plane vicinity the analytical
line-shape expressions obtained and studied in this paper refer to are observed in inelastic neutron scattering experiments
on spin-chain compounds \cite{Kohno_09,Kohno_10,Stone_03,Heilmann_78}. All cusp singularities at and near
both the gapped lower thresholds and lower thresholds found in this paper to occur in the thermodynamic limit,
indeed correspond to peaks shown in Fig. 4 of Ref. \cite{Kohno_09} for CuCl$_2$$\cdot$2N(C$_5$D$_5$) 
and in Fig. 5 of that reference for Cu(C$_4$H$_4$N$_2$)(NO$_3$)$_2$ at the finite values
of the magnetic field considered in these figures and suitable values of the exchange integral $J$.
Also in that reference such a correspondence was found within the spin-$1/2$ $XXX$ chain,
for finite-size systems.\\ \\
{\bf CRediT authorship contribution statement}\\

The three authors contributed equally to the formulation of the research goals and aims. J. M. P. C. extended a suitable dynamical theory to a subspace that accounts for the contribution from $n$-string states to the spin dynamical structure factors and prepared the original draft. T. \v{C}. and P. D. S. have designed the computer programs to solve the integral equations that define the phase shifts and spectra and T. \v{C}. has implemented them.\\ \\
{\bf Declaration of competing interest}\\

The authors declare that they have no known competing financial interests or personal relationships that could have appeared to influence the work reported in this paper.

\acknowledgements
J. M. P. C. would like to thank the Boston University's Condensed Matter Theory Visitors Program for support and
Boston University for hospitality during the initial period of this research. He acknowledges the support from
FCT through the Grants PTDC/FIS-MAC/29291/2017, SFRH/BSAB/142925/2018, and POCI-01-0145-FEDER-028887.
J. M. P. C. and T. \v{C}. acknowledge the support from FCT through the Grant UID/FIS/04650/2013.
T. \v{C}. gratefully acknowledges the support by the Institute for Basic Science in Korea (IBS-R024-D1).
P. D. S. acknowledges the support from FCT through the Grants UID/CTM/04540/2013 and UID/CTM/04540/2019.

J. M. P. C. and T. \v{C}. contributed equally to this work.

\appendix

\section{Gapless continuum spectra and energy gaps of the $n$-string states}
\label{A}

Within a $k$ extended zone scheme, the $S^{-+} (k,\omega)$ and $S^{+-} (k,\omega)$'s
spectra associated with lower $(k,\omega)$-plane continua, which for $S^{+-} (k,\omega)$
is shown in Fig. \ref{figure1}, read,
\begin{eqnarray}
& & \omega^{-+} (k) = - \varepsilon_{1} (q) - \varepsilon_{1} (q')
\hspace{0.20cm} {\rm where}\hspace{0.20cm} k = \iota\pi - q - q' \hspace{0.20cm}
{\rm and}\hspace{0.20cm}\iota = \pm 1
\nonumber \\
& & {\rm for}\hspace{0.20cm}q \in [-k_{F\downarrow},k_{F\downarrow}]
\hspace{0.20cm}{\rm and}\hspace{0.20cm}q' \in [-k_{F\downarrow},k_{F\downarrow}] \, ,
\label{dkEdPxxMP}
\end{eqnarray}
and
\begin{eqnarray}
& & \omega^{+-} (k) = \varepsilon_{1} (q) - \varepsilon_{1} (q') 
\hspace{0.20cm}{\rm where}\hspace{0.20cm} k = \iota\pi + q - q' \hspace{0.20cm}
{\rm and}\hspace{0.20cm}\iota = \pm 1
\nonumber \\
& & {\rm for}\hspace{0.20cm}\vert q\vert \in [k_{F\downarrow},k_{F\uparrow}] 
\hspace{0.20cm}{\rm and}\hspace{0.20cm}q' \in [-k_{F\downarrow},k_{F\downarrow}] \, ,
\label{dkEdPxxPM}
\end{eqnarray}
respectively. Here $\varepsilon_{1} (q)$ is the $1$-band energy dispersion given in Eq. (\ref{equA4}) of Appendix \ref{B}.
The spectrum $\omega^{xx} (k)$ of the transverse dynamical structure factor $S^{xx} (k,\omega)$ 
associated with the lower continuum in Fig. \ref{figure2}
results from combination of the two spectra $\omega^{-+} (k)$ and $\omega^{+-} (k)$
in Eqs. (\ref{dkEdPxxMP}) and (\ref{dkEdPxxPM}), respectively.

The spectrum $\omega^{zz} (k)$ associated with the lower continuum in Fig. \ref{figure3} is 
again within a $k$ extended zone scheme given by,
\begin{eqnarray}
& & \omega^{zz} (k) = \varepsilon_{1} (q) - \varepsilon_{1} (q') 
\hspace{0.20cm}{\rm where}\hspace{0.20cm} k = q - q'
\nonumber \\
& & {\rm for}\hspace{0.20cm}\vert q\vert \in [k_{F\downarrow},k_{F\uparrow}] 
\hspace{0.20cm}{\rm and}\hspace{0.20cm}q' \in [-k_{F\downarrow},k_{F\downarrow}] \, .
\label{dkEdPl}
\end{eqnarray}

The upper thresholds of the two-parametric spectra, Eqs. (\ref{dkEdPxxMP}) and (\ref{dkEdPxxPM}),
have the following one-parametric spectra for spin densities $m \in ]0,1[$,
\begin{eqnarray}
\omega^{+-}_{ut} (k) & = & 2\mu_B\,h - \varepsilon_{1} (k_{F\downarrow}-k) 
\hspace{0.20cm}{\rm where}\hspace{0.20cm} k = k_{F\downarrow} - q
\hspace{0.20cm}{\rm for}\hspace{0.20cm}k\in [0,k_{F\downarrow}] 
\hspace{0.20cm}{\rm and}\hspace{0.20cm}q \in [0,k_{F\downarrow}] \, ,
\nonumber \\
& = & \varepsilon_{1} (q) - \varepsilon_{1} (q') 
\hspace{0.20cm}{\rm where}\hspace{0.20cm} k = \pi + q - q'
\hspace{0.20cm}{\rm for}\hspace{0.20cm}k\in [k_{F\downarrow},\pi] 
\hspace{0.20cm}{\rm and}\hspace{0.20cm}v_s (q) = v_s (q')\hspace{0.20cm}
\nonumber \\
&& {\rm with} \hspace{0.20cm}q \in [-k_{F\uparrow},-k_{F\downarrow}]
\hspace{0.20cm}{\rm and}\hspace{0.20cm}q' \in [-k_{F\downarrow},0] \, ,
\label{Omxxut1}
\end{eqnarray}
and
\begin{equation}
\omega^{-+}_{ut} (k) = -2\varepsilon_{1} \left({\pi - k\over 2}\right) 
\hspace{0.20cm}{\rm where}\hspace{0.20cm} k = \pi - 2q
\hspace{0.20cm}{\rm for}\hspace{0.20cm}k\in [(k_{F\uparrow}-k_{F\downarrow}),\pi] 
\hspace{0.20cm}{\rm and}\hspace{0.20cm}q \in [-k_{F\downarrow},0] \, ,
\label{Omxxut2}
\end{equation}
respectively. The function $v_1 (q)$ is in Eq. (\ref{Omxxut1}) the $1$-band group velocity defined 
in Eq. (\ref{equA4B}) of Appendix \ref{B}.

The upper threshold spectrum $\omega^{xx}_{ut} (k)$ of the combined spectra, Eqs. (\ref{dkEdPxxMP}) and (\ref{dkEdPxxPM}),
is given by,
\begin{eqnarray}
\omega^{xx}_{ut} (k) & = & \omega^{+-}_{ut} (k)\hspace{0.20cm}{\rm for}\hspace{0.20cm} k \in [0,k^{xx}_{ut}]
\nonumber \\
& = & \omega^{-+}_{ut} (k)\hspace{0.20cm}{\rm for}\hspace{0.20cm} k \in [k^{xx}_{ut},\pi] \, ,
\label{Omxxutxx}
\end{eqnarray}
where the momentum $k^{xx}_{ut}$ is such that $\omega^{+-}_{ut} (k^{xx}_{ut}) = \omega^{-+}_{ut} (k^{xx}_{ut})$.

On the other hand, the one-parametric upper threshold spectrum associated
with the two-parametric longitudinal spectrum, Eq. (\ref{dkEdPl}), 
reads for $m \in ]0,1[$,
\begin{eqnarray}
\omega^{zz}_{ut} (k) & = & \varepsilon_{1} (q) - \varepsilon_{1} (q') 
\hspace{0.20cm}{\rm where}\hspace{0.20cm} k = q - q'
\hspace{0.20cm}{\rm for}\hspace{0.20cm}v_s (q) = v_s (q')\hspace{0.20cm}
{\rm and}\hspace{0.20cm}k\in [0,k_{F\uparrow}]\hspace{0.20cm}{\rm with}
\nonumber \\
&& q \in [k_{F\downarrow},k_{F\uparrow}]\hspace{0.20cm}{\rm and}\hspace{0.20cm}
q' \in [0,k_{F\downarrow}] \, ,
\nonumber \\
& = & 2\mu_B\,h - \varepsilon_{1} (k_{F\uparrow}-k) 
\hspace{0.20cm}{\rm where}\hspace{0.20cm} k = k_{F\uparrow} - q
\hspace{0.20cm}{\rm for}\hspace{0.20cm}k\in [k_{F\uparrow},\pi] 
\hspace{0.20cm}{\rm and}\hspace{0.20cm}q \in [-k_{F\downarrow},0] \, .
\label{Omlut}
\end{eqnarray}

At $k=0,k_{F\downarrow},\pi$ and $k=0,k_{F\uparrow}-k_{F\downarrow},\pi$,
the upper threshold spectra, Eqs. (\ref{Omxxut1}) and (\ref{Omxxut2}), respectively, are given by,
\begin{eqnarray}
& & \omega^{+-}_{ut} (0) = W_1^h = 2\mu_B\,h \, ; \hspace{0.30cm}
\omega^{+-}_{ut} (k_{F\downarrow}) = W_1 = 2\mu_B\,h + W_1^p
\, ; \hspace{0.30cm} \omega^{+-}_{ut} (\pi) = 0 
\nonumber \\
& & \omega^{-+}_{ut} (k_{F\uparrow}-k_{F\downarrow}) = 0
\, ; \hspace{0.30cm} \omega^{-+}_{ut} (\pi) = 2W_1^p \, .
\label{Omxxutlim}
\end{eqnarray}
Here $W_1 = W_1^p + W_1^h$, $W_1^p$, and $W_1^h$ are the
energy bandwidths of the full $1$-band, its occupied Fermi sea,
and unoccupied sea, respectively, in Eqs. (\ref{vares2limits}) and (\ref{W2um1}) 
of Appendix \ref{B}.

At $k=0,k_{F\uparrow},\pi$ the upper threshold spectrum $\omega^{zz}_{ut} (k)$ reads,
\begin{equation}
\omega^{zz}_{ut} (0) = 0 \, ; \hspace{0.30cm}
\omega^{zz}_{ut} (k_{F\uparrow}) = W_1 = 2\mu_B\,h + W_1^p
\, ; \hspace{0.30cm}
\omega^{zz}_{ut} (\pi) = W_1^h = 2\mu_B\,h  \, .
\label{Omlutlim}
\end{equation}

The spin dynamical structure factors's line shape near the lower thresholds
of the spectra, Eqs. (\ref{dkEdPxxMP}), (\ref{dkEdPxxPM}), and (\ref{dkEdPl}),
has the general form provided in Eq. (\ref{MPSs}).
In the case of (i) $S^{-+} (k,\omega)$ and (ii) $S^{+-} (k,\omega)$ and $S^{zz} (k,\omega)$
such lower thresholds refer to (i) a single $1$-branch line and 
(ii) two sections of a $1$-branch line, respectively.
Their spectra can be expressed in terms of the excitation
momentum $k$ or of the $1$-band momentum $q$ and are given by,
\begin{eqnarray}
\omega^{-+}_{lt} (k) & = & - \varepsilon_1 (k_{F\uparrow}-k) \hspace{0.2cm}{\rm and}
\hspace{0.20cm} k = k_{F\uparrow} - q\hspace{0.2cm}{\rm where}
\nonumber \\
k & \in & [(k_{F\uparrow}-k_{F\downarrow}),\pi]
\hspace{0.2cm}{\rm and}\hspace{0.20cm}
q \in ]-k_{F\downarrow},k_{F\downarrow}] \, ,
\label{OkPMRs}
\end{eqnarray}
\begin{eqnarray}
\omega^{+-}_{lt} (k) & = & \varepsilon_1 (k - k_{F\uparrow}) \hspace{0.2cm}{\rm and}
\hspace{0.20cm} k = k_{F\uparrow} + q\hspace{0.2cm}{\rm where}
\nonumber \\
k & \in & [0, (k_{F\uparrow}-k_{F\downarrow})]
\hspace{0.2cm}{\rm and}\hspace{0.20cm}
q \in [-k_{F\uparrow},-k_{F\downarrow}] \, ,
\nonumber \\
& = & - \varepsilon_1 (k_{F\uparrow}-k) \hspace{0.2cm}{\rm and}
\hspace{0.20cm} k = k_{F\uparrow} - q\hspace{0.2cm}{\rm where}
\nonumber \\
k & \in & [(k_{F\uparrow}-k_{F\downarrow}),\pi] 
\hspace{0.2cm}{\rm and}\hspace{0.20cm}
q \in [-k_{F\downarrow},k_{F\downarrow}] \, ,
\label{OkMPRs}
\end{eqnarray}
\begin{eqnarray}
\omega^{zz}_{lt} (k) & = & - \varepsilon_1 -k_{F\downarrow}[_{F\downarrow} - k) \hspace{0.2cm}{\rm and}
\hspace{0.20cm} k = k_{F\downarrow} - q \hspace{0.2cm}{\rm where}
\nonumber \\
k & \in &[0,2k_{F\downarrow}]\hspace{0.2cm}{\rm and}
\hspace{0.20cm} q \in [-k_{F\downarrow},k_{F\downarrow}] \, ,
\nonumber \\
& = & \varepsilon_1 (k - k_{F\downarrow}) \hspace{0.2cm}{\rm and}
\hspace{0.20cm} k = k_{F\downarrow} + q \hspace{0.2cm}{\rm where}
\nonumber \\
k & \in &[2k_{F\downarrow}),\pi]
\hspace{0.2cm}{\rm and}\hspace{0.20cm}
q \in [k_{F\downarrow},k_{F\uparrow}] \, .
\label{OkPMRsL}
\end{eqnarray}

The lower threshold's spectrum $\omega^{xx}_{lt} (k)$ of $S^{xx} (k,\omega)$ has the same expression 
as that of $S^{+-} (k,\omega)$,
\begin{eqnarray}
\omega^{xx}_{lt} (k) & = & \varepsilon_1 (k - k_{F\uparrow}) \hspace{0.2cm}{\rm and}
\hspace{0.20cm} k = k_{F\uparrow} + q\hspace{0.2cm}{\rm where}
\nonumber \\
k & \in & [0, (k_{F\uparrow}-k_{F\downarrow})]
\hspace{0.2cm}{\rm and}\hspace{0.20cm}
q \in [-k_{F\uparrow},-k_{F\downarrow}] \, ,
\nonumber \\
& = & - \varepsilon_1 (k_{F\uparrow}-k) \hspace{0.2cm}{\rm and}
\hspace{0.20cm} k = k_{F\uparrow} - q\hspace{0.2cm}{\rm where}
\nonumber \\
k & \in & [(k_{F\uparrow}-k_{F\downarrow}),\pi] 
\hspace{0.2cm}{\rm and}\hspace{0.20cm}
q \in [-k_{F\downarrow},k_{F\downarrow}] \, .
\label{OkxxRs}
\end{eqnarray}

Finally, the $k$ dependent expressions and the limiting values of the energy gaps in Eqs. (\ref{gapPMMP})-(\ref{gapL}) 
are provided. The energy gap $\Delta_{\rm gap}^{+-} (k)$ is finite and positive for $0<m<1$.
For spin densities $m\in ]0,\tilde{m}]$ where $\tilde{m}\approx 0.317$ it reads,
\begin{eqnarray}
\Delta_{\rm gap}^{+-} (k) & = & - 2\mu_B\,h + \varepsilon_{2} (k) + \varepsilon_{1} (k_{F\downarrow}-k) 
\hspace{0.20cm}{\rm for}\hspace{0.20cm}k\in [0,(k_{F\uparrow}-k_{F\downarrow})]
\nonumber \\
\Delta_{\rm gap}^{+-} (k) & = & 2\mu_B\,h - \varepsilon_{1} (k_{F\uparrow}-k) + \varepsilon_{1} (k_{F\downarrow}-k) 
\hspace{0.20cm}{\rm for}\hspace{0.20cm}[(k_{F\uparrow} - k_{F\downarrow}),{\tilde{k}}[
\nonumber \\
\Delta_{\rm gap}^{+-} (k) & = & 2\mu_B\,h - W_{2}\hspace{0.20cm}
{\rm for}\hspace{0.20cm}k\in ]{\tilde{k}},k_{F\downarrow}]
\nonumber \\
\Delta_{\rm gap}^{+-} (k) & = &  4\mu_B\,h - W_{2} - \varepsilon_{1} (k_{F\downarrow}-k) 
+ \varepsilon_{1} (q) - \varepsilon_{1} (k + q - \pi) 
\hspace{0.20cm}{\rm for}\hspace{0.20cm}k\in [k_{F\downarrow},2k_{F\downarrow}] \hspace{0.20cm}{\rm and}
\nonumber \\
& & q \in [-(k_{\bullet} - k_{F\uparrow} + k_{F\downarrow}),0]
\nonumber \\
\Delta_{\rm gap}^{+-} (k) & = & \varepsilon_{2} (k - 2k_{F\downarrow}) + \varepsilon_{1} (q) - \varepsilon_{1} (k + q - \pi) 
\hspace{0.20cm}{\rm for}\hspace{0.20cm}k\in [2k_{F\downarrow},\pi] \hspace{0.20cm} {\rm and}
\nonumber \\
& & {\rm for}\hspace{0.20cm}{\rm spin}\hspace{0.20cm}{\rm densities}\hspace{0.20cm}m\in ]0,\tilde{m}] \, .
\label{gapEXPm03}
\end{eqnarray}
For spin densities $m\in [\tilde{m},1[$ its expression is,
\begin{eqnarray}
\Delta_{\rm gap}^{+-} (k) & = & - 2\mu_B\,h + \varepsilon_{2} (k)  + \varepsilon_{1} (k_{F\downarrow}-k) 
\hspace{0.20cm}{\rm for}\hspace{0.20cm}k\in [0,{\tilde{k}}[
\nonumber \\
\Delta_{\rm gap}^{+-} (k) & = & 2\mu_B\,h - W_{2}\hspace{0.20cm}
{\rm for}\hspace{0.20cm}k\in ]{\tilde{k}},k_{F\downarrow}]
\nonumber \\
\Delta_{\rm gap}^{+-} (k) & = &  4\mu_B\,h - W_{2} - \varepsilon_{1} (k_{F\downarrow}-k) 
+ \varepsilon_{1} (q) - \varepsilon_{1} (k + q - \pi) 
\hspace{0.20cm}{\rm for}\hspace{0.20cm}k\in [k_{F\downarrow},2k_{F\downarrow}] \hspace{0.20cm}{\rm and}
\nonumber \\
& & q \in [-(k_{\bullet} - k_{F\uparrow} + k_{F\downarrow}),0]
\nonumber \\
\Delta_{\rm gap}^{+-} (k) & = & \varepsilon_{2} (k - 2k_{F\downarrow}) + \varepsilon_{1} (q) - \varepsilon_{1} (k + q - \pi) 
\hspace{0.20cm}{\rm for}\hspace{0.20cm}k\in [2k_{F\downarrow},\pi] \hspace{0.20cm} {\rm and}
\nonumber \\
& & q \in [-k_{F\downarrow},- (k_{\bullet} - k_{F\uparrow} + k_{F\downarrow})]
\nonumber \\
& & {\rm for}\hspace{0.20cm}{\rm spin}\hspace{0.20cm}{\rm densities}\hspace{0.20cm}m\in [\tilde{m},1[ \, .
\label{gapEXPm31}
\end{eqnarray}

In the above equations, the momentum $\tilde{k}$ is defined by the relations, Eq. (\ref{ktilde}),
and the momentum $k_{\bullet}$ satisfies the following equation,
\begin{equation}
v_1 (k_{\bullet}) = v_1 (k_{\bullet} - k_{F\uparrow} + k_{F\downarrow})
\hspace{0.20cm}{\rm where}\hspace{0.20cm}k_{\bullet}>k_{F\downarrow} \, .
\label{kbullet}
\end{equation}

On the other hand, for spin densities  $m\in ]0,\tilde{m}]$, the energy gap
$\Delta_{\rm gap}^{-+} (k) = \Delta^{-+} (k) - \omega^{-+}_{ut} (k)$ where
$\Delta^{-+} (k) = \Delta^{+-} (k)$ reads,
\begin{eqnarray}
\Delta_{\rm gap}^{-+} (k) & = & \varepsilon_{2} (k) \hspace{0.20cm}{\rm for}
\hspace{0.20cm}k\in [0,(k_{F\uparrow} - k_{F\downarrow})]
\nonumber \\
\Delta_{\rm gap}^{-+} (k) & = & 4\mu_B\,h - \varepsilon_{1} (k_{F\uparrow}-k) 
+ 2\varepsilon_{1} \left({\pi - k\over 2}\right)
\hspace{0.20cm}{\rm for}\hspace{0.20cm}
k\in [(k_{F\uparrow} - k_{F\downarrow}),{\tilde{k}}[
\nonumber \\
\Delta_{\rm gap}^{-+} (k) & = & 4\mu_B\,h - W_{2} - \varepsilon_{1} (k_{F\downarrow}-k) 
+ 2\varepsilon_{1} \left({\pi - k\over 2}\right)
\hspace{0.20cm}{\rm for}\hspace{0.20cm}k \in ]{\tilde{k}},2k_{F\downarrow}]
\nonumber \\
\Delta_{\rm gap}^{-+} (k) & = & \varepsilon_{2} (k - 2k_{F\downarrow}) 
+ 2\varepsilon_{1} \left({\pi - k\over 2}\right)
\hspace{0.20cm}{\rm for}\hspace{0.20cm}k\in [2k_{F\downarrow},\pi]  \, ,
\label{GgapSPM03}
\end{eqnarray}
whereas for $m\in [\tilde{m},1[$ it is given by,
\begin{eqnarray}
\Delta_{\rm gap}^{-+} (k) & = & \varepsilon_{2} (k) \hspace{0.20cm}{\rm for}
\hspace{0.20cm}k\in [0,{\tilde{k}}[
\nonumber \\
\Delta_{\rm gap}^{-+} (k) & = & 4\mu_B\,h - W_{2} - \varepsilon_{1} (k_{F\downarrow}-k) 
\hspace{0.20cm}{\rm for}\hspace{0.20cm}k \in ]{\tilde{k}},(k_{F\uparrow} - k_{F\downarrow})]
\nonumber \\
\Delta_{\rm gap}^{-+} (k) & = & 4\mu_B\,h - W_{2} - \varepsilon_{1} (k_{F\downarrow}-k) 
+ 2\varepsilon_{1} \left({\pi - k\over 2}\right)
\hspace{0.20cm}{\rm for}\hspace{0.20cm}k \in [(k_{F\uparrow} - k_{F\downarrow}),2k_{F\downarrow}]
\nonumber \\
\Delta_{\rm gap}^{-+} (k) & = & \varepsilon_{2} (k - 2k_{F\downarrow}) 
+ 2\varepsilon_{1} \left({\pi - k\over 2}\right)
\hspace{0.20cm}{\rm for}\hspace{0.20cm}k\in [2k_{F\downarrow},\pi]  \, .
\label{GgapSPM3pi}
\end{eqnarray}

The energy gap $\Delta_{\rm gap}^{+-} (k)$ is given by the constant energy scale $2\mu_B\,h - W_{2}$ 
where $W_{2}$ is the energy bandwidth of the $2$-band in Eqs. (\ref{vares2limits}) and (\ref{W2um1}) 
of Appendix \ref{B} for the following $k$ and spin density $m$ values and intervals,
\begin{eqnarray}
\Delta_{\rm gap}^{+-} (k) & = & 2\mu_B\,h - W_{2}
\nonumber \\
k & = & 0 \hspace{0.20cm}{\rm for}\hspace{0.20cm}m\in ]0,1[
\nonumber \\
k & = & k_{F\uparrow} - k_{F\downarrow}  \hspace{0.20cm}{\rm for}\hspace{0.20cm}m\in ]0,1/3]
\nonumber \\
k & \in & ]{\tilde{k}},k_{F\downarrow}] \hspace{0.20cm}{\rm for}\hspace{0.20cm}m\in ]0,1[ \, .
\label{gapEXPm03LIM}
\end{eqnarray}
The energy scale $2\mu_B\,h - W_{2}\geq 0$ has the following limiting values,
\begin{equation}
2\mu_B\,h - W_{2} = 0\hspace{0.20cm}{\rm for}\hspace{0.20cm}m\rightarrow 0
\hspace{0.50cm}{\rm and}\hspace{0.50cm}
2\mu_B\,h - W_{2} = J \hspace{0.20cm}{\rm for}\hspace{0.20cm}m\rightarrow 1 \, .
\label{gapOtherLIM}
\end{equation}

At $k=\pi$ the energy gap $\Delta_{\rm gap}^{+-} (k)$ reads,
\begin{equation}
\Delta_{\rm gap}^{+-} (\pi) = 4\mu_B\,h
\hspace{0.20cm}{\rm for}\hspace{0.20cm}m\in ]0,1[ \, .
\label{gapEXPm03LIMB}
\end{equation}
It has the following limiting values,
\begin{equation}
\Delta_{\rm gap}^{+-} (\pi) = 0\hspace{0.20cm}{\rm for}\hspace{0.20cm}m\rightarrow 0
\hspace{0.50cm}{\rm and}\hspace{0.50cm}
\Delta_{\rm gap}^{+-} (\pi) = 4J \hspace{0.20cm}{\rm for}\hspace{0.20cm}m\rightarrow 1 \, .
\label{gaPMpuL}
\end{equation}

The energy gap $\Delta_{\rm gap}^{-+} (k)$ has the following values at
$k=0, k_{F\uparrow}-k_{F\downarrow},\pi$,
\begin{eqnarray}
\Delta_{\rm gap}^{-+} (0) & = & 4\mu_B\,h - W_{2}
\hspace{0.20cm}{\rm for}\hspace{0.20cm}m\in ]0,1[
\nonumber \\
\Delta_{\rm gap}^{-+} (k_{F\uparrow}-k_{F\downarrow}) & = & 4\mu_B\,h 
\hspace{0.20cm}{\rm for}\hspace{0.20cm}m\in ]0,\tilde{m}]
\nonumber \\
\Delta_{\rm gap}^{-+} (\pi) & = & 4\mu_B\,h - 2W_1^p 
\hspace{0.20cm}{\rm for}\hspace{0.20cm}m\in ]0,1[ \, .
\label{gapMPkkk}
\end{eqnarray}

For small spin densities $m \in ]0,\bar{m}]$ where $\bar{m} \approx 0.276$ and the momentum intervals
given in Eq. (\ref{gapineq}) the inequality $\Delta_{\rm gap}^{-+} (k)<0$ holds. In such a $k$ intervals,
$\Delta_{\rm gap}^{-+} (k)$ equals the energy gap $\Delta_{\rm gap}^{xx} (k) = \Delta_{\rm gap}^{-+} (k)<0$. At $k=\pi$ 
such gaps thus have negative and positive values for spin densities $m<\bar{m}_0$ and $m>\bar{m}_0$, respectively,
where $\bar{m}_0 \approx 0.239$. The corresponding limiting values read,
\begin{eqnarray}
\Delta_{\rm gap}^{xx} (\pi) & = & \Delta_{\rm gap}^{-+} (\pi) = - \pi J\hspace{0.20cm}{\rm for}\hspace{0.20cm}m\rightarrow 0
\nonumber \\
& = & 0\hspace{0.20cm}{\rm for}\hspace{0.20cm}m=\bar{m}_0 \approx 0.239
\nonumber \\
& = & 4J \hspace{0.20cm}{\rm for}\hspace{0.20cm}m\rightarrow 1 \, .
\label{gaMPpuL}
\end{eqnarray}

\section{Some useful quantities}
\label{B}

In this Appendix some quantities needed for our study are defined and corresponding useful limiting behaviors are provided.
The quantum problem studied in this paper is described by the spin-$1/2$ $XXX$ chain in a longitudinal magnetic field,
Eq. (\ref{HXXX}), acting in the subspace considered in Sec. \ref{SECIIIA}. It involves a subset of Bethe ansatz equations.
That associated with the $1$-band is given by, 
\begin{eqnarray}
q_j & = & 2\arctan\left(\Lambda_1 (q_j)\right)
- {2\over L}\sum_{j'=1}^{N_{\uparrow}}\, N_{1}(q_{j'})\arctan
\left({\Lambda_1 (q_j)-\Lambda_1 (q_{j'})\over 2}\right) 
\nonumber \\
& - & {2\over L}\sum_{j'=1}^{N_{\uparrow}-N_{\downarrow}+N_{2}}\, N_{2}(q_{j'})
\left\{\arctan\left(\Lambda_1 (q_j)-\Lambda_{2} (q_{j'})\right) 
+ \arctan\left({\Lambda_1 (q_j)-\Lambda_{2} (q_{j'})\over 3}\right)\right\}
\nonumber \\
&& {\rm where} \hspace{0.5cm} j = 1,...,N_{\uparrow} \hspace{0.5cm}{\rm and}\hspace{0.5cm}N_{2} = 0,1 \, . 
\label{Taps}
\end{eqnarray}
That associated with the $2$-band reads,
\begin{eqnarray}
q_j & = & 2\arctan\left({\Lambda_{2} (q_j)\over 2}\right)
- {2\over L}\sum_{j'=1}^{N_{\uparrow}}\,N_{1}(q_{j'})\left\{\arctan\left(\Lambda_{2} (q_j)-\Lambda_1 (q_{j'})\right)
+ \arctan\left({\Lambda_{2} (q_j)-\Lambda_1 (q_{j'})\over 3}\right)\right\}
\nonumber \\
& & {\rm where} \hspace{0.5cm}  j = 1,...,N_{\uparrow}-N_{\downarrow}+N_{2} 
\hspace{0.5cm}{\rm and}\hspace{0.5cm}N_{2} = 0,1 \, .
\label{Tap2}
\end{eqnarray}
In these equations, $N_{1}(q_{j'})=1$ and $N_{2}(q_{j'})=1$ for occupied momentum values $q_{j'}$ and
$N_{1}(q_{j'})=0$ and $N_{2}(q_{j'})=0$ for unoccupied momentum values $q_{j'}$.

The subspace considered in Sec. \ref{SECIIIA} is spanned by excited energy eigenstates populated either by 
a number $N_1 = N_{\downarrow}$ of $1$-particles or by a number $N_1 = N_{\downarrow}-2$
of $1$-particles and a single $2$-particle. In the case of the latter class of states, the Bethe-ansatz equation, 
Eq. (\ref{Tap2}), does not include a third term, that in the case of the full Hilbert space involves
the spin rapidity differences $\Lambda_{2} (q_j)-\Lambda_{2} (q_{j'})$. Indeed,
when $\delta N_2 =1$ one has that such a term only contributes for $q_j = q_{j'}$
at which values it vanishes.

The $1$-band Bethe ansatz rapidity is real and associated with the rapidity function
$\Lambda_{1} (q_j)$ in the above Bethe-ansatz equations. In the case of general $n$-strings of length $n>1$,
the corresponding complex non-real Bethe ansatz rapidities have in the
thermodynamic limit the following form \cite{Takahashi_71},
\begin{eqnarray}
\Lambda_{n,l} (q_j) & = & \Lambda_n (q_j) + i (n+1-2l) \hspace{0.20cm}{\rm for}\hspace{0.20cm} l = 1,...,n
\hspace{0.20cm}{\rm and}\hspace{0.20cm}j = 1,...,L_n\hspace{0.20cm}{\rm where}
\nonumber \\
L_n & = & N_n + N^h_{n} \hspace{0.20cm}{\rm and}\hspace{0.20cm}
N^h_{n} = 2S + \sum_{n'=n+1}^{\infty}2(n'-n)N_{n'} \, .
\label{Lambda-jnl-ideal}
\end{eqnarray}
This general expression also applies for $n=1$, $\Lambda_{1,1} (q_j)$ being real and equal to $\Lambda_{1} (q_j)$. In the
general $n=1,...,\infty$ case, $\Lambda_n (q_j)$ are real rapidity functions defined by the set of coupled Bethe-ansatz
equations associated with the full Hilbert space not given here, $q_j$ such that $q_{j+1} - q_j = 2\pi/L$ 
are the corresponding $n$-band discrete momentum values whose number is $L_n$, $N_n$ is both the
number of $n$-band occupied momentum values and the number
of $n$-particles and thus of $n$-strings when $n>2$, $N_n^h$ is that of 
$n$-band unoccupied momentum values, and $S$ is the energy eigenstate's spin such that
$2S$ gives the number of unpaired physical spins $1/2$ that are not paired
and thus are not part of the $n$ particles's internal degrees of freedom.

In the present case of the subspace considered in Sec. \ref{SECIIIA}, the problem simplifies.
The $2$-band rapidity function $\Lambda_{2} (q_j)$ that appears in Eqs. (\ref{Taps}) and (\ref{Tap2}) 
is the real part of the following two Bethe ansatz complex rapidities associated with a $n$-string of length $n=2$,
\begin{equation}
\Lambda_{2,l} (q_j) = \Lambda_{2} (q_j) + i (3-2l) \hspace{0.20cm}{\rm for}\hspace{0.20cm} l = 1,2
\hspace{0.20cm}{\rm and}\hspace{0.20cm} j = 1,...,N_{\uparrow}-N_{\downarrow}+N_{2} \, .
\label{L2}
\end{equation} 
This expression refers to that given in Eq. (\ref{Lambda-jnl-ideal}) for $n=2$ and $l=1,2$.

The momentum values $q_j$ in Eqs. (\ref{Taps}), (\ref{Tap2}), and (\ref{L2}) are given by,
\begin{equation}
q_j = {2\pi\over L}\,I^{n}_j \hspace{0.20cm}{\rm for}\hspace{0.20cm} n = 1,2 \, ,
\label{q-j}
\end{equation} 
where the quantum numbers $I^{n}_j$ are either integers or half-odd integers according to the 
following boundary conditions \cite{Takahashi_71},
\begin{eqnarray}
I_j^{1} & = & 0,\pm 1,\pm 2,... \hspace{0.50cm}{\rm for}\hspace{0.15cm}N_{\uparrow}\hspace{0.15cm}{\rm odd} 
\nonumber \\
& = & \pm 1/2,\pm 3/2,\pm 5/2,... \hspace{0.50cm}{\rm for}\hspace{0.15cm}N_{\uparrow}\hspace{0.15cm}{\rm even} 
\nonumber \\
I_j^{2} & = & 0,\pm 1,\pm 2,... \hspace{0.50cm}{\rm for}\hspace{0.15cm}N_{2} = 1 \, .
\label{Ic-an}
\end{eqnarray}

In the thermodynamic limit, we often use continuous momentum variables $q$ that
replace the discrete $1$- and $2$-bands momenta $q_j$ such that $q_{j+1} - q_j=2\pi/L$.
They read $q\in[-k_{F\uparrow},k_{F\uparrow}]$ and $q \in [-(k_{F\uparrow}-k_{F\downarrow}),(k_{F\uparrow}-k_{F\downarrow})]$,
respectively. In that limit, the momenta $k_{F\downarrow}$ and $k_{F\uparrow}$ are given by,
\begin{equation}
k_{F\downarrow} = {\pi\over 2}(1-m)\, ;\hspace{0.20cm}k_{F\uparrow} = {\pi\over 2}(1+m)
\, ;\hspace{0.20cm}k_F = {\pi\over 2} \, ,
\label{kkk}
\end{equation}
for the spin-density interval, $m\in ]0,1[$ where $k_F = \lim_{m\rightarrow 0}k_{F\downarrow} = 
\lim_{m\rightarrow 0}k_{F\uparrow}$.

The energy dispersions $\varepsilon_1 (q)$ and $\varepsilon_{2} (q)$ that appear in the spectra of
the excited energy eigenstates are defined as,
\begin{equation}
\varepsilon_{1} (q) = {\bar{\varepsilon}_{1}} (\Lambda_1 (q)) 
\hspace{0.20cm}{\rm for}\hspace{0.20cm}q \in [-k_{F\uparrow},k_{F\uparrow}] 
\hspace{0.20cm}{\rm where}
\hspace{0.20cm}
{\bar{\varepsilon}_{1}} (\Lambda) = \int_{B}^{\Lambda}d\Lambda^{\prime}\,2J\eta_{1} (\Lambda^{\prime}) \, ,
\label{equA4}
\end{equation}
where $B$ is defined below and,
\begin{eqnarray}
\varepsilon_{2} (q) & = & 4\mu_B\,h + \varepsilon_{2}^0 (q)
\hspace{0.20cm}{\rm for}
\hspace{0.20cm}q \in [-(k_{F\uparrow} - k_{F\downarrow}),(k_{F\uparrow} - k_{F\downarrow})]
\hspace{0.20cm}{\rm where}
\nonumber \\
\varepsilon_{2}^0 (q) & = & {\bar{\varepsilon}}_{2}^0 (\Lambda_{2} (q)) \hspace{0.20cm}{\rm and}
\hspace{0.20cm}{\bar{\varepsilon}}_{2}^0 (\Lambda) = \int_{\infty}^{\Lambda}d\Lambda^{\prime}\,2J\eta_{2} (\Lambda^{\prime}) \, ,
\label{vare2}
\end{eqnarray}
respectively.

The corresponding $1$- and $2$-bands group velocities are given by,
\begin{equation}
v_1 (q) = {\partial\varepsilon_1 (q)\over\partial q} 
\hspace{0.20cm}{\rm and}\hspace{0.20cm}
v_{2} (q) = {\partial\varepsilon_{2} (q)\over\partial q} \, .
\label{equA4B}
\end{equation}

The distribution $2J\eta_{1} (\Lambda)$ appearing in Eq. (\ref{equA4}) is
the solution of the integral equation,
\begin{equation}
2J\eta_{1} (\Lambda) = {4J\,\Lambda\over (1 + \Lambda^2)^2} 
+ \int_{-B}^{B}d\Lambda^{\prime}\,G(\Lambda,\Lambda')\,2J\eta_{1} (\Lambda^{\prime}) \, .
\label{equA6}
\end{equation}
The kernel $G (\Lambda,\Lambda')$ appearing here is given by,
\begin{equation}
G(\Lambda,\Lambda') = - {1\over{2\pi}}\left({1\over{1+((\Lambda-\Lambda')/2)^2}}\right) \, .
\label{Gne1}
\end{equation} 

The values of the distribution $2J\eta_{2} (\Lambda)$ in Eq. (\ref{vare2}) are determined by
those of $2J\eta_{1} (\Lambda)$ as follows,
\begin{equation}
2J\eta_{2} (\Lambda) = {J\over 2}{\Lambda\over \left(1 + \left({\Lambda\over 2}\right)^2\right)^2} 
- \frac{1}{\pi} \int_{-B}^{B}d\Lambda^{\prime}\,
{2J\eta_{1} (\Lambda^{\prime})\over 1 +  \left(\Lambda - \Lambda^{\prime}\right)^2}
- \frac{1}{3\pi} \int_{-B}^{B}d\Lambda^{\prime}\,
{2J\eta_{1} (\Lambda^{\prime})\over 1 +  \left({\Lambda - \Lambda^{\prime}\over 3}\right)^2} \, ,
\label{eta2}
\end{equation}
where the distribution $2J\eta_{1} (\Lambda)$ is the solution of Eq.  (\ref{equA6}).

The rapidity distribution function $\Lambda_1 (q)$ where $q \in [-k_{F\uparrow},k_{F\uparrow}]$ in the argument of 
the auxiliary dispersion ${\bar{\varepsilon}_{1}}$ in Eq. (\ref{equA4}) is defined in terms of its $1$-band inverse 
function $q = q_1 (\Lambda)$ where $\Lambda \in [-\infty,\infty]$. The latter is defined by the equation,
\begin{equation}
q = q_1 (\Lambda) = 2\arctan (\Lambda) 
- \frac{1}{\pi} \int_{-B}^{B}d\Lambda^{\prime}\,2\pi\sigma (\Lambda^{\prime})\, \arctan \left({\Lambda -
\Lambda^{\prime}\over 2}\right) 
\hspace{0.20cm}{\rm for}\hspace{0.20cm} \Lambda \in [-\infty,\infty] \, .
\label{equA7}
\end{equation}

The rapidity distribution function $\Lambda_{2} (q)$ where
$q\in [-(k_{F\uparrow} - k_{F\downarrow}),(k_{F\uparrow} - k_{F\downarrow})]$
in the argument of the auxiliary dispersion ${\bar{\varepsilon}}_{2}^0$ in Eq. (\ref{vare2})
is also defined in terms of its $2$-band inverse function $q = q_2 (\Lambda)$ where 
$\Lambda \in [-\infty,\infty]$ as follows,
\begin{equation}
q_2 (\Lambda) = 2\arctan \left({\Lambda\over 2}\right) 
- \frac{1}{\pi} \int_{-B}^{B}d\Lambda^{\prime}\,2\pi\sigma (\Lambda^{\prime})
\arctan \left(\Lambda - \Lambda^{\prime}\right) 
- \frac{1}{\pi} \int_{-B}^{B}d\Lambda^{\prime}\,2\pi\sigma (\Lambda^{\prime})
\arctan \left({\Lambda - \Lambda^{\prime}\over 3}\right)
\hspace{0.20cm}{\rm for}\hspace{0.20cm} \Lambda \in [-\infty,\infty] \, .
\label{qtwoprime}
\end{equation}

The distribution $2\pi\sigma (\Lambda)$ in Eqs. (\ref{equA7}) and (\ref{qtwoprime}) is the solution of 
the integral equation,
\begin{equation}
2\pi\sigma (\Lambda) = {2\over 1 + \Lambda^2} 
+ \int_{-B}^{B}d\Lambda^{\prime}\,G(\Lambda,\Lambda')\,2\pi\sigma (\Lambda^{\prime}) \, ,
\label{equA10}
\end{equation}
whose kernel is given in Eq. (\ref{Gne1}). Such a distribution obeys the sum rule,
\begin{equation}
\frac{1}{\pi} \int_{-B}^{B}d\Lambda\,2\pi\sigma (\Lambda) = (1-m) \, .
\label{equA10B}
\end{equation}

The parameter $B = \Lambda_1 (k_{F\downarrow})$ appearing in the above equations has the limiting behaviors,
\begin{equation}
B = \Lambda_1 (k_{F\downarrow}) \hspace{0.20cm}{\rm with}\hspace{0.20cm}
\lim_{m\rightarrow 0} B = \infty\hspace{0.20cm}{\rm and}\hspace{0.20cm}
\lim_{m\rightarrow 1} B = 0 \, .
\label{QB-r0rs}
\end{equation}

Useful reference values of the rapidity functions $\Lambda_1 (q)$ and $\Lambda_{2} (q)$ are,
\begin{eqnarray}
\Lambda_1 (0) & = & 0\hspace{0.20cm}{\rm and}\hspace{0.20cm}
\Lambda_1 (\pm k_{F\uparrow}) = \pm\infty
\nonumber \\
\Lambda_{2} (0) & = & 0\hspace{0.20cm}{\rm and}\hspace{0.20cm}
\Lambda_2 (\pm (k_{F\uparrow} - k_{F\downarrow})) = \pm\infty \, .
\label{qtwoprimelimits}
\end{eqnarray}

The $1$-band energy dispersion,
\begin{equation}
\varepsilon_{1}^0 (q) = {\bar{\varepsilon}_{1}}^0 (\Lambda_1 (q))\hspace{0.20cm}{\rm where}
\hspace{0.20cm}{\bar{\varepsilon}_{1}}^0 (\Lambda) = \int_{\infty}^{\Lambda}d\Lambda^{\prime}\,2J\eta_{1} (\Lambda^{\prime}) \, ,
\label{equA11}
\end{equation}
whose zero-energy level is for $m>0$ shifted realtive to that of $\varepsilon_{1} (q)$ 
defines the spin density curve as follows,
\begin{equation}
h (m) = - {\varepsilon_{1}^0 (k_{F\downarrow})\over 2\mu_B}\vert_{m = 1 - 2k_{F\downarrow}/\pi} \in ]0,h_c[ \, .
\label{magcurve}
\end{equation}
Here $h_c=J/\mu_B$ is the critical field for fully polarized ferromagnetism achieved when $m\rightarrow 1$ and 
thus $k_{F\downarrow}\rightarrow 0$.

The $1$- and $2$-band energy dispersions $\varepsilon_{1} (q)$ and $\varepsilon_{2} (q)$, 
Eqs. (\ref{equA4}) and (\ref{vare2}), respectively, have limiting values,
\begin{eqnarray}
\varepsilon_{1} (0) & = & - W_{1}^p \, ; \hspace{0.50cm}
\varepsilon_{1} (\pm k_{F\downarrow}) = 0
\, ; \hspace{0.50cm}\varepsilon_{1} (\pm k_{F\uparrow}) = W_{1}^h = 2\mu_B\,h
\nonumber \\
\varepsilon_{2} (0) & = & W_{1}^h - W_{2} = 4\mu_B\,h - W_{2}
\, ; \hspace{0.50cm}\varepsilon_{2} (\pm (k_{F\uparrow} - k_{F\downarrow})) = 4\mu_B\,h \, ,
\label{vares2limits}
\end{eqnarray}
where $W_1 = W_{1}^p + W_1^h$,  $W_{1}^p$, and $W_1^h$ are the $1$-band
energy bandwidth, occupied Fermi sea energy bandwidth, and unoccupied sea energy bandwidth,
respectively, and $W_2$ is the $2$-band energy bandwidth. Such energy scales have
the limiting behaviors,
\begin{equation}
\lim_{m\rightarrow 0}W_{1} = W_1^p = {\pi\over2}\,J
\hspace{0.20cm}{\rm and}\hspace{0.20cm}
\lim_{m\rightarrow 0}W_{2} = 0 \, ,
\label{W2um0}
\end{equation}
and
\begin{equation}
\lim_{m\rightarrow 1}W_{1} = W_1^h = 2\mu_B\,h_c = 2J
\hspace{0.20cm}{\rm and}\hspace{0.20cm}
\lim_{m\rightarrow 1}W_{2} = J \, .
\label{W2um1}
\end{equation}

In the $m\rightarrow 0$ limit, the $2$-band does not exist in the ground state. In
that limit, it reduces to $q=0$ with $\varepsilon_{2} (0)=0$ for energy eigenstates for which $N_{2}=1$. 
In the same limit, the $1$-band energy dispersions and group velocity can be written as,
\begin{eqnarray}
\varepsilon_{1} (q) & = & \varepsilon_{1}^0 (q)  = - J {\pi\over 2} \cos q 
\hspace{0.50cm}{\rm and}\hspace{0.50cm}
v_{1} (q) = J{\pi\over 2} \sin q 
\nonumber \\
& & {\rm for}\hspace{0.20cm}q \in [-\pi/2,\pi/2]\hspace{0.20cm}{\rm and}\hspace{0.20cm}m\rightarrow 0 \, .
\label{varepsilonsulm0}
\end{eqnarray}

For $(1-m)\ll 1$, the $1$-band energy dispersions and group velocity, Eq. (\ref{equA4B}), behave as,
\begin{eqnarray}
\varepsilon_{1} (q) & = & - J\,(\cos q -1) + J\,(1-m)\sin q\,\arctan\left({1\over 2}\tan\left({q\over 2}\right)\right) 
\nonumber \\
\varepsilon_{1}^0 (q) & = & - J\,(\cos q + 1) + J\,(1-m)\sin q\,\arctan\left({1\over 2}\tan\left({q\over 2}\right)\right) 
\nonumber \\
v_{1} (q) & = & J\sin q  + J\,(1-m)\left\{{\sin q\over 1 + 3\cos^2\left({q\over 2}\right)}
+ \cos q\,\arctan\left({1\over 2}\tan\left({q\over 2}\right)\right)\right\}
\nonumber \\
& & {\rm for}\hspace{0.20cm}q \in \left[-{\pi\over 2}(1+m),{\pi\over 2}(1+m)\right]\hspace{0.20cm}{\rm and}\hspace{0.20cm}
(1-m)\ll 1 \, .
\label{varepsilonsulm1}
\end{eqnarray}

Also for $(1-m)\ll 1$, the behaviors of the $2$-band energy dispersion and group velocity are,
\begin{eqnarray}
\varepsilon_{2} (q) & = & 4J - {J\over 2}\,(1 + \cos q)  
+ {J\over 2}\,(1-m)\sin q\left\{\arctan\left(2\tan\left({q\over 2}\right)\right) 
+ \arctan\left({2\over 3}\tan\left({q\over 2}\right)\right)\right\}
\nonumber \\
\varepsilon_{2}^0 (q) & = & \varepsilon_{2} (q) - 4J
\nonumber \\
v_{2} (q) & = & {J\over 2}\sin q  + {J\over 2}\,(1-m)\sin q\left\{{1\over 1 + 3\sin^2\left({q\over 2}\right)}
+ {3\over 4 + 5\cos^2\left({q\over 2}\right)}\right\}
\nonumber \\
&& + {J\over 2}\,(1-m)\cos q\left\{\arctan\left(2\tan\left({q\over 2}\right)\right) 
+ \arctan\left({2\over 3}\tan\left({q\over 2}\right)\right)\right\}
\nonumber \\
& & {\rm for}\hspace{0.20cm}q \in [-\pi m,\pi m]\hspace{0.20cm}{\rm and}\hspace{0.20cm}(1-m)\ll 1 \, .
\label{varepsilons2ulm01}
\end{eqnarray}

From analysis of the above expressions, one finds that for $(1-m)\ll 1$ the
following equality holds,
\begin{equation}
v_{1} (k_{F\downarrow}) = v_{2} (k_{F\uparrow}-k_{F\downarrow}) = J{\pi \over 2} (1-m) \, .
\label{vvm1}
\end{equation}

The $1$-particle phase shifts play an important role in the dynamical properties. They are given by,
\begin{equation}
2\pi\,\Phi_{1,n}(q,q') = 2\pi\,\bar{\Phi }_{1,n} \left(\Lambda,\Lambda'\right) 
\hspace{0.20cm}{\rm where}\hspace{0.20cm}\Lambda = \Lambda_{1}(q)\hspace{0.20cm}
{\rm and}\hspace{0.20cm} \Lambda' = \Lambda_{n}(q') \, .
\label{Phi-barPhi}
\end{equation}
The rapidity phase shifts $2\pi\bar{\Phi }_{1,n}\left(\Lambda,\Lambda'\right)$ on the right-hand side of the above equality are functions of the rapidity-related variables $\Lambda$ for the $n$-bands. In the case of the excited energy eigenstates that span the subspace considered in Sec. \ref{SECIIIA}, the quantum number $n$ has the values $1$ and $2$. 
In units of $2\pi$, the corresponding $n=1,2$ rapidity phase shifts $2\pi\bar{\Phi }_{1,n}\left(\Lambda,\Lambda'\right)$ 
are defined by the following integral equations,
\begin{equation}
\bar{\Phi }_{1,1} (\Lambda,\Lambda') = {1\over \pi}\arctan\left({\Lambda-\Lambda'\over 2}\right) 
+ \int_{-B}^{B} d\Lambda''\,G (\Lambda,\Lambda'')\,{\bar{\Phi}}_{1,1} (\Lambda'',\Lambda') \, ,
\label{Phis1sn-m}
\end{equation}
and
\begin{equation}
\bar{\Phi }_{1,2}\left(\Lambda,\Lambda'\right) = 
{1\over \pi}\arctan(\Lambda-\Lambda') + {1\over \pi}\arctan\left({\Lambda-\Lambda'\over 3}\right) 
+ \int_{-B}^{B} d\Lambda''\,G (\Lambda,\Lambda'')\,{\bar{\Phi}}_{1,2} (\Lambda'',\Lambda') \, .
\label{Phis1s2-m}
\end{equation}
The kernel $G (r,r')$ in these equations is given in Eq. (\ref{Gne1}).

The phase shifts in units of $2\pi$ that appear in the expressions of the branch-line exponents,
Eqs. (\ref{expG+-}), (\ref{exps2pL}), (\ref{expsPM}), and (\ref{expszz}), are given by,
\begin{equation}
\Phi_{1,1}\left(\iota k_{F\downarrow},q\right) = \bar{\Phi }_{1,1} \left(\iota B,\Lambda_1 (q)\right) 
\hspace{0.20cm}{\rm and}\hspace{0.20cm}
\Phi_{1,2}\left(\iota k_{F\downarrow},q\right) = \bar{\Phi }_{1,2} \left(\iota B,\Lambda_{2} (q)\right) 
\hspace{0.20cm}{\rm where}\hspace{0.2cm}\iota = \pm 1 \, .
\label{Phis-all-qq}
\end{equation}

In the $m\rightarrow 0$ limit, the rapidity phase shift $\bar{\Phi }_{1,1} (\Lambda,\Lambda')$
(in units of $2\pi$) is given by,
\begin{eqnarray}
\bar{\Phi }_{1,1} (\Lambda,\Lambda') & = & {i\over 2\pi}\,
\ln\left({\Gamma \left({1\over 2} + i {(\Lambda - \Lambda')\over 4}\right)\Gamma \left(1 - i {(\Lambda - \Lambda')\over 4}\right)
\over\Gamma \left({1\over 2} - i {(\Lambda - \Lambda')\over 4}\right)\Gamma \left(1 + i {(\Lambda - \Lambda')\over 4}\right)}\right)
\hspace{0.2cm}{\rm for}\hspace{0.2cm}\Lambda \neq \iota\infty
\nonumber \\
& = & {\iota\over 2\sqrt{2}}
\hspace{0.2cm}{\rm for}\hspace{0.2cm}\Lambda= \iota\infty
\hspace{0.2cm}{\rm and}\hspace{0.2cm}\Lambda'\neq \iota\infty
\nonumber \\
& = &\iota\left({3\over 2\sqrt{2}} - 1\right)
\hspace{0.2cm}{\rm for}\hspace{0.2cm}\Lambda = \Lambda' = \iota\infty
\hspace{0.20cm}{\rm where}\hspace{0.2cm}\iota = \pm 1 \, ,
\label{Phis1sn-m0}
\end{eqnarray}
where $\Gamma (x)$ is the usual gamma function.

From the use of Eq. (\ref{Phis1sn-m0}), one finds that
in the $m\rightarrow 0$ limit for which $k_F = \pi/2$ the phase shift 
$\Phi_{1,1}\left(\iota \pi/2,q\right)=\lim_{m\rightarrow 0}\Phi_{1,1}\left(\iota k_{F\downarrow},q\right)$ 
reads (in units of $2\pi$),
\begin{eqnarray}
\Phi_{1,1}\left(\iota \pi/2,q\right) & = & {\iota\over 2\sqrt{2}}
\hspace{0.2cm}{\rm for}\hspace{0.2cm}q\neq \iota k_{F}
\nonumber \\
& = &\iota\left({3\over 2\sqrt{2}} - 1\right)
\hspace{0.2cm}{\rm for}\hspace{0.2cm}q= \iota k_{F}
\hspace{0.20cm}{\rm where}\hspace{0.2cm}\iota = \pm 1 \, .
\label{Phis-all-qq-0}
\end{eqnarray}

In the $m\rightarrow 0$ limit, the
phase shift $\Phi_{1,2}\left(\iota \pi/2,q\right)=\lim_{m\rightarrow 0}\Phi_{1,2}\left(\iota k_{F\downarrow},q\right)$ 
has in units of $2\pi$ the following value,
\begin{equation}
\lim_{m\rightarrow 0}\Phi_{1,2}\left(\iota \pi/2,q\right) = \Phi_{1,2}\left(\iota \pi/2,0\right) = {\iota\over\sqrt{2}} \, .
\label{Phis2-all-qq-0}
\end{equation}

In the opposite $m\rightarrow 1$ limit in which $k_{F\downarrow}\rightarrow 0$, 
the phase shifts $\Phi_{1,1}\left(\iota k_{F\downarrow},q\right)$
and $\Phi_{1,2}\left(\iota k_{F\downarrow},q\right)$ behave as,
\begin{eqnarray}
\lim_{m\rightarrow 1}\Phi_{1,1}(\iota k_{F\downarrow},q) & = & \Phi_{1,1}(0,q)  
= - {1\over\pi}\arctan\left({1\over 2}\tan\left({q\over 2}\right)\right) 
\nonumber \\
\lim_{m\rightarrow 1}\Phi_{1,2}(\iota k_{F\downarrow},q) & = & \Phi_{1,2}(0,q) = - {1\over\pi}\arctan\left(2\tan\left({q\over 2}\right)\right) 
- {1\over\pi}\arctan\left({2\over 3}\tan\left({q\over 2}\right)\right) \, . 
\label{PhiUinfm1qF}
\end{eqnarray}

The phase-shift related parameters $\xi_{1\,1}$ and $\xi_{1\,2}^{0}$ also appear 
in the expressions of the branch-line exponents,
Eqs. (\ref{expG+-}), (\ref{exps2pL}), (\ref{expsPM}), and (\ref{expszz}).
The $1$-band Fermi-points phase-shift parameter $\xi_{1\,1}$ is such that \cite{Carmelo_94},
\begin{eqnarray}
(\xi_{1\,1})^{\pm 1} & = & 1
+ \sum_{\iota=\pm 1} (\iota)^{1 \pm 1\over 2}\,\Phi_{1,1}\left(k_{F\downarrow},\iota k_{F\downarrow}\right) \, .
\label{x-aa}
\end{eqnarray}
It is also related to the phase shift $\Phi_{1,2} (k_{F\downarrow},q)$ in Eq. (\ref{Phis-all-qq})
as follows,
\begin{eqnarray}
\xi_{1\,1} & = & - \Phi_{1,2}(\pm k_{F\downarrow},(k_{F\uparrow}-k_{F\downarrow}))
\nonumber \\
& = & \Phi_{1,2}(\pm k_{F\downarrow},-(k_{F\uparrow}-k_{F\downarrow})) \, .
\label{xi1Phiss2}
\end{eqnarray}

From manipulations of the phase-shift integral equation, Eq. (\ref{Phis1sn-m}),
and of Eq. (\ref{x-aa}) one finds that the parameter $\xi_{1\,1}$ given by,
\begin{eqnarray}
\xi_{1\,1} & = & \xi_{1\,1} \left(B\right) \, .
\label{xi1all}
\end{eqnarray}
The function $\xi_{1\,1} (r)$ on the right-hand side of this equation at $\Lambda=B$ is the solution
of the integral equation,\begin{equation}
\xi_{1\,1} (\Lambda) = 1 + \int_{-B}^{B} d\Lambda'\,G (\Lambda,\Lambda')\,\xi_{1\,1} (\Lambda') \, ,
\label{xi-ss-qq}
\end{equation}
where the kernel $G (\Lambda,\Lambda')$ is given in Eq. (\ref{Gne1}).

The parameter $\xi_{1\,1}$ continuously increases upon increasing the spin density from $\xi_{1\,1}=1/\sqrt{2}$ as $m\rightarrow 0$ to
$\xi_{1\,1}=1$ for $m\rightarrow 1$, so that its limiting values are,
\begin{equation}
\lim_{m\rightarrow 0}\xi_{1\,1} = {1\over\sqrt{2}}\hspace{0.5cm}{\rm and}\hspace{0.5cm}
\lim_{m\rightarrow 1}\xi_{1\,1} = 1 \, .
\label{Limxiss}
\end{equation}

Finally, the parameter $\xi_{1\,2}^{0}$ that also appears in the momentum dependent
exponents is given by,
\begin{equation}
\xi_{1\,2}^{0} = 2\Phi_{1,2}(k_{F\downarrow},0) \, ,
\label{xis20}
\end{equation}
where the phase shift $\Phi_{1,2} (k_{F\downarrow},q)$ is defined in Eq. (\ref{Phis-all-qq}).
At $q=0$, it is such that $\Phi_{1,2}(\iota k_{F\downarrow},0)= \iota\,\Phi_{1,2}(k_{F\downarrow},0)$.
This justifies why $\iota\,\xi_{1\,2}^{0} = 2\Phi_{1,2}(\iota k_{F\downarrow},0)= \iota\,2\Phi_{1,2}(k_{F\downarrow},0)$
for $\iota=\pm 1$.

The parameter $\xi_{1\,2}^{0}$ continuously decreases upon increasing the spin density
from $\xi_{1\,2}^0=\sqrt{2}$ as $m\rightarrow 0$ to
$\xi_{1\,2}^0=0$ for $m\rightarrow 1$. Consitent,
it follows from Eqs. (\ref{Phis2-all-qq-0}) and (\ref{PhiUinfm1qF}) that, 
\begin{equation}
\lim_{m\rightarrow 0}\xi_{1\,2}^0 = \sqrt{2}\hspace{0.5cm}{\rm and}
\hspace{0.5cm}
\lim_{m\rightarrow 1}\xi_{1\,2}^0 = 0 \, .
\label{Limxis20}
\end{equation}


\end{document}